\documentclass[fleqn,usenatbib]{mnras}

\usepackage{amssymb}	
\usepackage{newtxtext,newtxmath}
\usepackage[T1]{fontenc}
\usepackage{graphicx}
\usepackage{comment}
\usepackage{multicol}
\usepackage{longtable}
\usepackage{array}
\usepackage{lscape}
\usepackage{lineno}
\usepackage{tablefootnote}
\usepackage{afterpage}
\usepackage{booktabs}
\usepackage{xcolor}
\usepackage{afterpage}

\DeclareRobustCommand{\VAN}[3]{#2}
\let\VANthebibliography\thebibliography
\def\thebibliography{\DeclareRobustCommand{\VAN}[3]{##3}\VANthebibliography}
\usepackage{adjustbox}

\usepackage{graphicx}	
\usepackage{amsmath}	
\usepackage{eso-pic}
 \usepackage{amssymb}	

\AddToShipoutPictureBG*{%
 \AtPageUpperLeft{%
   \hspace{0.75\paperwidth}%
  \raisebox{-3.5\baselineskip}{%
   \makebox[0pt][l]{\textnormal{DES-2023-0803}}
}}}%

\AddToShipoutPictureBG*{%
  \AtPageUpperLeft{%
    \hspace{0.75\paperwidth}%
    \raisebox{-4.5\baselineskip}{%
     \makebox[0pt][l]{\textnormal{FERMILAB-PUB-24-0452}}
}}}%

\newcommand{\orcid}[1]{\href{https://orcid.org/#1}{\includesvg[width=10pt]{orcid}}}
\title[TNOs and Centaurs from DESY6 measurements]{Year six photometric measurements of known Trans-Neptunian Objects and Centaurs by the Dark Energy Survey}

\author[F.~S.~Ferreira et. al.]{
\parbox{\textwidth}{
\Large
F.~S.~Ferreira$^{1,2}$\thanks{E-mail: felipheferreira@on.br},
J.~I.~B.~Camargo$^{1,2}$,
R.~Boufleur$^{1,2}$,
M.V.~Banda-Huarca$^{1,2}$,
A.~Pieres$^{2}$,
V.~F.~Peixoto$^{3,2}$,
M.~Assafin$^{3,2}$,
P.~H.~Bernardinelli$^{4}$,
H.W.~Lin$^{5}$,
F.~Braga-Ribas$^{6, 1, 2}$,
A.~Gomes-Junior$^{7,2}$,
R.~Vieira-Martins$^{1,2}$,
L.~N.~da Costa$^{2}$,
T. M. C. Abbott$^{8}$,
M. Aguena$^{2}$,
Sahar S. Allam$^{9}$,
O. Alves$^{5}$,
J. Annis$^{9}$,
D. Bacon$^{10}$,
D. Brooks$^{11}$,
D. L. Burke$^{12,13}$,
A. Carneiro Rosell$^{14,2}$,
J. Carretero$^{15}$,
S. Desai$^{16}$,
P. Doel$^{11}$,
S. Everett$^{17}$,
I. Ferrero$^{18}$,
J. Frieman$^{9,19}$,
J. García-Bellido$^{20}$,
M. Gatti$^{21}$
E. Gaztanaga$^{22,10,23}$,
G. Giannini$^{15,19}$,
D. Gruen$^{24}$,
R. A. Gruendl$^{25,26}$,
K. Herner$^{9}$,
S. R. Hinton$^{27,28}$,
D. L. Hollowood$^{29}$,
K. Honscheid$^{30}$,
D. J. James$^{31}$,
K. Kuehn$^{32,33}$,
S. Lee$^{17}$,
J. L. Marshall$^{34}$,
J. Mena-Fernández$^{35}$,
R. Miquel$^{36,15}$,
J. Myles$^{37}$,
A. Palmese$^{38}$,
A. A. Plazas Malagón$^{12,13}$,
M. E. S. Pereira$^{39}$,
S. Samuroff$^{40}$,
E. Sanchez$^{41}$,
D. Sanchez Cid$^{41}$,
I. Servila-Noarbe$^{41}$,
M. Smith$^{42}$,
E. Suchyta$^{43}$,
M. E. C. Swanson$^{25}$,
G. Tarle$^{5}$,
C. To$^{30}$,
D. L. Tucker$^{9}$,
J. de Vicente$^{41}$,
V. Vikram$^{44}$,
A. R. Walker$^{8}$,
N. Weaverdyck$^{45,46}$,
(DES Collaboration)
}
\vspace{0.5cm}
\\
\parbox{\textwidth}{
$^{1}$Observatório Nacional/MCTI, R. General José Cristino 77, CEP 20921-400 Rio de Janeiro - RJ, Brazil\\
$^{2}$Laboratório Interinstitucional de e-Astronomia - LIneA - and INCT do e-Universo. Av. Pastor Martin Luther King Jr, 126 Del Castilho, Nova América Offices, Torre 3000/sala 817 CEP: 20765-000, Brazil\\
$^{3}$Universidade Federal do Rio de Janeiro - Observatório do Valongo, Ladeira Pedro Antônio 43, CEP 20.080-090 Rio de Janeiro - RJ, Brazil\\
$^{4}$Department of Astronomy and the DIRAC Institute, University of Washington, 3910 15th Ave NE, Seattle, WA 98195, USA\\
$^{5}$Department of Physics, University of Michigan 450 Church Street Ann Arbor, MI 48109-1107, USA\\
$^{6}$Federal University of Technology - Paraná (UTFPR / DAFIS), Rua Sete de Setembro, 3165, CEP 80230-901, Curitiba, PR, Brazil\\
$^{7}$Federal University of Uberlândia (UFU), Physics Institute, Av. João Naves de Ávila 2121, Uberlândia, MG 38408-100, Brazil\\
$^{8}$Cerro Tololo Inter-American Observatory, NSF’s National Optical-Infrared Astronomy Research Laboratory, Casilla 603, La Serena, Chile\\
$^{9}$Fermi National Accelerator Laboratory, P. O. Box 500, Batavia, IL 60510, USA\\
$^{10}$Institute of Cosmology and Gravitation, University of Portsmouth, Portsmouth, PO1 3FX, UK\\
$^{11}$Department of Physics \& Astronomy, University College London, Gower Street, London, WC1E 6BT, UK\\
$^{12}$Kavli Institute for Particle Astrophysics \& Cosmology, P. O. Box 2450, Stanford University, Stanford, CA 94305, USA\\
$^{13}$SLAC National Accelerator Laboratory, Menlo Park, CA 94025, USA\\
$^{14}$Instituto de Astrofisica de Canarias, E-38205 La Laguna, Tenerife, Spain\\
$^{15}$Institut de Física d’Altes Energies (IFAE), The Barcelona Institute of Science and Technology, Campus UAB, 08193 Bellaterra (Barcelona) Spain\\
$^{16}$Department of Physics, IIT Hyderabad, Kandi, Telangana 502285, India\\
$^{17}$Jet Propulsion Laboratory, California Institute of Technology, 4800 Oak Grove Dr., Pasadena, CA 91109, USA\\
$^{18}$Institute of Theoretical Astrophysics, University of Oslo. P.O. Box 1029 Blindern, NO-0315 Oslo, Norway\\
$^{19}$Kavli Institute for Cosmological Physics, University of Chicago, Chicago, IL 60637, USA\\
$^{20}$Instituto de Fisica Teorica UAM/CSIC, Universidad Autonoma de Madrid, 28049 Madrid, Spain\\
$^{21}$Department of Physics and Astronomy, University of Pennsylvania, Philadelphia, PA 19104, USA\\
$^{22}$Institut d'Estudis Espacials de Catalunya (IEEC), 08034 Barcelona, Spain\\
$^{23}$Institute of Space Sciences (ICE, CSIC), Campus UAB, Carrer de Can Magrans, s/n, 08193 Barcelona, Spain\\
$^{24}$University Observatory, Faculty of Physics, Ludwig-Maximilians-Universität, Scheinerstr. 1, 81679 Munich, Germany\\
$^{25}$Center for Astrophysical Surveys, National Center for Supercomputing Applications, 1205 West Clark St., Urbana, IL 61801, USA\\
$^{26}$Department of Astronomy, University of Illinois at Urbana-Champaign, 1002 W. Green Street, Urbana, IL 61801, USA\\
$^{27}$School of Mathematics and Physics, University of Queensland, Brisbane, QLD 4072, Australia\\
$^{28}$Department of Physics, The Ohio State University, Columbus, OH 43210, USA\\
$^{29}$Santa Cruz Institute for Particle Physics, Santa Cruz, CA 95064, USA\\
$^{30}$Center for Cosmology and Astro-Particle Physics, The Ohio State University, Columbus, OH 43210, USA\\
$^{31}$Center for Astrophysics | Harvard \& Smithsonian, 60 Garden Street, Cambridge, MA 02138, USA\\
$^{32}$Australian Astronomical Optics, Macquarie University, North Ryde, NSW 2113, Australia\\
$^{33}$Lowell Observatory, 1400 Mars Hill Rd, Flagstaff, AZ 86001, USA\\
$^{34}$George P. and Cynthia Woods Mitchell Institute for Fundamental Physics and Astronomy, and Department of Physics and Astronomy, Texas A \& M University, College Station, TX 77843, USA\\
$^{35}$LPSC Grenoble - 53, Avenue des Martyrs 38026 Grenoble, France\\
$^{36}$Institució Catalana de Recerca i Estudis Avançats, E-08010 Barcelona, Spain\\
$^{37}$Department of Astrophysical Sciences, Princeton University, Peyton Hall, Princeton, NJ 08544, USA\\
$^{38}$Department of Physics, Carnegie Mellon University, Pittsburgh, Pennsylvania 15312, USA\\
$^{39}$Hamburger Sternwarte, Universität Hamburg, Gojenbergsweg 112, 21029 Hamburg, Germany\\
$^{40}$Department of Physics, Northeastern University, Boston, MA 02115, USA\\
$^{41}$Centro de Investigaciones Energéticas, Medioambientales y Tecnológicas (CIEMAT), Madrid, Spain\\
$^{42}$School of Physics and Astronomy, University of Southampton, Southampton, SO17 1BJ, UK\\
$^{43}$Computer Science and Mathematics Division, Oak Ridge National Laboratory, Oak Ridge, TN 37831\\
$^{44}$Argonne National Laboratory, 9700 S Cass Ave, Lemont, IL60439, USA\\
$^{45}$Department of Astronomy, University of California, Berkeley, 501 Campbell Hall, Berkeley, CA 94720, USA\\
$^{46}$Lawrence Berkeley National Laboratory, 1 Cyclotron Road, Berkeley, CA 94720, USA\\
}
}

\date{Accepted 2025. Received 2025; in original form 2024}

\pubyear{\the\year{}}

\begin{document}
\label{firstpage}
\pagerange{\pageref{firstpage}--\pageref{lastpage}}
\maketitle
\begin{abstract}
We identified known Trans-Neptunian Objects (TNOs) and Centaurs in the complete Dark Energy Survey (DES) year six catalog (DES Y6) through the Sky Body Tracker (SkyBoT) tool. We classified our dataset of 144 objects into a widely used 4-class taxonomic system of TNOs. No such previous classification was available in the literature for most of these objects. From absolute magnitudes and average albedos, an estimation of the diameters of all these objects is obtained. Correlations involving colours, orbital parameters, dynamical classes and sizes are also discussed. In particular, our largest reddest object has a diameter of $390^{+68}_{-53}$ km and our largest cold classical, $255^{+19}_{-17}$ km. Also, a weak correlation between colour and inclination is found within the population of resonant TNOs in addition to weak correlations between colour and phase slope in different bands. 
\end{abstract}




\section{Introduction}

Small bodies are seen as fundamental pieces to understand the history and evolution of the Solar System. Trans-Neptunian Objects (TNOs, located beyond the orbit of Neptune at $\sim 30$ AU from the Sun), in particular, are regarded as collisional and dynamical remnants of an evolved planetesimal disk and, due to their large heliocentric distances, they are also relatively unaltered relics of the Solar System formation, providing important constraints of how this formation process worked \citep{2020tnss.book...25M}. Thus, these icy bodies can provide essential information about the evolution of the early solar nebula and hints about other planetary systems around young stars as well \citep{barucci2008composition}. 

Centaurs are objects characterized by highly chaotic orbit, with a dynamical lifetime from less than 1 to about 100 Myr \citep{ tiscareno2003dynamics} and  with their dynamical evolution mostly perturbed by the Giant Planets. It is widely accepted  that they are a transient population between the distant, cold Kuiper Belt Objects (KBOs) and the rapidly sublimating Jupiter family comets (JFCs) in the hotter region \citep{2019ApJ...880...71L}. In fact, also according to \citet{2002AJ....123.1050F}, Centaurs can be considered as brighter proxies of the more distant TNO population.

Investigations of the surface properties of TNOs and Centaurs allow us to obtain important information about the chemical processes of the outer proto-planetary disk formation as reported by, for instance, \citet{2010A&A...510A..53P,2014ApJ...793L...2L,2019AJ....157...94M, 2022PSJ.....3....9B}, helping us to develop theories about the birth and growth of our planetary system. 

Because of their faint magnitudes, however, these objects are difficult to observe. Spectroscopy is a powerful tool to investigate TNOs surface composition, but only the brightest ones are suited for that technique. Even with the world’s largest telescopes, most of the TNOs and Centaurs are too faint for spectroscopic observations \citep{barucci2005trans} preventing, at least in the immediate future, compositional analysis with sufficient
quality from ground-based facilities (see \citet{2021jwst.prop.2418P}).
On the other hand, multiband photometry provides access to all those pieces of information, faint objects included, although much less accurately. 

Observations from photometric surveys have revealed a wide range of colours that, even thought cannot be used to derive precise surface composition, are suitable to segregate objects into groups with colour similarities \citep[see for instance][]{1998Natur.392...49T, 2003A&A...410L..29P, 2012A&A...546A..86P,  2015A&A...577A..35P,
2003ApJ...599L..49T, 2016AJ....152..210T, 2005AJ....130.1291B, barucci2008composition, 2008ssbn.book..181F, 2012ApJ...749...33F, 2015ApJ...804...31F, 2020AJ....160...46N}. 

The wide colour diversity of TNOs may be a result of physical processes acting on their surfaces and/or possible differences in composition \citep{2008ssbn.book...91D}. These icy bodies could be seen as a compilation of remnants transferred from different locations of the solar system during planetary migration. The dynamically cold objects are thought to have been formed in situ and have preserved their original orbits after the migration of Neptune, whereas many of the dynamically hot objects were perturbed/scattered by Neptune. Thus, their current locations carry records of years of interactions \citep{2012AREPS..40..467B, 2022ApJ...937L..22C}. Also, the current size distribution of these objects is relevant to understand scenarios involving accretion and collisional erosion processes they have undergone \citep{2008ssbn.book...71P}. It is important to note that the definition of cold classicals takes into consideration ranges of semi-major axis and inclination. In particular, the concept of free-inclination \citep[$i_{\rm free}$, see][]{2022ApJS..259...54H, 2021ARA&A..59..203G, 2022DPS....5410507K} may be adopted. We are using, instead, osculating elements only here.


Diameter and albedo are also important parameters to infer the composition and internal structure of TNOs. Given that TNOs are not spatially resolved in general, the determination of their sizes can not be measured directly \citep{barucci2004surface}. The majority of diameters and albedos of small bodies are obtained from observations of these objects in thermal and optical wavelengths \citep[see, for instance,]   []{2010A&A...518L.146M,2017ApJ...839L..15G}. Less frequently, but with an increasing number of measurements \citep[see, for instance][]{2020tnss.book..413O}{}{}, accurate sizes/shapes can be obtained from stellar occultations \citep[e.g.][]{2011Natur.478..493S,2012Natur.491..566O,2014Natur.508...72B,2017Natur.550..219O,2023Natur.614..239M}{}{}. Then, from absolute magnitudes, their albedos can be inferred (see discussion in Sect. 4).


In general, as a consequence of their large distances from the Sun, TNOs have surface temperatures that are low enough to retain, in addition to water ice, highly volatile ices such as carbon dioxide ($CO_2$), methane ($CH_4$), carbon monoxide ($CO$) and nitrogen ($N_2$)   \citep{2011Natur.478..493S,2012Natur.491..566O,2014Natur.508...72B,2017Natur.550..219O,2021PSJ.....2...10F}. Their surfaces also show signs of complex organic elements known as tholins. These materials are formed through processes of radiation of ultraviolet and cosmic rays, causing a reddish colouring in these distant bodies \citep{1979Natur.277..102S, 1993Icar..103..290K, 2021PSJ.....2...10F}.  It is worth mentioning, in this context, that volatile sublimation and space weathering at smaller distances from the Sun favours a positive correlation between albedo and size/perihelion distance \citep{2020tnss.book..153M}.

In this context, wide area, multiband deep sky surveys like DES 
provide invaluable information to a variety of Solar Systems studies from their accurate photometric measurements. Although originally not aimed at Solar System studies, DES has made key contributions to the photometric and astrometric/dynamical analysis of small bodies with heliocentric distances ranging from the inner Solar System to its outskirts \citep{2018AJ....156...81B, 2018AJ....156..273K,2019AJ....157..120B, 2020ApJS..247...32B,2020AJ....159..133K, 2020PSJ.....1...28B,2021PSJ.....2...59N, 2021PASP..133a4501H,  2022PSJ.....3..269P, 2023PSJ.....4..115B,2024MNRAS.527.6495C}, including the discoveries of a large scattered disk object \citep[2014 UZ224 - ][]{2017ApJ...839L..15G}, of a messenger from the hypothesized Oort cloud \citep[C/2014 UN271 -][]{2021ApJ...921L..37B}, of four L4 Neptunian Trojans in which one of them (2013 VX30, classified as a scattered disk object by SkyBoT, described later in the text) has shown a significantly similar colour to that of the reddest TNOs \citep{ 2016AJ....151...39G, 2019Icar..321..426L}, in addition to the compilation of one of the two largest catalogs of TNOs \citep{, 2022ApJS..258...41B} as well as the largest TNO colour and light curve catalog \citep{2023ApJS..269...18B}.

Following the series of these contributions, this paper concentrates on photometric measurements of TNOs and Centaurs. Sect.~3, briefly describes the identification of known TNOs and Centaurs in the DES observations. Absolute magnitudes, in Sect.~4, are determined from the DES photometry and used to obtain diameters and a taxonomic classification.

Different approaches to determine taxonomic classes for TNOs and Centaurs have been proposed by many authors. \citet{2025NatAs...9..230P} revealed the existence of three main spectral groups of TNOs corresponding to different surface compositions from spectroscopic observations from the James Webb Space Telescope. \citet{2012ApJ...749...33F} and \citet{2015ApJ...804...31F} found that the surfaces of dynamically excited small TNOs and Centaurs are well describe by two mixture models of two components. Later, \citet{2023PSJ.....4...80F} presented a new interpretation of KBO colours, again in two populations, that occupy different optical-NIR colour regions following the reddening curve, the curve of constant spectral slope through colour-space.  \citet{2013Icar..222..307D}, using modiﬁed K-means clustering technique applied to multiband optical photometry and optical albedos, found 10 surface types. \citet{2017AJ....154..101P} found three distinct TNO surface types that result from  g-, r-, and z-band photometry. \citet{2005AJ....130.1291B} and \citet{2008ssbn.book..181F} use a clustering process \citep[G-mode, see][]{1977CG......3...85C,2000Icar..146..204F} to justify and present four different classes. We adopt in this work the classification proposed by \citet{2005AJ....130.1291B} and \citet{2008ssbn.book..181F} (see Sect.~4), although an exact number of classes is not a consensus in the literature. A discussion about the
taxonomic scheme adopted here and correlations involving colours, orbital parameters, dynamical classes and sizes is also presented. 

\section{The Dark Energy Survey - DES}

DES is an international collaborative effort designed to investigate the nature of the dark energy, responsible to accelerate the expansion of our universe. Over six years of observations (2013 – 2019), the collaboration used night and half-night of observations equivalent to 575 full nights which imaged 5\,000 square degrees of the celestial southern hemisphere in five optical filters, \textit{grizY}. Besides the explored galaxies, the survey allowed the observation of more than 300\,000 objects within the Solar System (see Table~\ref{tab:numobjects}) and has discovered hundreds of TNOs \citep{2016MNRAS.460.1270D, 2022ApJS..258...41B}, being an astrometric and photometric treasure to the study of small bodies.

The observations within the survey were made with an extremely sensitive 570-Megapixel digital camera, The Dark Energy Camera - DECam \citep{2015AJ....150..150F}, a mosaic of 62 2k x 4k red-sensitive CCDs with a hexagonal pattern located on the prime focus of the Blanco 4-meters Telescope at Cerro Tololo Inter-American Observatory (CTIO), in Chile. The camera has an effective field of view of $\sim$3.0 square degrees and the CCDs have 15 $\mu m$  x  15 $\mu m$ pixels with a plate scale of 0.263 per pixel. The DES wide-area survey images have a nominal $10\sigma$ limiting magnitudes of  $g = 23.57$, $r = 23.34$, $i = 22.78$, $z =  22.10$ and $Y = 20.69$, with the final coadded depth being roughly one magnitude deeper \citep{2018PASP..130g4501M}.

\section{Data and Tools}

The basic pieces of information used in this paper are derived from photometric measurements from DES for known TNOs and Centaurs.

A first step in this study is to locate which single epoch CCDs, and respective catalogs, contain observations of a TNO or a Centaur. The next step is then, among all astrometric and photometric measurements from a given selected CCD/catalog, retrieve those of the target. Procedures and tools to accomplish these tasks are detailed below.

\subsection{Observational data}

Following the procedures adopted by \citet{2019AJ....157..120B}, we identified all CCD frames, acquired during the six years of observations with the DECam within DES, that could contain the image(s) of one or more known small Solar System object(s). We then searched, for each CCD obtained from that identification, the corresponding DES catalog containing astrometric and photometric measurements. These catalogs are a compilation under the name Year-6 (Y6) with full name \textsc{Y6A1\_FINALCUT\_OBJECTS} \citep[refer to Sects.~1 and 4.8 in][for more details about the finalcut catalog]{2018PASP..130g4501M}. The magnitude zero-points of each CCD were determined by the collaboration through the Forward Global Calibration Method \citep{2018AJ....155...41B}. Relevant pieces of information to
this work were the CCD unique identification, bandpass, flux measurements and their respective uncertainties, ICRS positions, exposure time and date and time of observation. 

The download of CCD frames from the DES database was necessary for an independent measurement of the FWHM of the target objects. One important point to be mentioned is that the image header key OPENSHUT is available from image headers only and not from DES tables accessible via the \textsf{easyaccess} tool \citep{2018ascl.soft12008C}. The OPENSHUT key contains the accurate UTC time when the shutter opens, a fundamental quantity when determining the positions of small bodies and fast varying photometry.

In fact, an image key header that contains time and is readily accessible via \textsf{easyaccess} is DATE-OBS. It differs from the respective OPENSHUT value by a fraction of a second in most cases.

The identification of a Solar System object in a given DES catalog (see also Sect. 3.4) requires the knowledge of the mean observation time (that is, the time shown by OPENSHUT plus half of the exposure time). Due the large FOV (field of view) of DECam, the shutter takes around 1s to cover (or uncover) the FOV. A small correction to the mean observation time due to this fact is applied \citep[see][]{2015AJ....150..150F,2019AJ....157..120B}.

\subsection{Object Search in the DES database}

The identification of all known small bodies in the DES images/catalogs from the six years of observations is done through the Sky Body Tracker – SkyBoT \citep{2006ASPC..351..367B}. This project provides a virtual observatory developed by the \textit{Institut de Mécanique Céleste et de Calcul des Éphémérides} (IMCCE, Paris Observatory), which seeks and identifies every Solar System object in an astronomical image by knowing the pointing, observing date and site and FOV. 

A cone-search with SkyBoT delivers an output in text or VOTable format containing, among others, J2000 equatorial coordinates, name and number (if numbered), estimated V magnitude and dynamical classification. The number and class of known objects identified so far in DES are shown in Table~\ref{tab:numobjects}. The dynamical classification is that provided by SkyBoT.\footnote{The dynamical classification used by the SkyBoT can be readily accessed from \url{https://ssp.imcce.fr/webservices/skybot/}.} It should be also stressed  there are no overlaps between the different dynamical
classifications. For instance, an object classified as Classical, will not appear in any other dynamical group (resonants, SDO, etc). In particular, we have no objects from the Haumea family.

 The dynamical parameters of asteroids in SkyBoT come from the ASTORB (The Asteroidal Orbital Elements, Lowell Observatory). If an object is not in ASTORB or if SkyBoT was not updated when an object was included in ASTORB, then this object is unknown to SkyBoT. To avoid frequent and time consuming operations related to the queries on the DES database and on SkyBoT, the number of Solar System objects presented here is that obtained on 2022/MAY/31, close to the date this work started. In this context, the current numbers of known objects today are certainly larger than those shown in Table~\ref{tab:numobjects}.

\begin{table}
\centering
\begin{tabular}{lr}
\hline
Dyn. Class        & Number  \\
\hline
Main Belt         & $366\,706$ \\
Hungaria          & $12\,199$\\
NEAs              & $9\,400$ \\
Mars-Crossers     & $8\,213$ \\
TNOs              & $758$  \\
Centaurs          & $129$    \\
Trojans (Jupiter) & $1\,684$   \\
Comets            & $423$   \\
\hline
\textbf{TOTAL}    & $399\,512$ \\
\hline
\end{tabular}
\caption{Number of objects that could have DES images identified by SkyBoT. These figures may change each time the SkyBoT database
is updated.}
\label{tab:numobjects}
\end{table}

\subsection{TNOs and Centaurs}

The number of known such objects found by SkyBoT within the DES footprint amounted to 758 TNOs and 129 Centaurs. This, however, does not mean that all of them were effectively detected or identified. In fact, objects whose magnitudes are too faint to have been detected by a single-epoch image or whose positional uncertainties are too large do not appear in this work. 

We select in a first step, for photometric studies, all TNOs/Centaurs with at least three measurements in each {\it griz} filter. Then, the absolute magnitudes in each filter from Eq.~\ref{eq:magred} were calculated as a way to attenuate systematic issues in colours from rotational effects (see discussion later in the text). With the values for the absolute magnitudes, it is possible to obtain colour indices for these bodies from which a taxonomic classification can be inferred.

\subsection{Photometric measurements}

In this work, photometric data comes exclusively from the single epoch DES Year 6 (Y6, last season of survey) catalogs. A description of image processing and determination of magnitudes is given by \citet{2018PASP..130g4501M}. It is also important to emphasize that these measurements result from the same instrument and went through the same and well-defined reduction process \citep[see][]{2018PASP..130g4501M}.

The identification of an object present in the catalog is made by comparing the observed positions from DES with those predicted by the small body and planetary ephemerides from the JPL. The calculated (ephemeris) positions were obtained with the help of the SPICE/NAIF system \citep[][]{1996P&SS...44...65A,2018P&SS..150....9A}.

Our DECam single exposures were processed by DES Data Management, in a set known as "Final Cut", since the data is the final selection to coadd Y6. The photometry in this work is the PSF photometry (the only exception is mentioned below) resulting from SourceExtractor in single epoch exposures \citep[see section 3 in][DR1 paper]{2019PhRvD.100b3541A}, with improved calibration \citep[see section 3.1 in][]{2021ApJS..255...20A} over "First Cut" \citep[see again][]{2018PASP..130g4501M}. Single exposures are typically 90 s (wide area survey) for griz and 45 s for Y (DR2 paper). More information as the typical seeing, limiting magnitude and sky brightness can be accessed in table 1 of the DR2 paper, as well as details about the photometry in single exposures.

In addition to wide area survey, where most of our data comes from, this work also profits from observations from the supernova program, where longer (up to 400 sec) exposure times were used \citep[see][]{2015AJ....150..172K, 2016SPIE.9910E..1DD}. 

The impact of trailing from long exposure times may be generally estimated by measuring the FWHM of targets in different exposure times. This can be seen in Tables~\ref{tab:centseeing} and \ref{tab:kboseeing}. It is interesting to note in those tables that the values of the FWHM are similar regardless the exposure time.


One object, however, required special attention: the Centaur 2009 HH36. It has an orbital period of 45.58 year and combines the shortest orbital period with the longest exposure time along our objects (Table~\ref{tab:taxonclass}). In fact, a few hundreds of seconds of exposure time for this centaur implies a trail with size comparable to the DECam pixel scale \citep[$0.263^{"}$/pixel, see][]{2021ApJS..255...20A}.  All other objects were checked and their differential displacements with respect to the stars represent just a small fraction of the pixel size during their respective exposure times in each filter.

In order to avoid loss of flux in the case of 2009 HH36 due to the use of PSF photometry, we used the results from MAG\_APER\_8, that is, flux measurement in a circular aperture of 5.84 arcseconds in diameter \citep[see ][] {2018ApJS..239...18A}. That aperture is large enough to contain all the flux from the object. All images that contributed to a measurement of the magnitude of 2009 HH36 were visually inspected to search for the presence of contaminating flux sources and none was found.

\begin{table}
\centering
\begin{tabular}{cccccc}
\hline
& \multicolumn{4}{c}{Centaurs} & \\
Exposure time (s) & \textit{g} & \textit{r} & \textit{i} & \textit{z} & \#measurements\\
\cline{2-5} \\
&  \multicolumn{4}{c}{Median FWHM} & \\
\hline
$t \leq 90$ & 1.16 & 1.01 & 0.93 & 0.87 & 432 \\
$t > 90$ & 1.07 & 1.17 & 1.15 & 0.92 & 254 \\
\hline
\end{tabular}
\caption{Column 1: exposure time interval; columns 2 to 5: median FWHM as obtained from all target observations with exposure time intervals given by column 1. FWHM were determined with the astrometric package PRAIA \citep{2023P&SS..23805801A}; column 6: number of measurements used in the determination of the median FWHM.}
\label{tab:centseeing}
\end{table}

\begin{table}
\centering
\begin{tabular}{cccccc}
\hline
& \multicolumn{4}{c}{TNOs} & \\
Exposure time (s) & \textit{g} & \textit{r} & \textit{i} & \textit{z} & \#measurements\\
\cline{2-5} \\
&  \multicolumn{4}{c}{Median FWHM} & \\
\hline
$t \leq 90$ & 1.10 & 0.98 & 0.91 & 0.90 & 7110 \\
$90< t\leq 300$ & 1.21 & 1.10 & 1.02 & 0.99 & 1844 \\
$t > 300$ & 1.01 & 1.03 & 0.99 & 0.91 & 1607 \\
\hline
\end{tabular}
\caption{Same as~\ref{tab:centseeing}}.
\label{tab:kboseeing}
\end{table}

\section{Results} \label{sec:colours}

As previously mentioned, an exact number of taxonomic classes for TNOs and Centaurs is not a consensus in the literature and this work does not aim at finding it. The one we adopt here is that by \citet{2005AJ....130.1291B, 2008ssbn.book..181F}. Further discussion on this is provided later in the text.

\subsection{Colours from Absolute Magnitudes}

Given the observational cadence of the DES, it is difficult to obtain the rotational light curve of all objects. As a consequence, colours are not reliably obtained from the direct combination of multiband photometric measurements because it is not possible to assure that observations in different bands are taken in the same (or nearly the same) rotational phase. A simple approach to overcome this problem involves the determination of the absolute magnitude, the solar phase slope and the consideration, as described in Subsect.~\ref{subsec:agabeta}, of the amplitude of an unknown rotational light curve. 


The absolute magnitude $H$ is given by Eq.~\ref{eq:mgab} \citep[e.g.,][]{2003MNRAS.345..981H}.

\begin{equation}
H = m - 5\log(R\Delta) +2.5\log \phi(\alpha),
\label{eq:mgab}
\end{equation}

\noindent where $m$ is the apparent magnitude at the date of observation, $R$ is the heliocentric distance of the object, $\Delta$ is the observer distance of the object and $\phi(\alpha)$ is the phase function such that, for $\alpha=0^\circ$ (opposition), $\phi(0)=1$. $H$ is the observed magnitude with $\alpha=0^\circ$ and $R=\Delta=1~AU$. 

Such a determination is straightforward from least-squares, for objects with three or more observations, from the more simple Eq.~\ref{eq:magred}.

\begin{equation}
H_{f}+\alpha\beta_{f}=m_{f}-5\log({\rm R}\Delta)
\label{eq:magred}  
\end{equation}

\noindent where $m_{f}$ is the observed magnitude at a given filter \textit{f}, $\alpha$ is the phase angle, $\beta_{f}$ is the slope coefficient and $H_{f}$ is the absolute magnitude at the given filter. The only unknowns are, therefore, $H_{f}$ and $\beta_{f}$. The second member of Eq.~\ref{eq:magred} is known as reduced magnitude (object's magnitude at a distance of 1 AU from the observer and the Sun for a given phase angle). Phase angles and distances (heliocentric and from the observer) are directly determined from small body ephemerides provided by the JPL.

A linear function was preferred instead of more complex models such as phase curve fit, e.g., \textit{H-G} model \citep{1989aste.conf..524B}, $H-G_{1}-G_{2}$ \citep{2010Icar..209..542M} or $H-G_{12}$ \citep{2010Icar..209..542M, 2016P&SS..123..117P} to describe the behaviour of the reduced magnitude as a function of  the phase angle. Among others, the suitability of a simpler function are due because of the small range of phase angles that distant objects are seen from the ground. Other authors
\citep[e.g.][]{2002AJ....124.1757S,2012ApJ...749...10O,2018MNRAS.481.1848A} also opted for a linear model when studying more distant objects \citep[see also discussion in][]{2009A&A...494..693S}.

\subsubsection{Diameter and albedo}

Here, it is interesting to estimate the relative error between albedo and diameter, since it quantifies the importance of both accurate magnitudes and diameters, such as those obtained from DES and stellar occultations, respectively. In fact, the latter can determine, with kilometer accuracy, shape and dimension projected onto the sky plane of the occulting body \citep[see, for instance,][]{2020tnss.book..413O, 2023MNRAS.526.6193A, 2024MNRAS.527.3624P}. 

It can be demonstrated that \citep[e.g.][]{1997Icar..126..450H, 1916ApJ....43..173R}\footnote{See also \url{https://mathscinotes.com/wp-content/uploads/2016/08/Jackpot.pdf}, section 4.2.}

\begin{equation}
D=\frac{2~{\rm AU}\times10^{0.2(m_{\sun}-H)}}{\sqrt{\rho}},\label{eq:Diam}
\end{equation}

where $D$ is the equivalent diameter\footnote{Equivalent diameter is the diameter of a sphere with the same surface area of the object.} of the body in km, $m_{\sun}$ is the apparent magnitude of the Sun in a given band, $H$ is the absolute magnitude of the body in the same band, $\rho$ is the geometric albedo also in the same band and AU is the astronomical unit in km (149\,597\,870.7 km). If we consider the band \textit{V} and the apparent magnitude of the Sun in it ($m_{\sun}=-26.77$), we arrive at the well-known relation ($D$ in km).

\begin{equation}
D=\frac{1324.206\times10^{-0.2 H_{V}}}{\sqrt{\rho_{V}}}.\label{eq:Dro}
\end{equation}

A simple hypothesis about Eq.~\ref{eq:Dro} gives us very relevant information. The assumption is that ${H_{V}}$ is accurate enough so that $\sigma_{H_{V}}$ can be neglected when compared to $\sigma_{\rho{_V}}$ and $\sigma_D$. In this case, the relationship between the uncertainty in the albedo and that in the diameter is given by

\begin{equation}
\frac{\sigma_{\rho{_V}}}{\rho_V}=2\frac{\sigma_D}{D}.\label{eq:err_relat}
\end{equation}

In other words, a given relative uncertainty in the diameter leads to twice that value to the relative uncertainty in albedo.


\subsection{$H_f$, $\beta_f$ and uncertainties}\label{subsec:agabeta}

In order to obtain values for absolute magnitudes and slopes, in addition to realistic uncertainties to them, effects due to an unknown rotation has to be taken into consideration. In this context, we adopted a similar procedure to the one describe in works
like \citet{2018MNRAS.481.1848A} and \citet{2022PSJ.....3..269P}. A brief
description of such a procedure is given next:
\begin{itemize}
     \item For each object, and at a given filter, we determine the respective reduced magnitudes (see Eq.~\ref{eq:magred}).
     \item To each reduced magnitude, the value of the function $y(\phi)=A\times sin(\phi)$ is determined for $\phi$ randomly chosen in the interval [0:$2\pi$[, with $A=0.15$ \citep[see, for instance,][]{2009A&A...505.1283D, 2006AJ....131.2314L}, and added to the reduced magnitude (that is, a displacement along the $y$ axis). In this paper, therefore, we are adopting a sine wave as a model to the shape of an unknown rotational light curve to the objects. The most frequent values in such a distribution are clearly those close to the maximum/minimum of the sine wave \citep[see, for instance,][]{2023LPICo2851.2552P}.
    \item A random value from the interval [$-\sigma$;$+\sigma$] is added to the previous one, where $\sigma$ is the uncertainty of the observed magnitude ($m_{f}$) as provided by the DES catalog.
    \item An absolute magnitude $H_f$ and a slope $\beta_f$, along with their uncertainties, are determined by weighted least squares and error propagation.
    \item Steps two and three are repeated 20\,000 times.
    \item The final values $H_f$ and $\beta_f$ of each object is determined by the mean of the 20\,000 values. The final uncertainties of the absolute magnitudes and slopes are given by the standard deviations of the respective 20\,000 measurements.
\end{itemize}

The adopted value of 0.3 mag to the amplitude of the sinusoidal distribution is a conservative one based on the typical (0.22 mag) of the rotational amplitudes of TNOs and Centaurs listed in \citet{2008ssbn.book..129S}. 

The determination of colours from the steps listed above results from a specific scenario, i.e., unfavourable conditions to directly relate magnitudes taken in different filters. \citet{2021PSJ.....2...40S} present a similar scenario but take a different approach \citep[see also][]{2023ApJS..269...18B}. However, both discussions involve the use of colours from absolute magnitudes and the consequent determination of phase slopes, taking into account the importance of considering the effects of an unknown rotational light curve on the observed magnitudes.


In brief, \citet{2021PSJ.....2...40S} derives the full probability distribution \textit{P} of observing an apparent magnitude \textit{m} in a particular band given a true magnitude ${m_p}$ and its (assumed Gaussian) uncertainty $\sigma$. The peak-to-peak variation of an unknown rotational light curve, assumed to have a sinusoidal shape, is also modelled in \textit{P}. From it, and
considering the H-G phase function by \cite{1989aste.conf..524B}, \citet{2021PSJ.....2...40S} derive colour, phase parameter, absolute magnitude and amplitude of a rotational light curve for 1\,000 Jupiter Trojans.

Also, \citet{2021PSJ.....2...40S} reports a comparison between their own amplitudes and those from different sources. Although a general broad consistency, well measured rotational amplitudes from the Kepler space telescope data are typically 50\% larger than theirs. Those authors believe that, from further simulations, their amplitudes should be thought as the smaller of a light curve with two amplitudes.

Our procedure, considering the phase function it adopts, saves one variable with respect to \citet{2021PSJ.....2...40S}, namely, the rotational amplitude of the object. As mentioned before, we adopted the value of 0.3 mag instead of determining an amplitude for each object. The determined absolute magnitudes and slopes should not be affected by this under the reasonable hypothesis that rotational phase and observation date are not correlated. We opted for this because of the low number of observations -- typically, 3 to 8 -- in each band available to most of the objects presented here. This simpler procedure provides uncertainties that tend to be larger than those that would have been obtained if we had adopted the full probability distribution \textit{P} from \citet{2021PSJ.....2...40S}. The reliability of the measurements obtained here and their respective uncertainties can be inferred from the results presented later in the text and from comparisons with the literature.

\subsection{Clustering algorithm}\label{sec:cluster_algo}

The segregation of objects into four taxonomic classes, following \citet{2005AJ....130.1291B} and \citet{2008ssbn.book..181F}, requires a clustering algorithm.
The algorithm \textit{PAM} - Partition Around Medoids \citep[see][for more details]{kaufman2009finding} - is chosen to cluster our TNOs and Centaurs from their colours. Here, uncertainties in the colours were not used and
the elimination of outliers to determine the clusters' boundaries were
made iteratively and described later in the text.

Given an observational data set, \textit{PAM} aims at finding \textit{k} medoids in it. Medoids are representative objects of each cluster in such a way that the sum of distances given by a previously chosen metric, to all the objects in the cluster, is minimal (that is, if any of the medoids is replaced by another data point then the minimal
sum of distances is violated). Also note that a medoid is always a member of the data set (which is not always true if we choose a point that is not representative of the cluster’s barycenter).

Here, it is important to highlight that this work does not intend to define any new taxonomy, but to present a classification for TNOs and Centaurs that is coherent with that already existing in the literature.
As a consequence, we set $k=4$. As a sanity check, the determination of the optimal value of \textit{k} with the help of \textit{PAM}, using 3 colours (the same number of colours we have in this work) from a set of 67 objects in \citet{2008ssbn.book..181F}, returned $k=4$.

It should be also stressed that the numerical procedure to segregate TNOs and Centaurs into groups used by works like \citet{2005AJ....130.1291B,2008ssbn.book..181F,2010A&A...510A..53P} is the G-mode \citep[see ][]{2000Icar..146..204F}. This method segregates a set of \textit{N} elements into $N_c$ clusters through the $V$ variables that describes each of the elements. It is assumed that each variable in each cluster follows a Gaussian distribution. The method is unsupervised but the user has to define a critical parameter (confidence level) \textit{q}. The larger the value of \textit{q}, the more general is the classification \citep[see][]{1992EM&P...59..141G, 2000Icar..146..204F, 2013scpy.conf...48H}. \textit{PAM}, however, is ready to use and to produce useful plots in a simple way from the statistical package R \citep{R}. In addition, it is also expected that the description of general physical characteristics of the surfaces of small bodies be strongly dependent on their intrinsic features (like colours) and not on the algorithm to group them.

In this way, we started with a list of 217 TNOs and Centaurs with absolute magnitudes in all {\it griz} bands. However, the determination of clusters is made after some extra filtering whose intervals, although somewhat arbitrary, aim at keeping a large number of objects while avoiding the presence of interlopers in the clustering process.

First, only objects whose absolute magnitudes in {\it g}, {\it r}, {\it i} and {\it z} are used (they all have, at least, 3 measurements in each band). Second, any object whose $H_g-H_r$, $H_g-H_i$ or $H_g-H_z$ colours are outside the interval $[-2.5,+2.5]$ and whose slopes in any of the four bands are outside the interval $[-0.5,+0.5]$ are eliminated. The final step comes from the clustering process with \textit{PAM} adopting, as mentioned earlier, $k=4$ (four clusters).

In fact, colours and phase slopes, as shown later in the text, have different values in different taxonomic classes. Therefore, filtering based on mean and standard deviation of the full set may lead to systematic eliminations of the border representatives in extreme (bluest and reddest) classes. 

The colour interval is wide and includes 95\% of the unfiltered sample, as values outside its limits are not frequent \citep[see, for instance,][]{2012ApJ...749...10O, 2019ApJS..243...12S, 2017AJ....154..101P}. These works either have their colours in the Sloan Digital Sky Survey (SDSS) or scale their photometry to it. Note that the DES filters are similar to those of the SDSS \citep{2018PASP..130g4501M}, so our comparison is generally valid. Conversions from DES to SDSS colours can be found in \citet{2024MNRAS.527.6495C}.

The slope interval is more restrictive (represents 78\% of the sample) and privileges mainly the inclusion of negative slopes. Its upper limit can be considered large enough from different works in the literature \cite[see, for instance and for a broader discussion on phase slopes, ][]{2012ApJ...749...10O,2018MNRAS.481.1848A,2022A&A...667A..81A,2023PSJ.....4...75D,2024P&SS..25105970A}. The exclusion of objects with colours and slopes outside the mentioned intervals intends to avoid bad measurements and, consequently, introducing interlopers in the clustering process. Of course, objects outside those intervals may not necessarily have bad photometric measurements, but such an analysis is out of the scope of this work. 

It is hard to find a physical explanation for negative slopes but they correspond, most probably, to insufficient sampling of the phase curve profile \citep[see][]{2023PSJ.....4...75D}. However, allowing objects with negative slopes into our final sample is important so as to include those for which the absolute value of the slope is smaller than its respective uncertainty.

\begin{table}
\centering
\begin{tabular}{cccc}
\hline
   &    $H_{g}-H_{r}$ & $H_{g}-H_{i}$  & $H_{g}-H_{z}$  \\
\hline
\textit{BB} &   0.45($\pm 0.07$) & 0.5($\pm 0.1$) & 0.7($\pm 0.2$) \\
\textit{BR} &   0.57($\pm 0.08$) & 0.8($\pm 0.1$) & 0.9($\pm 0.1$) \\
\textit{IR} &   0.7($\pm 0.1$) & 1.0($\pm 0.1$) & 1.2($\pm 0.1$) \\
\textit{RR} &   0.99($\pm 0.05$) & 1.2($\pm 0.1$) & 1.54($\pm 0.08$) \\
\hline
\end{tabular}
\caption{Mean colours and standard deviations in each class.}
\label{tab:meancolourstaxo}
\end{table}

\begin{table}
\centering
\begin{tabular}{ccccc}
\hline
\multicolumn{4}{c}{Reflectance} \\
\hline
   &    $r$ band & $i$ band & $z$ band \\
\hline
\textit{BB} &   1.01($\pm 0.07$) & 0.98($\pm 0.09$) & 1.1($\pm 0.2$) \\
\textit{BR} &   1.13($\pm 0.09$) & 1.2($\pm 0.1$) & 1.4($\pm 0.1$) \\
\textit{IR} &   1.3($\pm 0.1$) & 1.6($\pm 0.1$) & 1.8($\pm 0.2$) \\
\textit{RR} &   1.66($\pm 0.08$) & 1.9($\pm 0.2$) & 2.5($\pm 0.2$) \\
\hline
\end{tabular}
\caption{Mean reflectances and standard deviations in each class and each band. Reflectance is normalized to 1 in the $g$ band.}
\label{tab:meanreflectancetaxo}
\end{table}

\subsection{Formation of clusters}

We started with a set of 144 objects that survived the above filtering, from which \textit{PAM}, with a fixed number k = 4,  delivered a first segregation into 4 groups. Once the members of each group are set, we calculate a low resolution reflectance spectrum to each object, normalized at the \textit{g} band (Eq.~\ref{eq:refg}) \citep{2008ssbn.book...91D}


\begin{equation}
    \overline{E_f} = 10^{-0.4[(H_f - H_g)-(m_{f\sun} - m_{g\sun})]} \label{eq:refg}
\end{equation}
, where $H_f$ and $m_{f\sun}$ are the absolute magnitude of the target and the apparent magnitude of the Sun in the same filter, respectively. Note that PAM just segregates the objects into groups. In order to assign a taxonomic class to each of them, their colours must be investigated. One way of doing this is obtaining the mean reflectance spectra of each group, as given by Fig.~\ref{fig:reflectance}.  Table~\ref{tab:filterssunmag} shows the adopted central wavelength for DES filters along with the respective magnitudes of the Sun.

An elimination of reflectances (and, consequently, of objects), at the $2\sigma$ level is made in each group to each wavelength. The clustering is then recalculated and the groups, reorganized. This process was repeated until there were no more changes in the groups. 

\begin{figure}
\begin{center}
\includegraphics[scale=0.5]{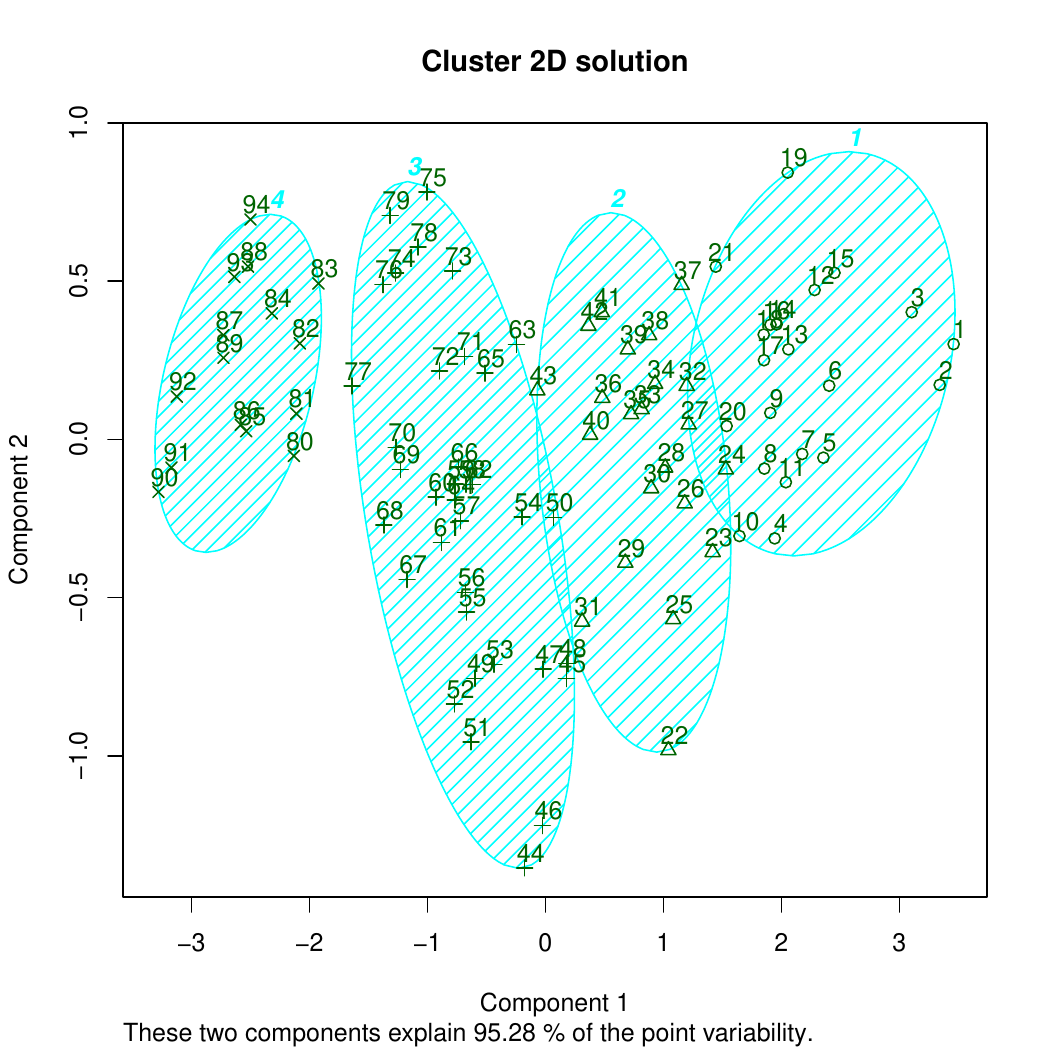}
\caption{Clusters obtained with \textit{PAM}. The axes in principal components are just a comfortable way to present the clusters and they do not participate in their determination. The principal components are derived from colours.}
\label{fig:pca1}
\end{center}
\end{figure}

A total of 94 objects remained in the end from the iterative $2\sigma$ clipping procedure, see Table~\ref{tab:taxonclass}. The elimination process may have been strict, but it provides a higher confidence in the characterization of each group as provided by Tables~\ref{tab:meancolourstaxo} and \ref{tab:meanreflectancetaxo}. In addition, from Table~\ref{tab:meancolourstaxo}, it is possible to later segregate into the already populated clusters the 50 objects that did not survive the elimination process. Figure~\ref{fig:pca1} shows the clusters, where the numbers in each one follow the same order as that in Table~\ref{tab:taxonclass}. Principal components are only used as a tool to help the visualization of the clusters.





\subsubsection{Taxonomic Classification}

Taxonomic classification, by itself, is not enough to provide meaningful information about mineralogical properties of the surfaces of individual objects. Yet, it is still helpful to infer about a general distribution of compositions over different dynamical classes \citep[see for instance][for further considerations]{barucci2008composition,2010A&A...510A..43C}. In this context, the  of TNOs and Centaurs play an important role to tell about the surface properties and/or different physical processes that affects the evolution of these faint and icy primitive bodies. 


\citet{2010Msngr.141...15B} mention that detailed information about the surface composition of TNOs can be only acquired from spectroscopy and that multiband photometry, from visible to near-IR wavelengths, can be used to tie together the different spectral wavelength ranges. 

A taxonomic classification showing 4 groups of different reflectances for TNOs and Centaurs is presented by \citet{2005AJ....130.1291B} and \citet{2008ssbn.book..181F}. These 4 groups - \textit{BB, BR, IR} and \textit{RR} - represent objects showing colours with respect to the Sun ranging from neutral (\textit{BB}) to very red (\textit{RR}).

The spectra of objects in the \textit{BB} (bluest) group are typically flat and bluish in the near-infrared. The $H_2O$ absorption bands seem generally stronger than the other groups. The reddest objects, classified as \textit{RR}, present a small percentage of ice on their surface and indicate the presence of complex organic material. Objects in the \textit{BR} group, an intermediate class between \textit{BB} and \textit{IR} group, have a small rate of $H_2O$ ice on their surface, while some of the objects in the \textit{IR} group seem to contain hydrous silicates on the surface \citep{2008ssbn.book..181F,2010Msngr.141...15B}.

\begin{figure}
\begin{center}
\includegraphics[scale=0.7]{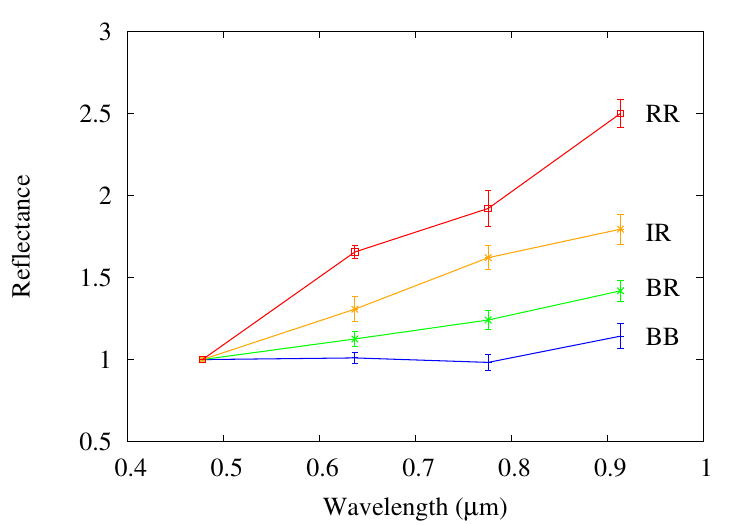}
\caption{Average reflectance in each taxonomic group normalized in g band. Number of defining elements in each group is 21 (\textit{BB}), 22 (\textit{BR}), 36 (\textit{IR}) and 15 (\textit{RR}), see also Table~\ref{tab:taxonclass}. Error bars length is 1$\sigma$.}
\label{fig:reflectance}
\end{center}
\end{figure}

The labeling of each class determined in this work (\textit{BB}, \textit{BR}, \textit{IR} and \textit{RR}) is, as already mentioned, obtained from Fig.~\ref{fig:reflectance}. Comparisons with the literature, presented in Table~\ref{tab:taxonclass}, show that our results are consistent with classification from \citet{2005AJ....130.1291B} and \citet{2008ssbn.book..181F}.



\begin{table}
\centering
\begin{tabular}{ccccc}
\hline
  & g        & r        & i        & z       \\
\hline
$\lambda_c$ (nm) & 475      & 635      & 775      & 925  \\
Sun (mag.) & -26.5071 & -26.9587 & -27.0478 & -27.0542 \\
\hline
\end{tabular}
\caption{Central Wavelength\tablefootnote{\url{https://noirlab.edu/science/programs/ctio/filters/Dark-Energy-Camera}} (nanometers) and solar magnitudes\tablefootnote{\url{ https://cdcvs.fnal.gov/redmine/projects/descalibration/wiki}} adopted for each DES filter.}
\label{tab:filterssunmag}
\end{table}


\begin{figure*}
\begin{center}
\includegraphics[scale=0.6]{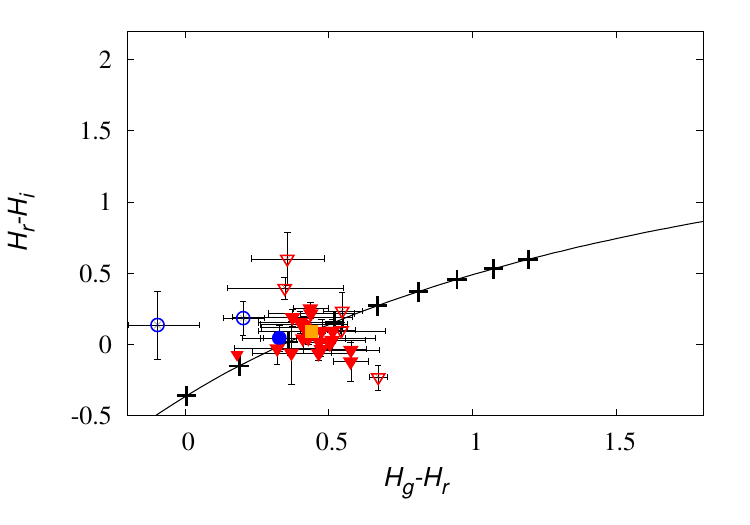}\includegraphics[scale=0.6]{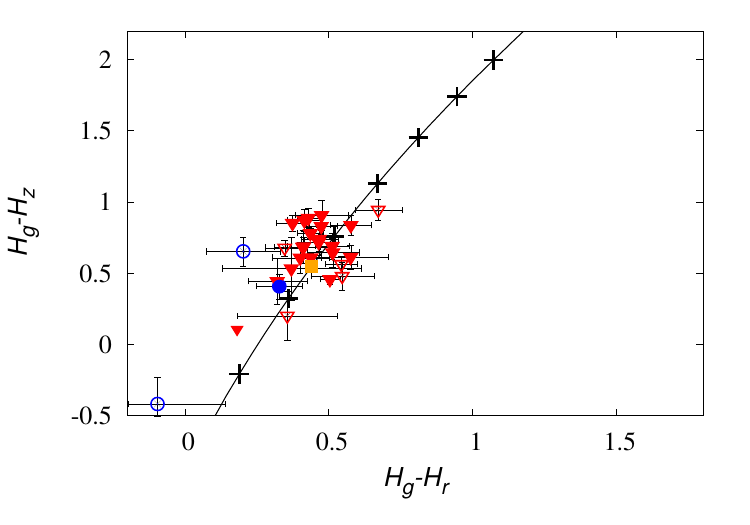}

\includegraphics[scale=0.6]{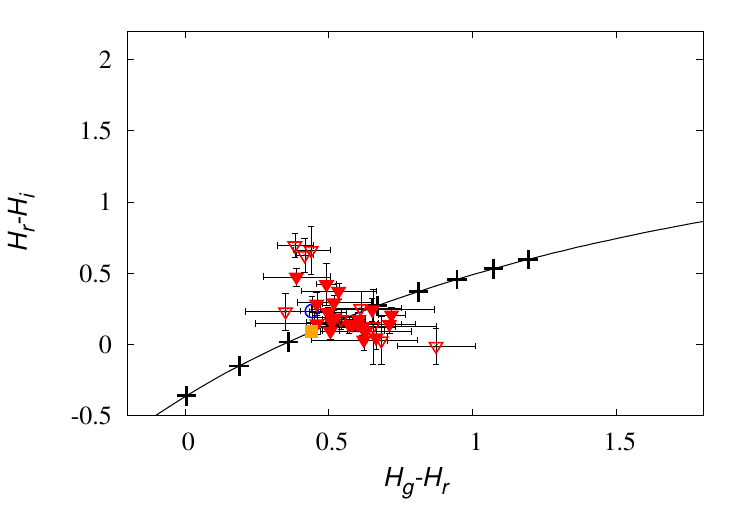}\includegraphics[scale=0.6]{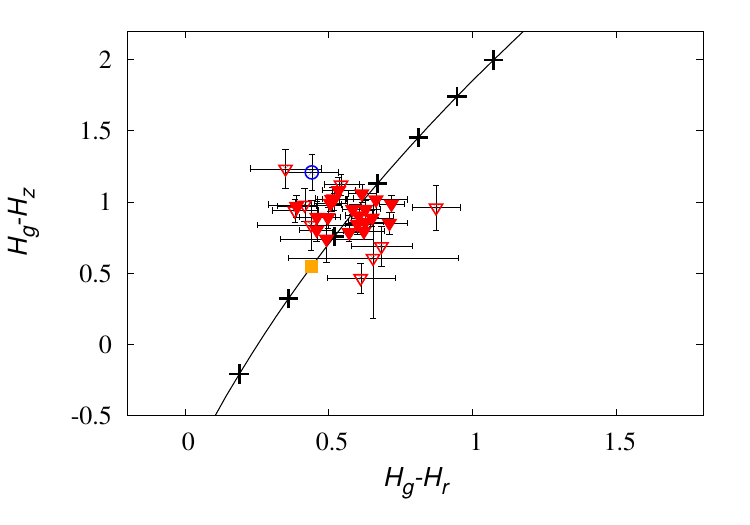}

\includegraphics[scale=0.6]{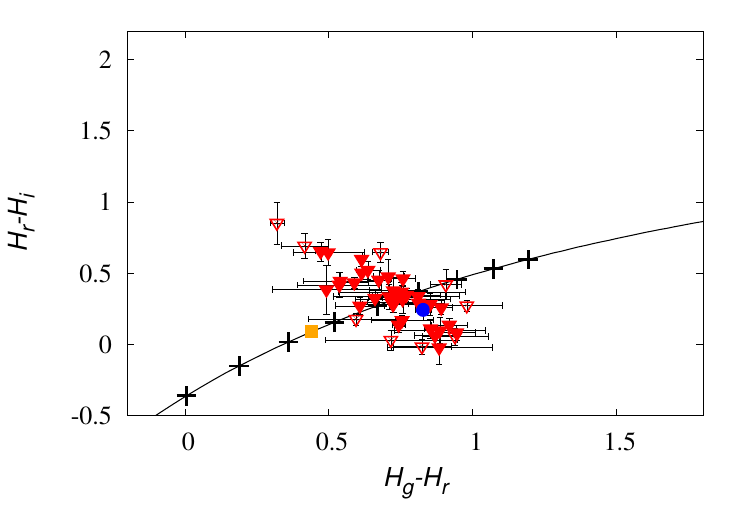}\includegraphics[scale=0.6]{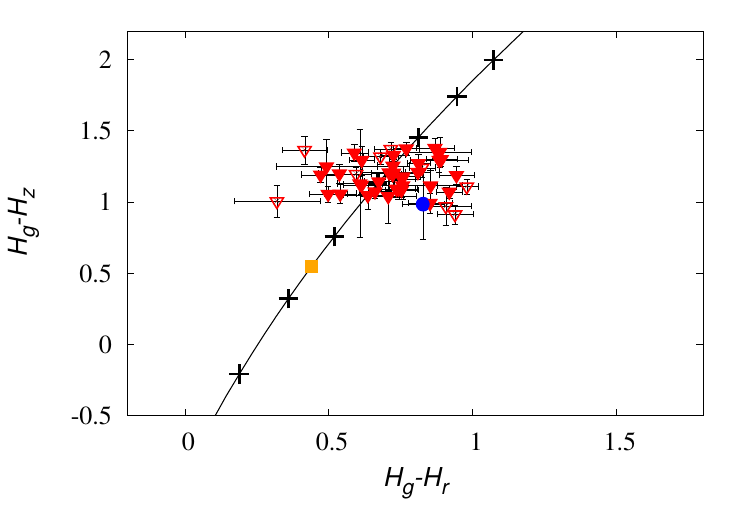}

\includegraphics[scale=0.6]{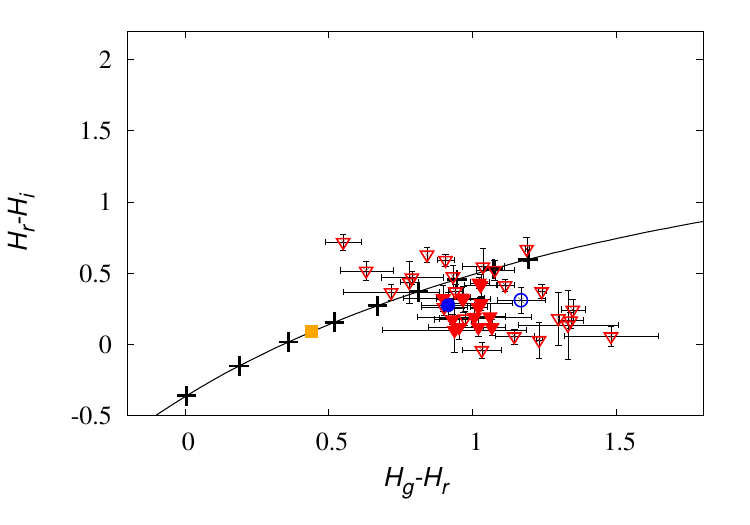}\includegraphics[scale=0.6]{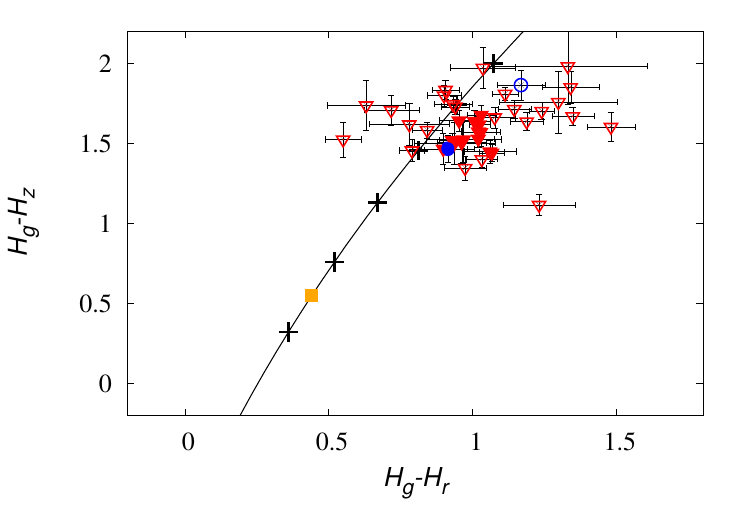}

\caption{Left column: $(H_g-H_r)\times(H_r-H_i)$ plot of the objects in Table~\ref{tab:taxonclass}. Solid markers are those class-defining (first 94 objects in Table~\ref{tab:taxonclass}). Empty markers are the remaining 50 objects. Triangles represent KBOs. Circles represent Centaurs. The thin curve (reddening line, see Sect. 4.5) shows the locus of objects with linear reflectivity spectrum. 
The solid square on the reddening line indicates the locus of the solar colours. From top to bottom: \textit{BB}, \textit{BR}, \textit{IR} and \textit{RR} groups are represented. Marks on the reddening line range from -02\% up to 22\% per 100nm. A mark is shown each 4 units. Right column: Same as left for colours $(H_g-H_r)\times(H_g-H_z)$. Error bars length is 1$\sigma$.}
\label{fig:redd_line}
\end{center}
\end{figure*}

\begin{figure*}
\begin{center}
\includegraphics[scale=1.2]{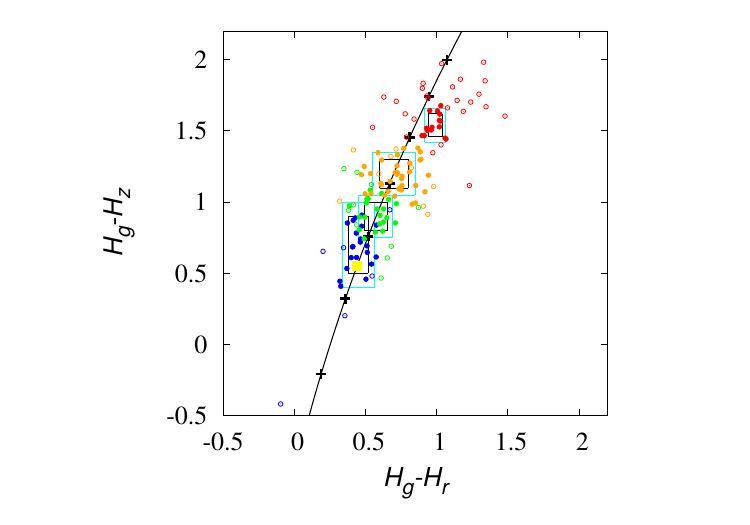}
\caption{Distribution of the taxonomic classes in a colour-colour diagram: \textit{BB} (blue), \textit{BR} (green), \textit{IR} (orange) and \textit{RR} (red). Solid circles represent the defining objects, open circles otherwise (see Table~\ref{tab:taxonclass}). The thin curve (reddening line) shows the locus of objects with linear reflectivity spectrum. Marks on the reddening line range from -02\% up to 22\% per 100nm. Boxes indicate the locus of each class, with side lengths of $2\sigma$ (black) and $3\sigma$ (cyan) (see Table~\ref{tab:meancolourstaxo}).}
\label{fig:general}
\end{center}
\end{figure*}

\subsection{colour-colour diagrams}

Figure~\ref{fig:redd_line} shows colour-colour diagrams along with the reddening line, defined as the reddening per 100 nm, in percentage as given by Eq.~\ref{eq:redd_line}

\begin{equation}
    S(\lambda_1,\lambda_2)=10^4\times\frac{\overline{E_f}_{2}-\overline{E_f}_{1}}{\lambda_2-\lambda_1},
    \label{eq:redd_line}
\end{equation}

\noindent where $\lambda_{1,2}$ are given in nanometers ($\lambda_{2}>\lambda_{1}$) and $\overline{E_f}_{1,2}$ are given by Eq.~\ref{eq:refg}.

It is seen in Fig.~\ref{fig:redd_line} that the objects in the \textit{BB}, \textit{BR} and \textit{IR} groups tend to follow, on average, the reddening line, whereas those in the \textit{RR} group have a preference to be located below it. These results are supported by previous works, and those for the redder ones indicate slope changes towards the near-IR~\citep[see, for instance,][]{2001A&A...378..653B,2001A&A...380..347D,2002A&A...389..641H,2004A&A...417.1145D}. The behaviours just mentioned of the taxonomical classes with respect to the reddening line are also highlighted by Fig.~\ref{fig:general}. It can be compared to figure 1 from \citet{2023arXiv231003998M}, where the spectral slope of their reddest object for wavelengths shorter than 1$\mu m$ also stands out when compared to those of the other (and less red) ones.



\subsection{Phase slopes}

The values of the phase slopes (see Eq.~\ref{eq:magred}) are listed in Table~\ref{tab:taxonclass} and synthesized in Tables~\ref{tab:def_slopes} and \ref{tab:ndef_slopes} for each taxonomic class. 

It is interesting to note from Tables~\ref{tab:def_slopes} and \ref{tab:ndef_slopes} that the slopes in the \textit{g} band become gradually less steep from the bluer to redder objects, in agreement with the tentative correlations reported by \citet{2023PSJ.....4...75D} between slope and colour. Also, \citet{2019MNRAS.488.3035A,2024A&A...685A..29A} reports that their data show steeper phase curves in the \textit{R} filter for the redder objects and that the bluer objects have steeper phase curves in the \textit{V} filter. The values shown in Tables~\ref{tab:def_slopes} and \ref{tab:ndef_slopes} (\textit{BB} column) agree with the results for the \textit{V} filter 
and are not in contradiction with those for the \textit{R} filter if our \textit{z} band measurements (\textit{RR} column) are considered.

\begin{figure*}
\begin{center}
    \includegraphics[scale=0.5]{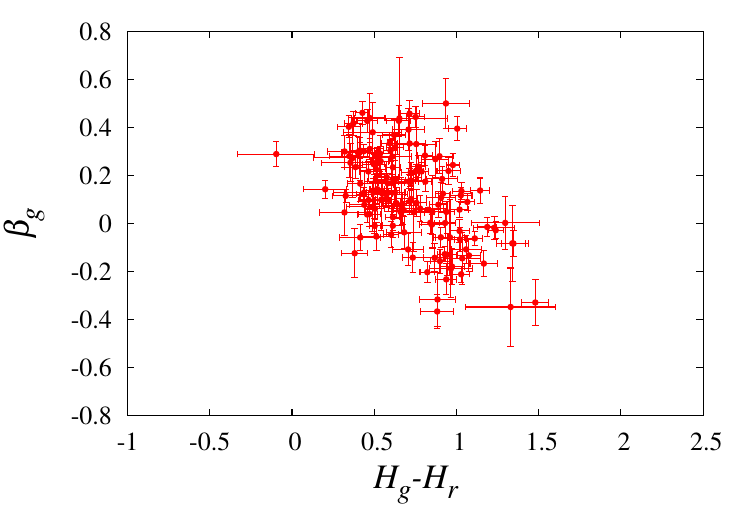}\includegraphics[scale=0.5]{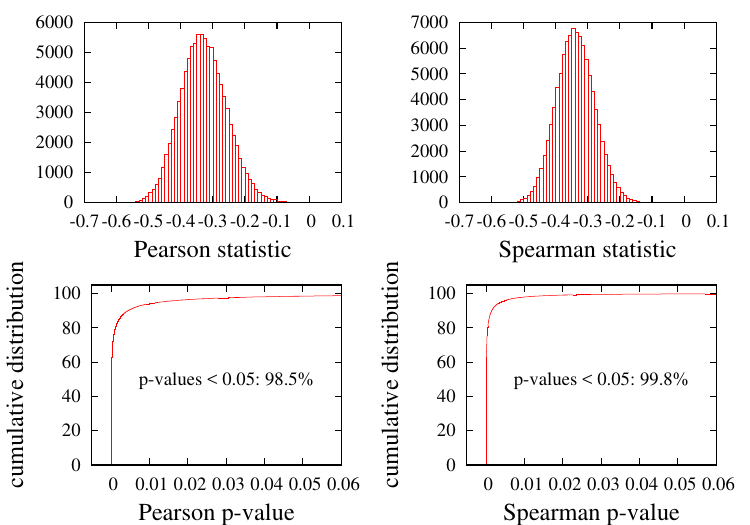}
    \includegraphics[scale=0.5]{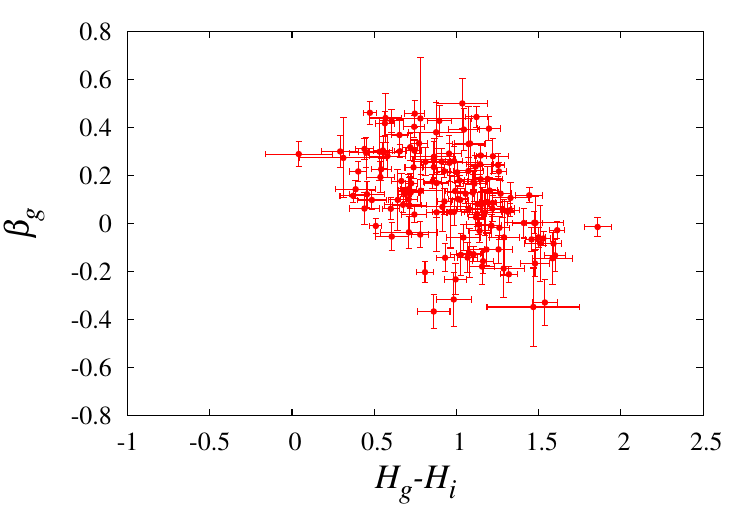}\includegraphics[scale=0.5]{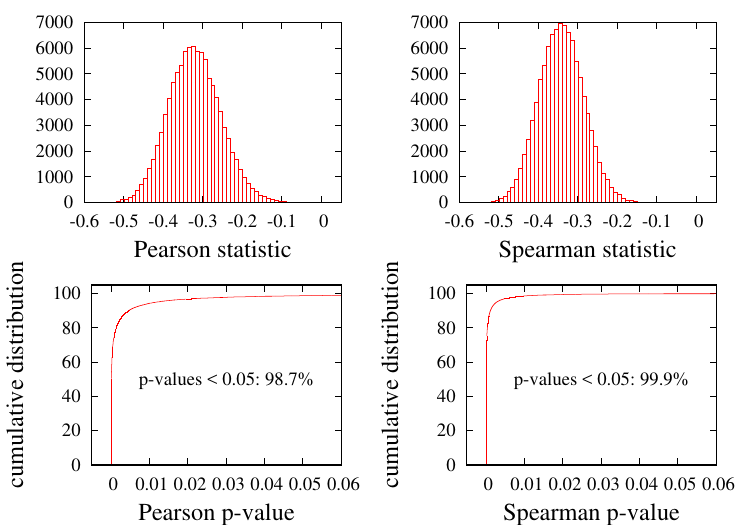}
    \includegraphics[scale=0.5]{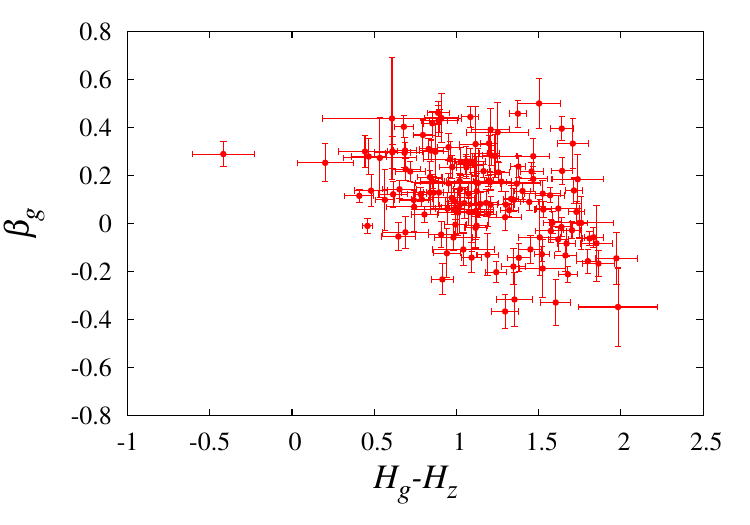}\includegraphics[scale=0.5]{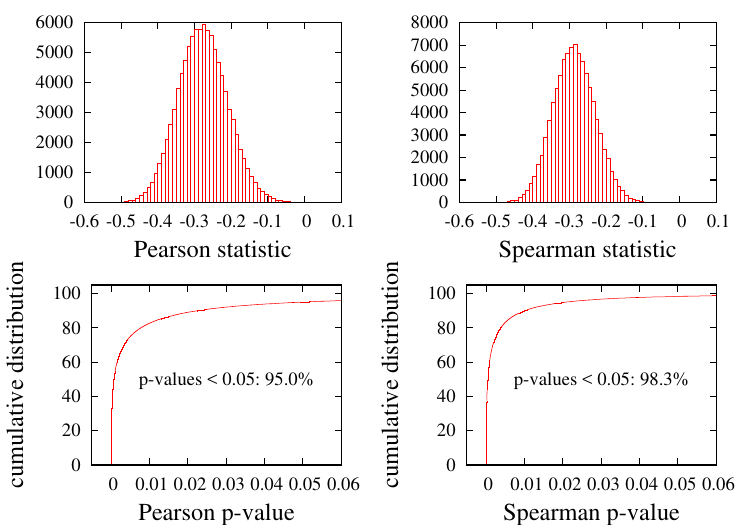}
    \includegraphics[scale=0.5]{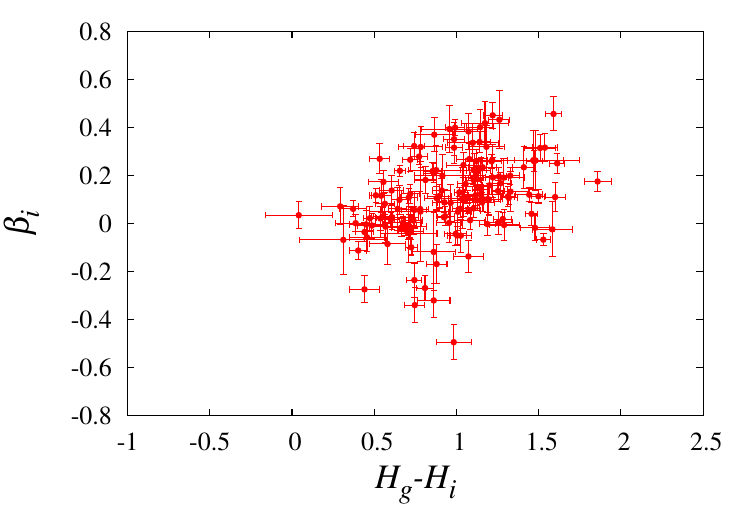}\includegraphics[scale=0.5]{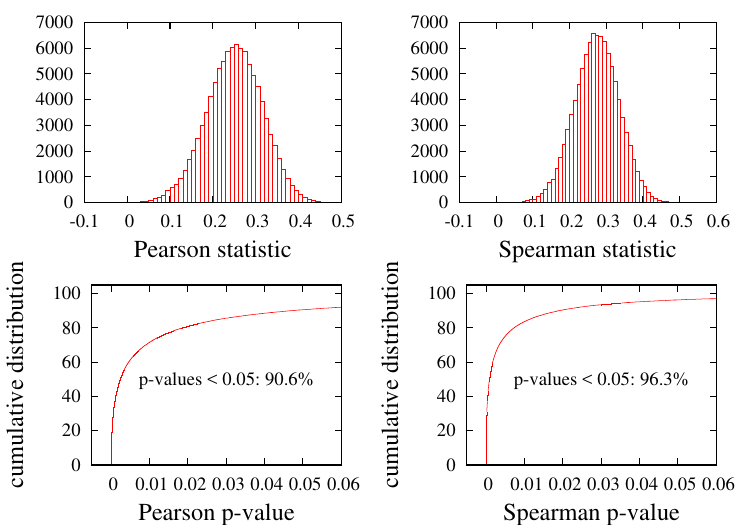}
    \includegraphics[scale=0.5]{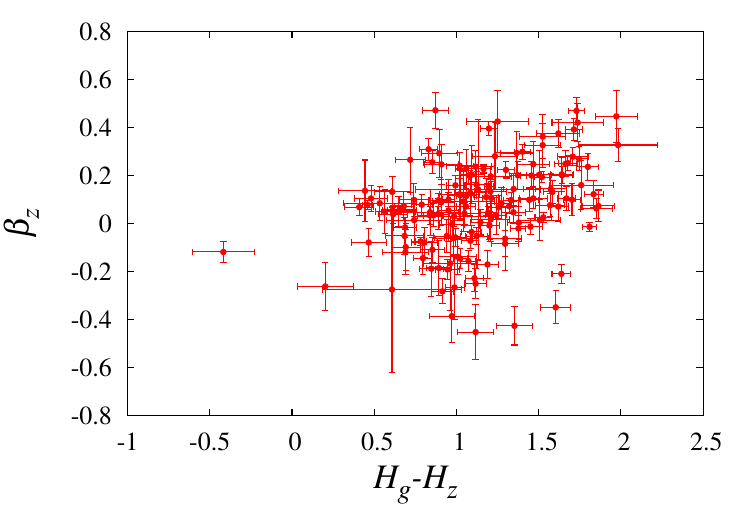}\includegraphics[scale=0.5]{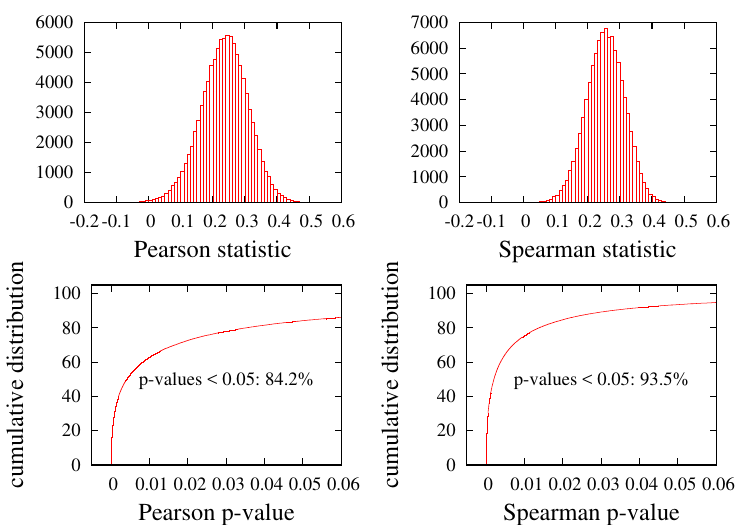}\\
\caption{Colour $\times$ slope in the \textit{g}, \textit{r}, \textit{i} and \textit{z} bands for the objects in Table~\ref{tab:taxonclass}. Histograms of Spearman and Pearson statistics along with the respective cumulative p-values in the right column were obtained by resampling the points in the left column 100\,000 times. Each point is resampled in the colour and slope directions following a gaussian distribution with mean given by the point's colour/slope values and respective standard deviations in each coordinate. Error bars length is 1$\sigma$}.
\label{fig:gz_bg}
\end{center}
\end{figure*}

Figure~\ref{fig:gz_bg} indicates a weak correlation between colours and slopes to a specific combination of bands. Combinations of colour/slope other than those shown by Fig.~\ref{fig:gz_bg} were investigated and no correlation has been found.

\begin{table*}
\centering
\begin{tabular}{ccccc}
\hline
& \multicolumn{4}{c}{Slopes - Defining objects} \\
Band & \textit{BB} & \textit{BR} & \textit{IR} & \textit{RR} \\
\hline
\textit{g} & $0.23\pm0.03~(0.14)$ & $0.19\pm0.03~(0.13)$ & $0.10\pm0.03~(0.18)$ & $0.07\pm0.05~(0.21)$ \\
\textit{r} & $0.16\pm0.02~(0.11)$ & $0.12\pm0.03~(0.13)$ & $0.11\pm0.02~(0.13)$ & $0.16\pm0.03~(0.12)$ \\
\textit{i} & $0.04\pm0.02~(0.10)$ & $0.08\pm0.02~(0.11)$ & $0.12\pm0.03~(0.19)$ & $0.09\pm0.03~(0.11)$ \\
\textit{z} & $0.10\pm0.03~(0.14)$ & $0.02\pm0.03~(0.15)$ & $0.04\pm0.03~(0.17)$ & $0.14\pm0.03~(0.10)$ \\
\hline
\end{tabular}
\caption{Average values of phase slopes, standard deviations of the averages and standard deviations of the measurements (inside parenthesis) for each taxonomic class and considering the class-defining objects only.}
\label{tab:def_slopes}
\end{table*}

\begin{table*}
\centering
\begin{tabular}{ccccc}
\hline
& \multicolumn{4}{c}{Slopes - All objects} \\
Band & \textit{BB} & \textit{BR} & \textit{IR} & \textit{RR} \\
\hline
\textit{g}  & $0.23\pm0.03~(0.13)$ & $0.17\pm0.02~(0.14)$ & $0.10\pm0.03~(0.19)$ & $0.01\pm0.03~(0.18)$ \\
\textit{r}  & $0.15\pm0.02~(0.10)$ & $0.10\pm0.02~(0.14)$ & $0.11\pm0.02~(0.15)$ & $0.14\pm0.02~(0.15)$ \\
\textit{i}  & $0.06\pm0.03~(0.14)$ & $0.08\pm0.02~(0.14)$ & $0.11\pm0.03~(0.20)$ & $0.16\pm0.02~(0.15)$ \\
\textit{z}  & $0.07\pm0.03~(0.15)$ & $0.01\pm0.03~(0.15)$ & $0.03\pm0.03~(0.19)$ & $0.17\pm0.03~(0.18)$ \\
\hline
\end{tabular}
\caption{Average values of phase slopes, standard deviations of the averages and standard deviations of the measurements (inside parenthesis) for each taxonomic class and considering all objects.}
\label{tab:ndef_slopes}
\end{table*}

\subsection{Estimation of diameters}

We combine the absolute magnitudes of the TNOs presented here with median values of albedo, representative of different dynamical classes, from \citet{2020tnss.book..153M}
(see table~1 of that work). Since their albedos are mostly given in the V band, we convert DES absolute magnitudes from the $g_{\rm DES}$ band into the V band using the transformation equations from \citet{2021ApJS..255...20A}. The variances of the obtained absolute V magnitudes are determined from the error expansion of the respective transformation equations and by adding quadratically the rms value provided for the respective ones.

Equivalent diameters ({\it D}) are finally computed from Eq.~\ref{eq:Diam}, taking the apparent magnitude of the Sun in the V band (-26.77) from \citet{2018ApJS..236...47W}. Confidence intervals $[D_{\rm min}, D_{\rm max}]$ for the diameters are inferred by determining the pairs (${H_v}_{\rm min}$, ${\rho_v}_{\rm max}$) and (${H_v}_{\rm max}$, ${\rho_v}_{\rm min}$), where ${H_v}_{\rm min}=H_v + \sigma_{H_v}$, ${H_v}_{\rm max}=H_v - \sigma_{H_v}$, 
${\rho_v}_{\rm min}=\rho_v - \sigma_{\rho_v}$ and ${\rho_v}_{\rm max}=\rho_v + \sigma_{\rho_v}$. The adopted values for $\sigma_{\rho_v}$ are those of the standard deviation of the mean as given by table 1 in \citet{2020tnss.book..153M}. All diameters and uncertainty intervals are also shown in Table~\ref{tab:taxonclass}.

\begin{figure}
\begin{center}
\includegraphics[scale=0.52]{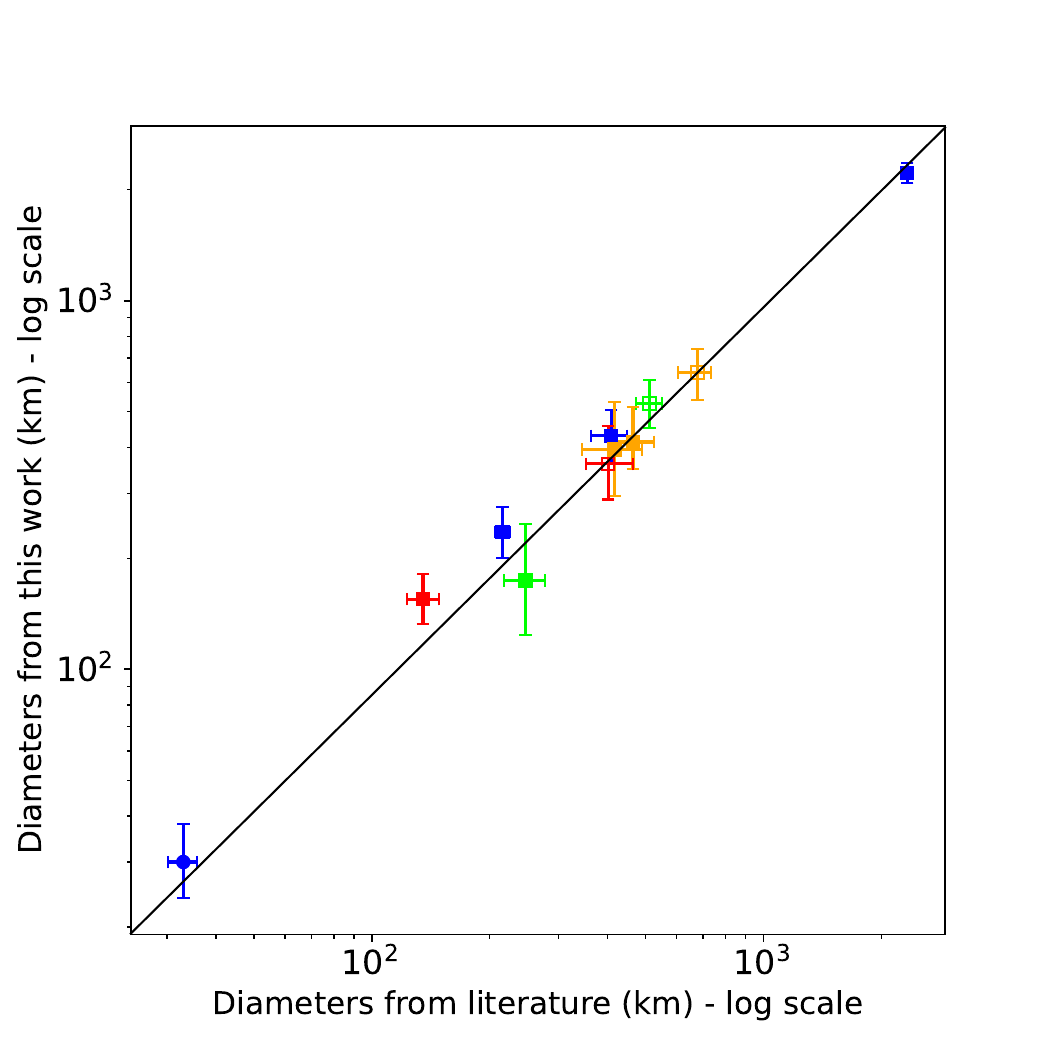}
\caption{Comparison of diameters and respective errors found in the literature and those from this work. The largest diameter is that from Eris.}
\label{fig:diameters}
\end{center}
\end{figure}

An interesting quality verification of these values is given by Fig.~\ref{fig:diameters}, where a good agreement between our measurements and those from the literature are seen. This is another indication that the absolute magnitudes here determined are meaningful and also that the diameters provided by Table~\ref{tab:taxonclass}, within their uncertainties, are reliable. The objects that appear in those plots are (smallest to largest diameter): 2009 HH36, 1999 OX3, 2007 RW10, 2060 Chiron, 2003 QW90, 2005 QU182, 2005 TB190, 2001 QF298, 2004 TY364, 2005 RN43 and Eris. All diameters external to this work result from a search at the Virtual European Solar and Planetary Access \citep[VESPA,][ - TNOs are Cool]{2018P&SS..150...65E}\footnote{\url{https://vespa.obspm.fr/planetary/data/display/?&service_id=ivo://padc.obspm.planeto/tnosarecool/q/epn_core&service_type=epn}}, exceptions made to 2009 HH36 \citep{2013ApJ...773...22B} and 2003 QW90 \citep{2021PSJ.....2...10F}.

\subsubsection{Special cases}

Objects 2009 HH36, Eris and 2005 TB190 also appear in \citet{2020tnss.book..153M} so their individual albedos in the V band were taken directly from that paper. For Eris, we also use the geometric albedo in the V band obtained by \citet{2011Natur.478..493S} from a stellar occultation to calculate its diameter. Both results are very similar and presented in Table ~\ref{tab:taxonclass}. 

Object 2007 RW10 is a binary object and also appears in \citet{2020tnss.book..153M}. Its individual albedo is, however, given in the R band. In this context, its diameter is calculated by transforming its absolute magnitude from the $r_{\rm DES}$ band into the R band using again the transformation equations from \citet{2021ApJS..255...20A}. The apparent magnitude of the Sun in the R band (-26.97) is taken from \citet{2018ApJS..236...47W}.

\subsubsection{Dynamical and spectral classes and albedos}

Although the determination of diameters presented here profits from a correspondence between albedo and dynamical classes, the correlation between albedo with spectral class presented by \citet{2014ApJ...782..100F, 2014ApJ...793L...2L} could have also been considered.

In addition to the above correlation, \citet{2014ApJ...793L...2L} (see its table 2) also shows median albedos with their respective confidence intervals as function of dynamical classes. Those values agree with the ones adopted in this work from \citet{2020tnss.book..153M}. 

In any case, the reader may opt to use the
spectral class - albedo correlation to estimate diameters
from the absolute magnitudes presented here. In this case,
a value of 5\% for blue objects (\textit{BB}, \textit{BR})
and 15\% for red objects (\textit{IR}, \textit{RR}) \citep[see figure 2 in ][]{2014ApJ...793L...2L} is a
valid choice.

\subsection{Relations between colours and orbital elements}

\begin{figure}
\begin{center}
    \includegraphics[scale=0.65]{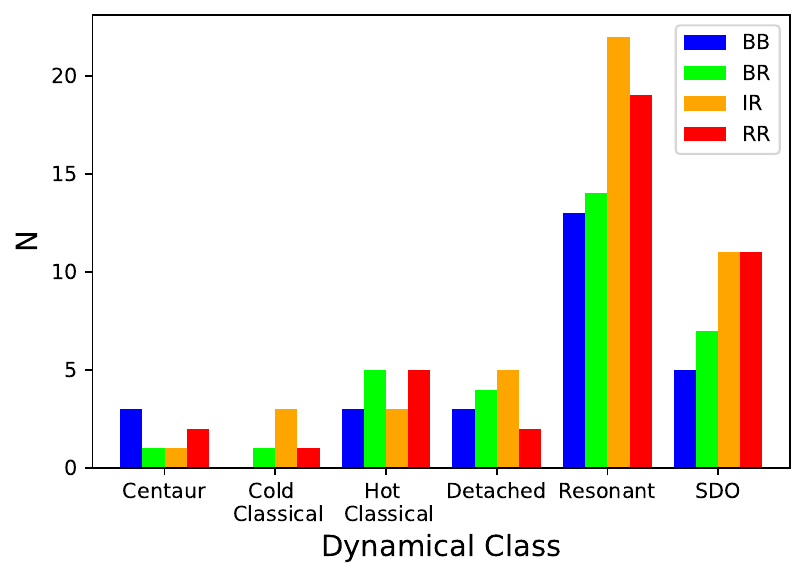}
\caption{Distribution of taxonomic classes within each dynamical class.}
\label{fig:distribution}
\end{center}
\end{figure}

We searched, in the full set of 144 objects, for possible correlations involving colours, orbital parameters and sizes.  Figure~\ref{fig:distribution} shows the distribution of the taxonomic classes within each dynamical class and should be understood mostly as a feature of our observational data and unknown sampling biases rather than an actual physical scenario. It should be noted that the dynamical classification from the SkyBoT (adopted here), do not segregate between hot and cold classicals. In this context,
the definitions from \citet{2022ApJS..259...54H} and \citet{2023PSJ.....4..160M} have been used: $42.5 < a < 45$ AU and $i < 4^{\rm o}$ or $45 < a < 47$ AU and $i < 6^{\rm o}$. Although the inclination, strictly speaking, should be understood in the context of \citet{2022ApJS..259...54H} (free inclination), we use osculating elements. All our cold classicals, exception made to the TNO 2014 NB66 (BR type), belong to our reddest (IR and RR) objects. The few Centaur objects are dominated by two groups  (\textit{BB} and \textit{RR}), and the \textit{IR} objects, in agreement with \citet{2008ssbn.book..181F} and \citet{2010A&A...510A..53P}, are most frequent in the resonant class.
We note that, also as shown by Fig.~\ref{fig:distribution}, our SDOs are dominated by the reddest classes (\textit{IR} and \textit{RR}). This is in disagreement with \citet{2021AJ....162...19A}, which found a paucity of Very Red objects in the Scattered Disk.

\begin{figure*}
\begin{center}
    \includegraphics[scale=1.1]{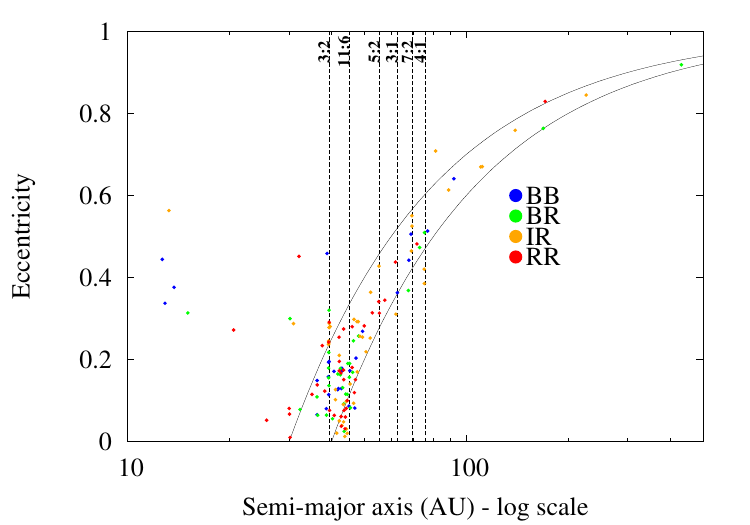}
\caption{Eccentricity as a function of the semi-major for the objects in Table~\ref{tab:taxonclass}. Some resonances with Neptune are indicated by the dashed lines. The continuous lines show constant \textit{q} of 30 and 40 AU.}
\label{fig:axe}
\end{center}
\end{figure*}

A plot of semi-major axis $\times$ eccentricity is given by Fig.~\ref{fig:axe}. Taking into account the interval (13 -- 75 UA) in semi-major, in which most of the objects are located, we find no correlation between the taxonomic classes and heliocentric distance. It is possible to see that, however, the few objects with $a<20$ AU tend to be blue.

\begin{figure*} 
\includegraphics[scale=0.90]{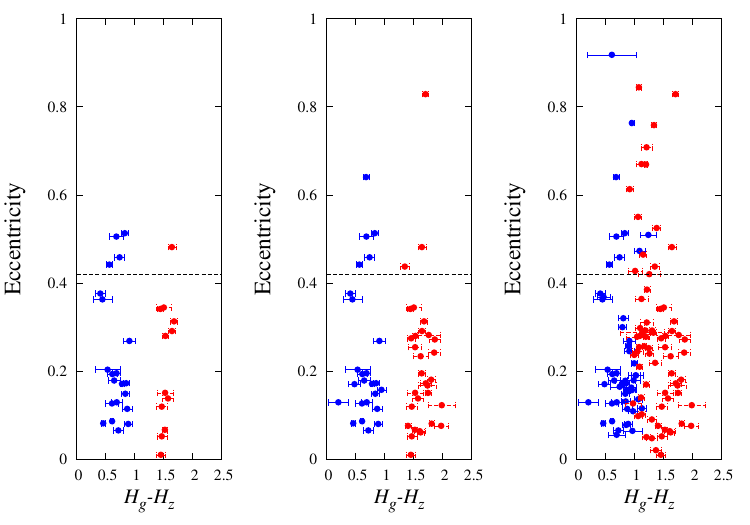}
\caption{colour $\times$ eccentricity for selected objects (excluding Centaurs with $T_j<3$ and any object with $i<5^{\rm o}$). Left: defining (first 94 objects in table~\ref{tab:taxonclass}) \textit{BB} (blue) and \textit{RR} (red) objects. Middle: \textit{BB} and \textit{RR} objects, defining and non-defining. Right: \textit{BB} plus \textit{BR} (blue) and \textit{IR} plus \textit{RR} (red), defining and non-defining. Dashed line at $e=0.42$ \citep[see ][]{2021AJ....162...19A}.}
\label{fig:Ecc}
\end{figure*}

Figure~\ref{fig:Ecc} shows plots of eccentricity as a function of colours for three sets of objects: (i) defining \textit{BB} and \textit{RR}, (ii) defining and non-defining \textit{BB} and \textit{RR}, and (iii) defining and non-defining \textit{BB} plus \textit{BR} and defining and non-defining \textit{IR} plus \textit{RR} (i.e., considering two classes only). \citet{2021AJ....162...19A} finds that their reddest objects are strongly limited to $e<0.42$ so we adopt it for the sake of our analysis.

We note that, for the leftmost and middle plots of Fig.~\ref{fig:Ecc}, 80\%-81\% of the bluest population has $e<0.42$, whereas 90\%-92\% of the reddest population has eccentricities below that limit. The same fraction of blue and red objects in the rightmost plot, however, shows $e<0.42$. We can interpret these results as an agreement, although it is difficult to take a limiting value for $e$ from our own data, with those from \citet{2021AJ....162...19A} when the two extreme classes (\textit{BB} and \textit{RR}) are concerned. On the other hand, there is no evident limit when two classes are formed from \textit{BB} plus \textit{BR} (blue) and \textit{IR} plus \textit{RR} (red).

\begin{figure*} 
\includegraphics[scale=0.90]{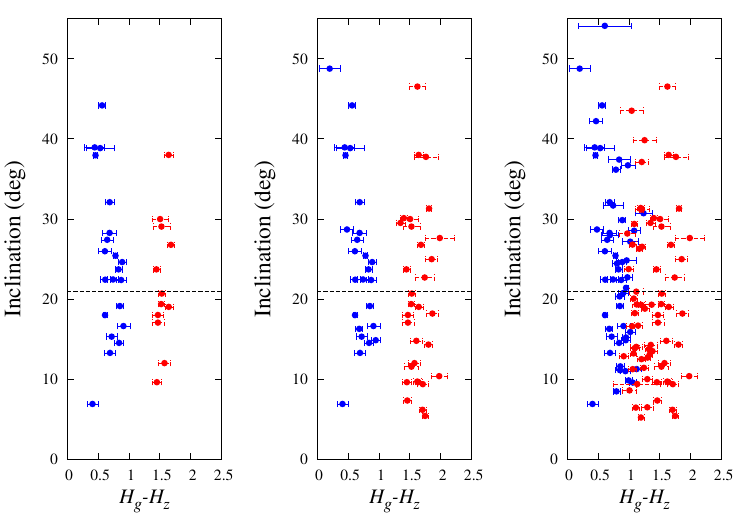}
\caption{colour $\times$ inclination for selected objects (excluding Centaurs with $T_j<3$ and any object with $i<5^{\rm o}$). Left: defining (first 94 objects in table~\ref{tab:taxonclass}) \textit{BB} (blue) and \textit{RR} (red) objects. Middle: \textit{BB} and \textit{RR} objects, defining and non-defining. Right: \textit{BB} plus \textit{BR} (blue) and \textit{IR} plus \textit{RR} (red), defining and non-defining. Dashed line at $i=21^{\rm o}$ \citep[see ][]{2021AJ....162...19A}.}
\label{fig:Inc}
\end{figure*}

When inclinations are concerned, we note that 33\%-39\% of the blue objects have $i<21^{\rm o}$, whereas 58\%-64\% of the red objects have inclinations below that limit \citep[adopted from ][]{2019AJ....157...94M,2021AJ....162...19A}. It is interesting to note that these results hold for all three plots in Fig.~\ref{fig:Inc}.

It is important to emphasize here that we can not infer strong confinement limits in eccentricity and/or inclination as
\citet{2019AJ....157...94M} and \citet{2021AJ....162...19A} did, but we can
say that the fraction of red objects in our data confined in the region ($e<0.42$, $i<21^{\rm o}$) is clearly larger than that of our blue ones (with the exception mentioned above when we consider only two classes for the eccentricity).

A correlation between colour and inclination for classical TNOs is reported by various works \citep[see, for instance][]{2000Natur.407..979T, 2002ApJ...566L.125T, 2015A&A...577A..35P}. Our data contain 21 classical TNOs spreading over the four taxonomic classes from which a colour-inclination correlation is not clear. 


We checked all dynamical classes from our full dataset individually and the presence of a weak correlation was found from both Pearson and Spearman tests only for the resonant objects, 68 in total, from our dataset. This can be seen  in Fig. ~\ref{fig:inccolour}. It is important to emphasize that such correlation is due to the resonant \textit{BB}. We also checked the colour-inclination correlation for three subsets of resonant objects: (i) excluding the 7:4 resonance; (ii) excluding low inclination ($i<5^{\rm o}$) objects; and (iii) excluding low inclination and and low eccentricity ($e<0.09$) objects. In all cases, a weak but clear correlation is present as long as the resonant \textit{BB} are kept in the sample.




Figure~\ref{fig:DiamClass} shows the distribution of diameters of our bluest (\textit{BB}) and reddest (\textit{RR}) objects as a function of their $H_g-H_z$ colours. It is clear from this distribution that the reddest objects are concentrated around smaller diameters, with a small fraction of objects larger than $\sim 250$ km, as compared to the neutral (bluest) ones. 
Also, we highlight that no large (D $\gtrsim400$ km, error bars not included) objects were found in the Cold Classical population, which is in line with \citet{2022DPS....5410507K}. Also, {\citet{2012ApJ...749...33F} suggest that larger objects are dominated by other processes that have little effect at smaller ones, like the onset of differentiation or cryovolcanic processes, both of which would have the effect of covering or removing non-icy materials from the surface. This could be a justification for the absence of objects with D $\gtrsim 400$ km among our reddest ones.

\begin{figure*}
    \includegraphics[scale=0.35]{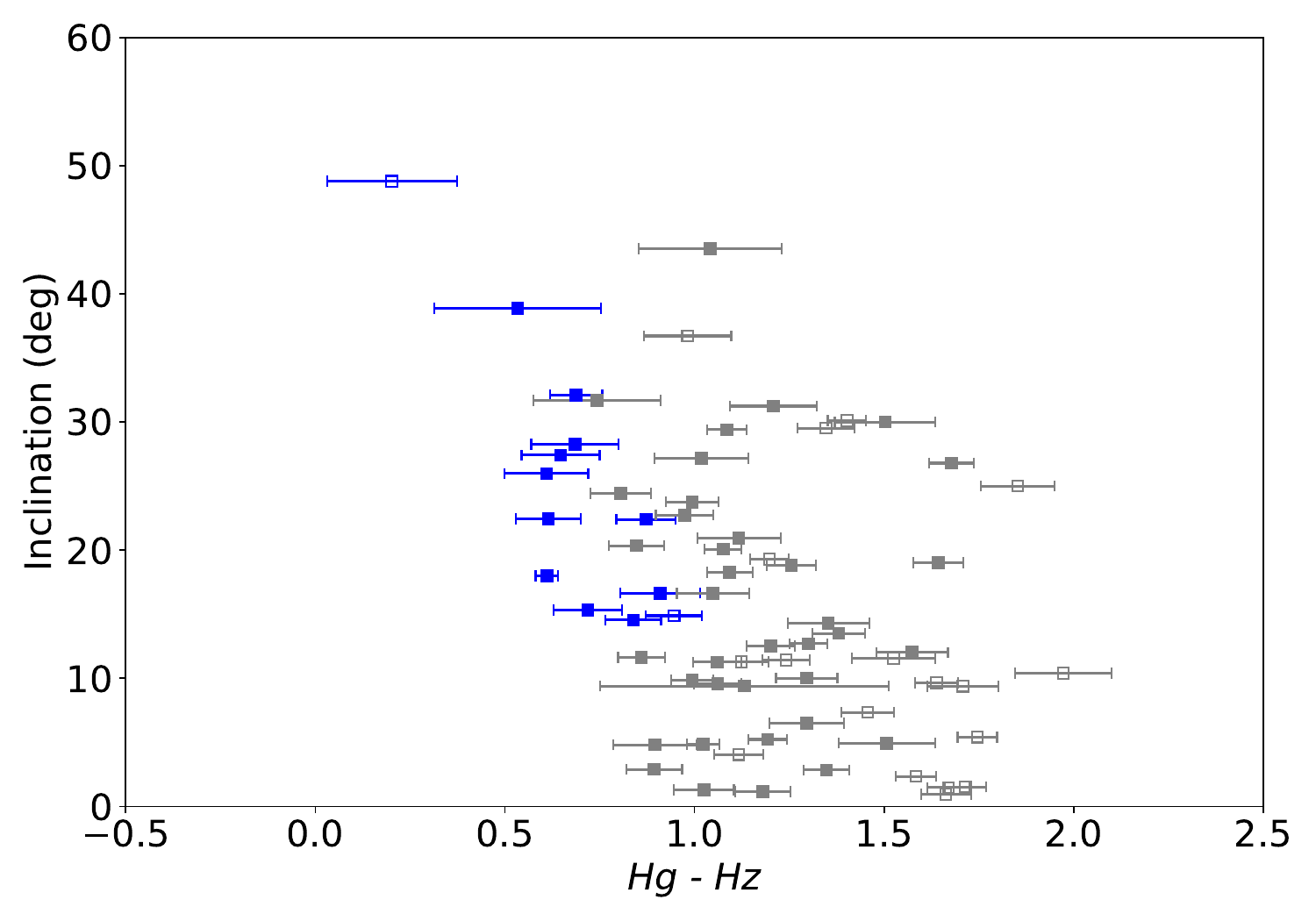}
    \includegraphics[scale=0.68]{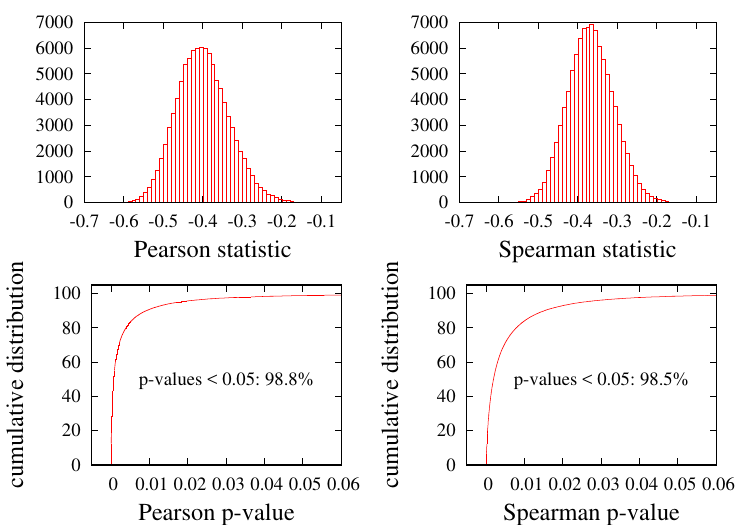}
\caption{Left panel: colour $\times$ inclination correlation, considering all 68 resonant objects in Table~\ref{tab:taxonclass}. Solid squares: objects among the first 94 objects in that table. Open squares: remaining ones. Blue squares represent the BB-type resonant objects and the gray ones represent the other taxonomic classes (\textit{BR, IR} and \textit{RR}). Upper right panel: results from Pearson and Spearman tests obtained as in Fig.~\ref{fig:gz_bg}. Lower right panel: results from Pearson and Spearman tests obtained as in Fig.~\ref{fig:gz_bg}. Error bars length is 1$\sigma$}. \label{fig:inccolour}
\end{figure*}

\begin{figure}
\begin{center}
\includegraphics[scale=0.65]{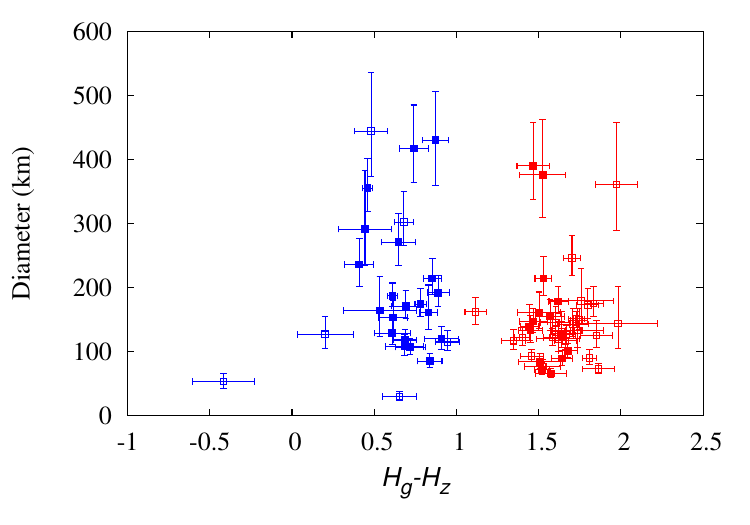}
\caption{colour$\times$diameter for object in classes \textit{BB} (blue) and \textit{RR} (red). Solid markers are those class-defining (first 94 objects in Table~\ref{tab:taxonclass}). Empty markers are the remaining 50 objects. Eris (diameter $2221_{-137}^{+148}$ km), for convenience, is not shown. 92.5\% of the \textit{RR} objects have diameters smaller than 400 km and 90\% have diameters smaller than 250km. In both cases, the $1\sigma$ uncertainty was taken into consideration. In the case of the \textit{BB} objects, the percentiles are 81.5\% and 66.7\% (Eris included).}
\label{fig:DiamClass}
\end{center}
\end{figure}

\subsection{A few words about the adopted taxonomic scheme}

It is reasonable to look for classes that meaningfully reflect spectroscopic characteristics of its members. At the same time, it is clear that broad band photometry with few colours just can not compete with an spectroscopic study of the surfaces of small bodies. In principle, such a difficulty of refinement may suggest
that a large number of classes would be meaningless.

\citet{2017A&A...604A..86M} found that their spectroscopic results for a set of 43 TNOs and Centaurs are in good agreement with the bluest (\textit{BB}) and reddest (\textit{RR}) objects in the taxonomic classification by \citet{2005AJ....130.1291B, 2008ssbn.book..181F}. Objects in the other two intermediate classes, \textit{BR} and \textit{IR}, however, show intermixed mean spectra indicating that at
least part of the objects show no particular preference to one of those classes.

Choosing the taxonomy by \citet{2005AJ....130.1291B, 2008ssbn.book..181F} in this context allows us to, at least, be confident that our bluest and reddest objects have surface properties showing reliable correspondence with those obtained from spectroscopic results. In addition, relevant information is gathered about the objects in the intermediate classes, leading to a possible improved discrimination of surfaces populating them, as well as increasing the number of objects in all classes of such a taxonomy scheme. 





As mentioned earlier, this work is neither aimed at advocating for any specific method to classify TNO/Centaur surfaces through their optical colours nor to present any new such method. Although the use of the 4-class taxonomic system can be ultimately considered a choice in this work, the data set itself provides some support for, at least, a minimum number of classes. Figure~\ref{fig:hghz_hist} is helpful in this context.

It shows the histograms for the colours $H_g-H_r$, $H_g-H_i$ and $H_g-H_z$. It is important
to highlight that these plots reflect the colour distribution of the data without influence of whatever procedure to separate the objects into classes although, obviously, the class limits obtained here must agree with such distribution.

The following can be noted in those histograms (see Table~\ref{tab:meancolourstaxo}):

\begin{enumerate}
    \item the colour limit between \textit{IR} and \textit{RR} for colour $H_g-H_r$ is close to the minimum of the $H_g-H_r$ histogram,
    \item the colour limit between the \textit{BR} and \textit{IR} for colour $H_g-H_i$ is close to the minimum of the $H_g-H_i$ histogram, and
    \item the colour limits between the classes \textit{BB}-\textit{BR} and \textit{IR}-\textit{RR} for colour $H_g-H_z$ are close to the two minima of the $H_g-H_z$ histogram.
\end{enumerate}
}

\begin{figure*}
\begin{center}
\includegraphics[scale=0.5]{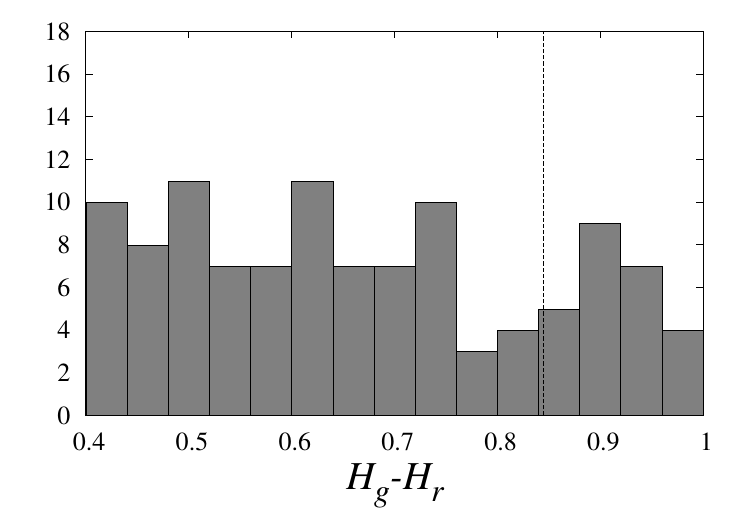}\includegraphics[scale=0.5]{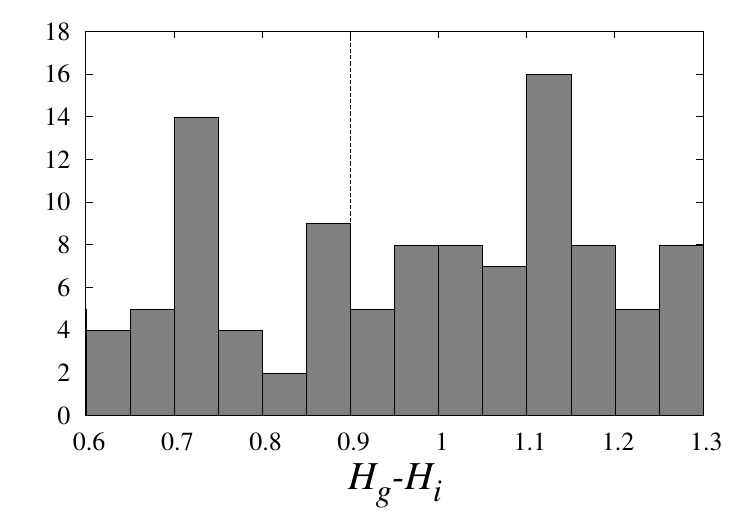}\\
\includegraphics[scale=0.5]{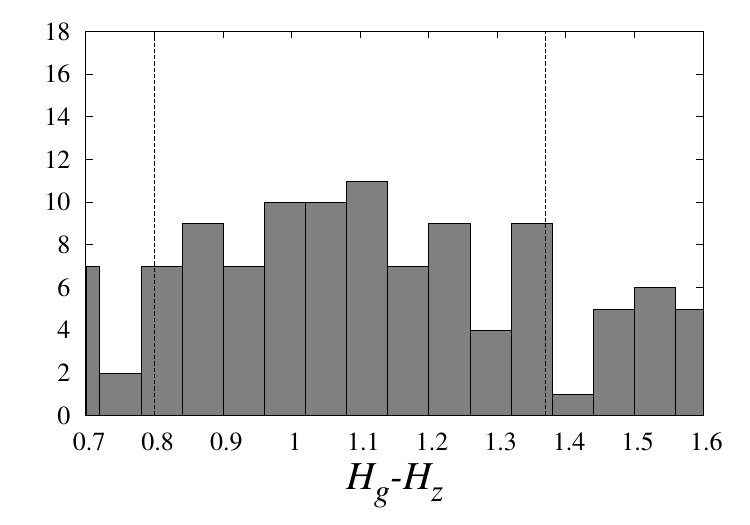}
\caption{Histogram of the $H_g-H_r$, $H_g-H_i$ and $H_g-H_z$ colours for the 144 objects that passed the filtering process (see Sect.~\ref{sec:cluster_algo}). The dashed lines indicate the limits between taxonomic classes as obtained from  Table~\ref{tab:meancolourstaxo}. We recall that Table~\ref{tab:meancolourstaxo} considers only the first 94 (defining) objects in Table~\ref{tab:taxonclass}}.
\label{fig:hghz_hist}
\end{center}
\end{figure*}

The histogram of the $H_g-H_z$ colours shows two gaps, indicating three groups of objetcs: one group is identified with the \textit{BB} class ($H_g-H_z<0.83$), another group is identified with the \textit{RR} class ($H_g-H_z>1.39$) and an intermediate group that comprises the \textit{BR} and \textit{IR} classes. The histogram of the $H_g-H_i$ colours shows a minimum close to the limits between the \textit{BR} and \textit{IR} classes, but one may argue that conclusions assigning gaps to a given class name (\textit{BB}, \textit{BR}, \textit{RR} or \textit{RR}) require a previous knowledge of the clustering results. In this way, it is possible to say from the histogram of the $H_g-H_z$ colours that three groups are present and an analysis combining the pieces of information from all the three colours is, however, a more suitable approach to discuss the number of classes. Although, as previously mentioned, this work has no intention to advocate in favour of any specific method of class finding and the eventual number of classes, our choice for the four-class scheme agrees with the behaviour of the colour distribution of our data as given by Fig.~\ref{fig:hghz_hist}.

In general, it is clear that there are, at least, two classes (a bluer one and a redder one). One may want
to say that classes BB/BR compose the bluer one and that classes IR/RR compose the redder one.

Due to the intermixed spectra found for classes BR and
IR, it would be difficult for the present work to derive part of its results without a discrimination between the two extreme classes (\textit{BB} and \textit{RR}) and those two intermediate (\textit{BR} and \textit{IR}) ones. Interlopers in the BB and RR classes make it difficult to derive conclusions, for instance from the rightmost panel in Fig.~\ref{fig:Ecc}. Also, results from Fig. \ref{fig:DiamClass} profit from the use of \textit{BB} and \textit{RR} objects only.

Based on observations from the Colours of the Outer
Solar System Origins Survey  \citep[Col-OSSOS, see][]{2019ApJS..243...12S}, the existence of two types surfaces of TNOs was proposed by \citet{2023PSJ.....4...80F}. In this new taxonomy, TNOs are assigned into BrightIR or FaintIR classes based on their reflectivity in the NIR following the reddening curve. The BrightIR class is closer to the reddening curve and explores the full range of optical and NIR surfaces, while the FaintIR objects are limited in optical colour and are less bright in the NIR.

We searched in the literature objects that we classified with the \citet{2005AJ....130.1291B} taxonomy that were also classified following the taxonomy by \citet{2023PSJ.....4...80F}, see Table~\ref{tab:Barucci_Fraser}. It is interesting to note the overlap of BB-type with BrightIR and RR-type with FaintIR.

In addition, 1999 RB216 has ambiguous classifications (BR-IR) within the scheme by \citet{2005AJ....130.1291B} (see Table~\ref{tab:taxonclass}). In the scheme by \citet{2023PSJ.....4...80F} \citep[see ][]{2023PSJ.....4..160M} this TNO has an unclassified surface type. This could be related to the intermixed mean spectra found for classes \textit{BR} and \textit{IR} in the four-colour taxonomy \citep{2017A&A...604A..86M}.

\begin{table*}
\centering
\begin{tabular}{cccc}
\hline
Object & Taxonomy found in this work & Surface Class based on & Reference \\
& based on \citet{2005AJ....130.1291B} & \citet{2023PSJ.....4...80F} & \\
\hline
2001 QF298  &   BB   &  BrightIR      & 1    \\
2013 SA100  &   BR   &  BrightIR      & 1, 2 \\
1999 RB216  &   IR   &  unclassified  & 1    \\
2005 SA278  &   BB   &  BrightIR      & 1    \\
2015 RU245  &   IR   &  FaintIR       & 2    \\   
2003 QW90   &   RR   &  FaintIR       & 1    \\
2015 RT245  &   RR   &  FaintIR       & 1, 2 \\
\hline
\end{tabular}
\caption{Objects classified in this work with \citet{2005AJ....130.1291B} taxonomic scheme which were also found in literature with a possible surface type based on \citet{2023PSJ.....4...80F}. The last column represents the reference where these objects with a possible taxonomy based on \citet{2023PSJ.....4...80F} were found: 1: \citet{2023PSJ.....4..160M}. 2: \citet{2023PSJ.....4..200P}.}
\label{tab:Barucci_Fraser}
\end{table*}

\citet{2015A&A...577A..35P} makes a reanalysis of colours of 354 Centaurs and KBOs, twelve of them are in common with this work: 1999 OX3, 2001 QO297, 2001 QF298, 2002 PA149, 2003 QW90, 2004 TY364, 2005 PU21, 2007 UM126, 2005 TB190, Chiron, Eris and 1999 RB216. We searched all the references, provided by the authors, from which their photometry was obtained. In all cases, their results are compatible to the taxonomical classification this work presents for them.
 
\citet{2025NatAs...9..230P} revealed the existence of three main spectral groups of TNOs corresponding to different surface compositions from spectroscopic data using the James Webb Space Telescope - bowl, double-dip and cliff. These three classes have distinctive spectral features of water ice absorption and of complex and aliphatic organics.

It is interesting to note in \citet{2025NatAs...9..230P} (hereafter PnA) that, to wavelengths shorter than $\sim 1.25\mu{\rm m}$ (see figure 3 of that paper), slopes become increasingly steeper in the reflectance spectra of bowl-, double-dip and cliff-type TNOs respectively. Table~\ref{tab:Barucci_Pinilla} lists the objects in common between this work and PnA, and a correspondence between both classifications is seen. Such a correspondence, in addition to the results in PnA, may provide rough indications of features in the near-infrared from optical spectrophotometry. For instance, it does not seem likely that a blue object in this study be classified as a cliff. It should be also noted that Figs.~\ref{fig:Ecc} and~\ref{fig:Inc} show that our redder population is generally less excited than our bluer one, another feature that agrees with PnA.

\begin{table*}
\centering
\begin{tabular}{lcc}
\hline
Object & Taxonomy found in this work & Surface Type based on \\
 & based on \citet{2005AJ....130.1291B} & \citet{2025NatAs...9..230P} \\
\hline
2007 RW10   &  BR  &  Bowl \\
2004 TY364  &  BR  &  Double-Dip \\
2005 QU182  &  IR  &  Double-Dip \\
2005 TB190  &  IR  &  Double-Dip \\
2005 RN43   &  IR  &  Double-Dip \\
1999 OX3    &  RR  &  Cliff  \\
2005 PU21   &  RR  &  Cliff  \\
\hline
\end{tabular}
\caption{Objects classified in this work with \citet{2005AJ....130.1291B} taxonomic scheme which were also classified by \citet{2025NatAs...9..230P} surface types. Note that 2004 TY364, although classified as a BR type in this work,
also has redder (RR and, mainly, IR) classifications by other works.
}
\label{tab:Barucci_Pinilla}
\end{table*}

\section{Conclusions}

The high quality of the DES photometric measurements lead to reliable colours of TNOs and Centaurs. Despite the absence of accurate rotational information, a taxonomic classification for those objects has been determined based on procedures well established and accepted in the  literature. Out of 144 objects presented here, we have not found in the literature a taxonomic classification available in the scheme proposed by \citet{2005AJ....130.1291B} for 132 of them.

As objects are discovered by DES, their taxonomic classification may be  readily determined from the values presented here and sources with unusual features may be identified. It is interesting to note that part of the analysis presented here involved only the bluest (\textit{BB}) and reddest (\textit{RR}) colours. This is so because the separation between these colours makes the discrimination of intrinsic characteristics easier and, probably, subject to less contamination. In particular, it seems clear that TNOs have neutral and very red colours with respect to the Sun, with possibly two intermediate classes with more mixed spectroscopic properties.

We checked correlations between colour and orbital parameters. A weak but clear correlation between colour and inclination was found for the resonant objects, in which the presence of the bluest objects (\textit{BB}) is relevant. In addition, weak but clear correlations between colour and phase slopes are found.

The fraction of red objects in the region ($e<0.42$, $i<21^{\rm o}$) \citep[see these limits in  ][]{2019AJ....157...94M, 2021AJ....162...19A} is larger than the fraction of blue objects in the same region. It is interesting to note that, in our sample, the eccentricity limit holds only when the extreme classes (\textit{BB} and \textit{RR}) are considered. The limit in inclination, on the other hand, also holds when two classes (\textit{BB} plus \textit{BR} - blue and \textit{IR} plus \textit{RR} - red) are considered.

Our reddest and bluest TNOs have different surface compositions and there is dynamical evidence \citep[see, for instance][]{2020AJ....160...46N} for different origins. Spectroscopic results \citep[see][]{2017A&A...604A..86M} are found to be in good agreement with the bluest and reddest objects in the taxonomic classification scheme by \citet{2005AJ....130.1291B, 2008ssbn.book..181F}, which we also use in this work. Therefore, our \textit{BB} and \textit{RR} objects are a relevant contribution to models of dynamical evolution of the Solar System. It is interesting to note, however, that we have a mix of colours for $13\lesssim a\lesssim 75$ AU, where most of our objects are placed.

With respect to their sizes, useful diameters for all objects were determined with
the help of absolute magnitudes. A trend showing that our bluest objects tend to harbour larger diameters, as compared to our reddest objects, has been found. This agrees with the hypothesis that larger objects may have modified or resurfaced their surfaces from processes that are not present in the smaller ones.

The Dark Energy Survey was designed to understand, based on its six years of ground-based multi-band observations, the nature of the dark energy that accelerates the expansion of the Universe. It is clear, however, that DES contributions go well beyond that\footnote{See publications in \url{https://dbweb8.fnal.gov:8443/DESPub/app/PB/pub/pbpublished}} and Solar System science is one of the examples. 
DES is a genuinely important predecessor of the unprecedented Legacy Survey of the Space and Time (LSST; \citealt{2009arXiv0912.0201L}; \citealt{ 2019ApJ...873..111I}), the 10-year survey of the Southern Hemisphere sky that will take place at the Vera C. Rubin Observatory and will image the visible sky from Cerro Pachón twice every week, enlightening numerous areas in Astronomy in visible wavelengths, Solar System in particular.

\section*{Acknowledgements}

\textbf{Author contribution: }F.S.F. and J.I.B.C. made significant contributions to the development of this work, including paper writing and figure production. R.C.B. contributed to data analysis and critical software development. M.V.B.-H., A.P and V.F.-P. contributed to the construction and validation of the catalog with astrometric and photometric information of known Solar System small bodies from DES observations and dynamical information from SkyBoT. M.A. contributed to paper revision and useful discussions and suggestions that significantly improved the text. P.H.B, H.W.L. (DES internal reviewers), F.B-R., A.G-J. and R.V.-M. provided useful discussions and suggestions that significantly improved the text. L.N.C. coordinates big data software, data storage and high performance hardware projects that supported this work. The remaining authors have made contributions to this paper that include, but are not limited to, the construction of DECam and other aspects of collecting the data; data processing and calibration; developing broadly used methods, codes, and simulations; running the pipelines and validation tests; and promoting the science analysis.

 F.S.F. acknowledges a CAPES support. This study was financed in part by Coordenação de Aperfeiçoamento de Pessoal de Nível Superior - Brazil (CAPES) - Finance Code 001. This research used computational resources from the Interinstitucional Laboratory of e-Astronomy (LIneA) with financial support from the National Institute of Science and Technology (INCT) of the e-Universo (Process number 465376/2014-2). The following grants are acknowledged: J.I.B.C.:  CNPq 305917/2019-6, 306691/2022-1 and FAPERJ 201.681/2019; R.C.B.: FAPERJ E26/202.125/2020; M.V.B.H.: FAPERJ E-26/ 200.480/2020; A.P.: CNPq 163708/2022-3; V.F.P.: CNPq, PIBIC/ON (process 143944/2022-3); M.A.: CNPq 427700/2018-3, 310683/2017-3, 473002/2013-2; P.H.B. University of Washington College of Arts and Sciences, Department of Astronomy, and the DiRAC Institute; F.B.R.: grant 314772/2020-0; R.V.M.: CNPq 307368/2021-1. The author acknowledges an anonymous referee for helping improve the text.

This research has made use of the VESPA portal and services (https://vespa.obspm.fr) funded by European Union's under grant agreement No 871149". Except where explicitly mentioned, data are distributed under Etalab Open License 2.0 (compliant with CC-BY 2.0 license).
The Europlanet 2024 Research Infrastructure project has received funding from the European Union's Horizon 2020 research and innovation programme under grant agreement No 871149.
The Europlanet 2020 Research Infrastructure project has received funding from the European Union's Horizon 2020 research and innovation programme under grant agreement No 654208.
This work used the EGI Infrastructure with the dedicated support of IN2P3-IRES and CESNET-MCC.
VESPA has first been designed in the frame of Europlanet-RI JRA4 work package (IDIS activity).
Additional funding was provided in France by the Action Spécifique Observatoire Virtuel and Programme National de Planétologie / INSU. 

This paper has gone through internal review by the DES collaboration.

Funding for the DES Projects has been provided by the U.S. Department of Energy, the U.S. National Science Foundation, the Ministry of Science and Education of Spain, 
the Science and Technology Facilities Council of the United Kingdom, the Higher Education Funding Council for England, the National Center for Supercomputing 
Applications at the University of Illinois at Urbana-Champaign, the Kavli Institute of Cosmological Physics at the University of Chicago, 
the Center for Cosmology and Astro-Particle Physics at the Ohio State University,
the Mitchell Institute for Fundamental Physics and Astronomy at Texas A\&M University, Financiadora de Estudos e Projetos, 
Funda{\c c}{\~a}o Carlos Chagas Filho de Amparo {\`a} Pesquisa do Estado do Rio de Janeiro, Conselho Nacional de Desenvolvimento Cient{\'i}fico e Tecnol{\'o}gico and 
the Minist{\'e}rio da Ci{\^e}ncia, Tecnologia e Inova{\c c}{\~a}o, the Deutsche Forschungsgemeinschaft and the Collaborating Institutions in the Dark Energy Survey. 

The Collaborating Institutions are Argonne National Laboratory, the University of California at Santa Cruz, the University of Cambridge, Centro de Investigaciones Energ{\'e}ticas, 
Medioambientales y Tecnol{\'o}gicas-Madrid, the University of Chicago, University College London, the DES-Brazil Consortium, the University of Edinburgh, 
the Eidgen{\"o}ssische Technische Hochschule (ETH) Z{\"u}rich, 
Fermi National Accelerator Laboratory, the University of Illinois at Urbana-Champaign, the Institut de Ci{\`e}ncies de l'Espai (IEEC/CSIC), 
the Institut de F{\'i}sica d'Altes Energies, Lawrence Berkeley National Laboratory, the Ludwig-Maximilians Universit{\"a}t M{\"u}nchen and the associated Excellence Cluster Universe, 
the University of Michigan, NSF's NOIRLab, the University of Nottingham, The Ohio State University, the University of Pennsylvania, the University of Portsmouth, 
SLAC National Accelerator Laboratory, Stanford University, the University of Sussex, Texas A\&M University, and the OzDES Membership Consortium.

Based in part on observations at Cerro Tololo Inter-American Observatory at NSF's NOIRLab (NOIRLab Prop. ID 2012B-0001; PI: J. Frieman), which is managed by the Association of Universities for Research in Astronomy (AURA) under a cooperative agreement with the National Science Foundation.

The DES data management system is supported by the National Science Foundation under Grant Numbers AST-1138766 and AST-1536171. 

The DES participants from Spanish institutions are partially supported by MICINN under grants ESP2017-89838, PGC2018-094773, PGC2018-102021, SEV-2016-0588, SEV-2016-0597, and MDM-2015-0509, some of which include ERDF funds from the European Union. IFAE is partially funded by the CERCA program of the Generalitat de Catalunya. Research leading to these results has received funding from the European Research
Council under the European Union's Seventh Framework Program (FP7/2007-2013) including ERC grant agreements 240672, 291329, and 306478. We  acknowledge support from the Brazilian Instituto Nacional de Ci\^encia
e Tecnologia (INCT) do e-Universo (CNPq grant 465376/2014-2).

This manuscript has been authored by Fermi Research Alliance, LLC under Contract No. DE-AC02-07CH11359 with the U.S. Department of Energy, Office of Science, Office of High Energy Physics.

\section*{Data Availability}

All data from Table~\ref{tab:taxonclass} are currently under consideration for publication at the Strasbourg Astronomical Data Center (CDS, see \url{https://cds.unistra.fr/}).  It should be also available as machine-readable table.  

\onecolumn
\begin{landscape}
\setlength
\tabcolsep{2.35pt}
\scriptsize \begin{center}
\begin{longtable}{llcccccccccccccccc}

\hline 
\multicolumn{1}{c}{N} &
\multicolumn{1}{l}{Object} & \multicolumn{1}{c}{$Hg$} & \multicolumn{1}{c}{$\beta_g$} & \multicolumn{1}{c}{$\beta_r$} & \multicolumn{1}{c}{$\beta_i$} & \multicolumn{1}{c}{$\beta_z$} & \multicolumn{1}{c}{$Hg-Hr$} & \multicolumn{1}{c}{$Hg-Hi$} & \multicolumn{1}{c}{$Hg-Hz$} & \multicolumn{1}{l}{Dyn. Class} & \multicolumn{1}{c}{Diam. (km)} &
\multicolumn{1}{c}{This Work} & \multicolumn{1}{c}{[1]} &
\multicolumn{1}{c}{[2]} &
\multicolumn{1}{c}{[3]} & \multicolumn{1}{c}{[4]}  \\ 
\hline
\endfirsthead

\multicolumn{3}{c}%
{{\bfseries \tablename\ \thetable{} -- continued from previous page}} \\
\hline 
\multicolumn{1}{c}{N} &
\multicolumn{1}{c}{Object} & \multicolumn{1}{c}{$Hg$} & \multicolumn{1}{c}{$\beta_g$} & \multicolumn{1}{c}{$\beta_r$} & \multicolumn{1}{c}{$\beta_i$} & \multicolumn{1}{c}{$\beta_z$} & \multicolumn{1}{c}{$Hg-Hr$} & \multicolumn{1}{c}{$Hg-Hi$} & \multicolumn{1}{c}{$Hg-Hz$} & \multicolumn{1}{c}{Dyn. Class} &\multicolumn{1}{c}{Diam. (km)} & \multicolumn{1}{c}{This Work}& \multicolumn{1}{c}{[1]} &
\multicolumn{1}{c}{[2]} & \multicolumn{1}{c}{[3]} & \multicolumn{1}{c}{[4]} \\ 
\hline
\endhead
\endfoot

\endlastfoot
1	&	 136199 Eris	&	-0.79394	($\pm 0.07528$)	&	0.09808	($\pm	0.25213$)	&	0.07163	($\pm	0.25544$)	&	0.06074	($\pm	0.17728$)	&	0.04952	($\pm	0.18036$)	&	0.543	($\pm	0.110$)	&	0.643	($\pm	0.107$)	&	0.565	($\pm	0.114$)	&	   Detached	        &	$2221_{-137}^{+148}$, $2221_{-187}^{+148}$ &                BB		&	BB	&	BB	&		&	BB	\\
2	&	 2010 TJ	&	5.53982	($\pm	0.14384$)	&	0.29991	($\pm	0.13438$)	&	0.19426	($\pm	0.13947$)	&	0.07153	($\pm	0.15380$)	&	0.13612	($\pm	0.25686$)	&	0.320	($\pm	0.206$)	&	0.293	($\pm	0.224$)	&	0.444	($\pm	0.324$)	&	   Detached	        &	$291_{-57}^{+91}$ &	                 BB		&		&		&		&		\\
3	&	 2001 QF298	&	5.53246	($\pm	0.09742$)	&	0.29852	($\pm	0.23679$)	&	0.19902	($\pm	0.09962$)	&	0.26912	($\pm	0.12557$)	&	0.47137	($\pm	0.14845$)	&	0.414	($\pm	0.118$)	&	0.533	($\pm	0.123$)	&	0.872	($\pm	0.156$)	&	   Resonant (3:2)	&	$430_{-70}^{+76}$ &	                 BB		&	BB	&		&	BB	&	BB	\\
4	&	 2013 RP98*	&	7.40231	($\pm	0.07621$)	&	0.22567	($\pm	0.09138$)	&	0.34171	($\pm	0.10924$)	&	0.11560	($\pm	0.13618$)	&	-0.09845	($\pm	0.19814$)	&	0.511	($\pm	0.116$)	&	0.542	($\pm	0.122$)	&	0.692	($\pm	0.177$)	&	   Hot Classical	&	$171_{-20}^{+25}$ &	                 BB		&		&		&		&		\\
5	&	 2013 UF31*	&	8.09176	($\pm	0.07970$)	&	0.30394	($\pm	0.07019$)	&	0.11381	($\pm	0.06992$)	&	0.02994	($\pm	0.07201$)	&	-0.01640	($\pm	0.08362$)	&	0.412	($\pm	0.116$)	&	0.553	($\pm	0.120$)	&	0.688	($\pm	0.138$)	&	   Resonant (3:2)	&	$118_{-13}^{+16}$ &	                 BB		&		&		&		&		\\
6	&	 2014 SO373*	&	7.23976	($\pm	0.15824$)	&	0.29655	($\pm	0.13266$)	&	0.16157	($\pm	0.10073$)	&	0.17332	($\pm	0.09401$)	&	0.13179	($\pm	0.13038$)	&	0.400	($\pm	0.198$)	&	0.556	($\pm	0.185$)	&	0.610	($\pm	0.220$)	&	   Resonant (5:3)	&	$129_{-18}^{+22}$ &	                 BB		&		&		&		&		\\
7	&	 2014 TB86	&	6.84360	($\pm	0.07690$)	&	0.41719	($\pm	0.10056$)	&	0.08067	($\pm	0.09960$)	&	0.08172	($\pm	0.09132$)	&	0.25496	($\pm	0.09738$)	&	0.373	($\pm	0.112$)	&	0.565	($\pm	0.117$)	&	0.853	($\pm	0.109$)	&	   Hot Classical	&	$214_{-25}^{+31}$ &	                 BB		&		&		&		&		\\
8	&	 2014 QA442	&	5.68230	($\pm	0.14541$)	&	-0.05429	($\pm	0.11717$)	&	0.00004	($\pm	0.08470$)	&	0.02651	($\pm	0.10506$)	&	0.04887	($\pm	0.12349$)	&	0.513	($\pm	0.179$)	&	0.607	($\pm	0.198$)	&	0.647	($\pm	0.206$)	&	   Resonant (7:4)	&	$271_{-36}^{+44}$ &	                 BB		&		&		&		&		\\
9	&	 2015 PF312	&	6.45139	($\pm	0.05107$)	&	0.30061	($\pm	0.05381$)	&	0.25433	($\pm	0.04923$)	&	0.21917	($\pm	0.04622$)	&	0.05303	($\pm	0.03368$)	&	0.437	($\pm	0.068$)	&	0.656	($\pm	0.065$)	&	0.611	($\pm	0.060$)	&	   Resonant (11:6)	&	$187_{-16}^{+20}$ &	                 BB		&		&		&		&		\\
10	&	 2013 QP95*	&	7.74242	($\pm	0.05856$)	&	0.12226	($\pm	0.05895$)	&	-0.01617	($\pm	0.06033$)	&	-0.02547	($\pm	0.04773$)	&	-0.07487	($\pm	0.03408$)	&	0.435	($\pm	0.091$)	&	0.688	($\pm	0.078$)	&	0.782	($\pm	0.068$)	&	   SDO	                &	$174_{-19}^{+25}$ &	                 BB		&		&		&		&		\\
11	&	 2013 RQ109*	&	8.23591	($\pm	0.12223$)	&	0.19321	($\pm	0.13125$)	&	0.17525	($\pm	0.08999$)	&	0.02098	($\pm	0.09897$)	&	0.03374	($\pm	0.07580$)	&	0.576	($\pm	0.143$)	&	0.538	($\pm	0.169$)	&	0.839	($\pm	0.146$)	&	   Resonant (4:3)	&	$85_{-10}^{+12}$ &	                 BB		&		&		&		&		\\
12	&	 2013 UE31*	&	6.70687	($\pm	0.40096$)	&	0.27306	($\pm	0.33276$)	&	0.14883	($\pm	0.20583$)	&	-0.06850	($\pm	0.28874$)	&	0.08448	($\pm	0.14106$)	&	0.369	($\pm	0.482$)	&	0.312	($\pm	0.531$)	&	0.534	($\pm	0.441$)	&	   Resonant (2:1)	&	$164_{-40}^{+54}$ &	                 BB		&		&		&		&		\\
13	&	 2016 SA56	&	7.44205	($\pm	0.14072$)	&	0.43923	($\pm	0.20291$)	&	0.30282	($\pm	0.15005$)	&	-0.01304	($\pm	0.12885$)	&	0.24630	($\pm	0.16775$)	&	0.474	($\pm	0.183$)	&	0.569	($\pm	0.190$)	&	0.909	($\pm	0.210$)	&	   Resonant (19:9)	&	$120_{-16}^{+19}$ &	                 BB		&		&		&		&		\\
14	&	 2016 SJ57	&	7.67754	($\pm	0.07861$)	&	0.21672	($\pm	0.08344$)	&	0.18157	($\pm	0.06830$)	&	-0.11235	($\pm	0.07114$)	&	0.26501	($\pm	0.27215$)	&	0.464	($\pm	0.103$)	&	0.403	($\pm	0.102$)	&	0.719	($\pm	0.182$)	&	   Resonant (4:3)	&	$107_{-11}^{+13}$ &	                 BB		&		&		&		&		\\
15	&	 2012 WD36*	&	6.89609	($\pm	0.08294$)	&	0.31022	($\pm	0.08968$)	&	0.41056	($\pm	0.08231$)	&	-0.03538	($\pm	0.08225$)	&	0.30903	($\pm	0.09167$)	&	0.473	($\pm	0.108$)	&	0.441	($\pm	0.108$)	&	0.831	($\pm	0.111$)	&	   Detached	        &	$161_{-27}^{+43}$ &	                 BB		&		&		&		&		\\
16	&	 2014 XB48*	&	7.63211	($\pm	0.14438$)	&	0.29415	($\pm	0.12444$)	&	0.11191	($\pm	0.10770$)	&	-0.00378	($\pm	0.11494$)	&	-0.05191	($\pm	0.15572$)	&	0.409	($\pm	0.199$)	&	0.451	($\pm	0.205$)	&	0.685	($\pm	0.231$)	&	   Resonant (7:2)	&	$108_{-14}^{+18}$ &	                 BB		&		&		&		&		\\
17	&	 2013 RO124*	&	7.53653	($\pm	0.05261$)	&	0.46101	($\pm	0.09708$)	&	0.23375	($\pm	0.06971$)	&	0.02174	($\pm	0.10047$)	&	0.03940	($\pm	0.12783$)	&	0.428	($\pm	0.078$)	&	0.474	($\pm	0.079$)	&	0.889	($\pm	0.134$)	&	   SDO	                &	 $192_{-21}^{+27}$ &	                 BB		&		&		&		&		\\
18	&	 2010 TY53	&	5.86531	($\pm	0.09339$)	&	0.09790	($\pm	0.07533$)	&	0.08043	($\pm	0.06939$)	&	-0.00861	($\pm	0.10386$)	&	0.09818	($\pm	0.13778$)	&	0.466	($\pm	0.129$)	&	0.486	($\pm	0.180$)	&	0.741	($\pm	0.172$)	&	   SDO	                &	$417_{-53}^{+68}$ &	                 BB		&		&		&		&		\\
19	&	 2060 Chiron	&	5.86356	($\pm	0.09885$)	&	0.11535	($\pm	0.05570$)	&	0.06482	($\pm	0.08063$)	&	0.06090	($\pm	0.06757$)	&	0.06845	($\pm	0.07174$)	&	0.327	($\pm	0.162$)	&	0.372	($\pm	0.160$)	&	0.409	($\pm	0.179$)	&	   Centaur	&	$236_{-35}^{+40}$ &	                 BB		&	BB	&		&	BB	&		\\
20	&	 2014 QS441*	&	5.81261	($\pm	0.05328$)	&	-0.01010	($\pm	0.06385$)	&	0.15697	($\pm	0.05234$)	&	0.11716	($\pm	0.05754$)	&	0.07723	($\pm	0.03992$)	&	0.503	($\pm	0.069$)	&	0.510	($\pm	0.073$)	&	0.459	($\pm	0.064$)	&	   Hot Classical	&	$355_{-36}^{+46}$ &	                 BB		&		&		&		&		\\
21	&	 2014 WC536	&	7.60884	($\pm	0.13716$)	&	0.12156	($\pm	0.10782$)	&	0.13956	($\pm	0.15195$)	&	-0.05753	($\pm	0.11455$)	&	0.03818	($\pm	0.07601$)	&	0.575	($\pm	0.258$)	&	0.456	($\pm	0.213$)	&	0.615	($\pm	0.172$)	&	   Resonant (3:2)	&	$153_{-22}^{+27}$ &	                 BB		&		&		&		&		\\
22	&	 2014 OB394	&	6.73129	($\pm	0.06412$)	&	-0.04612	($\pm	0.10586$)	&	0.08842	($\pm	0.06640$)	&	0.05996	($\pm	0.14293$)	&	0.08770	($\pm	0.19349$)	&	0.603	($\pm	0.091$)	&	0.780	($\pm	0.105$)	&	0.907	($\pm	0.158$)	&	   SDO	&	$289_{-33}^{+42}$ &	                 BR		&		&		&		&		\\
23	&	 2013 TE172*	&	7.66070	($\pm	0.09444$)	&	0.42728	($\pm	0.13007$)	&	0.13958	($\pm	0.09885$)	&	0.11383	($\pm	0.09860$)	&	0.09707	($\pm	0.05556$)	&	0.650	($\pm	0.147$)	&	0.896	($\pm	0.149$)	&	0.891	($\pm	0.113$)	&	   SDO	&	$190_{-25}^{+32}$ &	                 BR		&		&		&		&		\\
24	&	 2014 SN363*	&	7.61910	($\pm	0.17640$)	&	0.07633	($\pm	0.23269$)	&	0.18157	($\pm	0.11847$)	&	-0.01193	($\pm	0.09711$)	&	0.12165	($\pm	0.20377$)	&	0.663	($\pm	0.213$)	&	0.713	($\pm	0.206$)	&	1.019	($\pm	0.247$)	&	   Resonant (11:6)	&	$115_{-17}^{+21}$ &	                 BR		&		&		&		&		\\
25	&	 2013 RD109*	&	8.61443	($\pm	0.06608$)	&	0.17465	($\pm	0.05832$)	&	0.21956	($\pm	0.07559$)	&	0.21806	($\pm	0.07087$)	&	-0.10940	($\pm	0.10454$)	&	0.710	($\pm	0.124$)	&	0.855	($\pm	0.108$)	&	0.854	($\pm	0.149$)	&	   SDO	&	$125_{-15}^{+19}$ &	                 BR		&		&		&		&		\\
26	&	 2014 NB66	&	6.12518	($\pm	0.05103$)	&	0.09186	($\pm	0.10797$)	&	0.24101	($\pm	0.11722$)	&	0.08763	($\pm	0.11039$)	&	-0.26592	($\pm	0.27117$)	&	0.717	($\pm	0.088$)	&	0.927	($\pm	0.087$)	&	0.989	($\pm	0.115$)	&	   Cold Classical	&	$255_{-17}^{+19}$ &	                 BR		&		&		&		&		\\
27	&	 2013 SO102*	&	8.05735	($\pm	0.07414$)	&	0.06123	($\pm	0.09041$)	&	-0.07443	($\pm	0.13397$)	&	0.01514	($\pm	0.06201$)	&	0.15878	($\pm	0.07811$)	&	0.505	($\pm	0.117$)	&	0.601	($\pm	0.103$)	&	0.994	($\pm	0.113$)	&	   Resonant (3:2)	&	$123_{-13}^{+16}$ &	                 BR		&		&		&		&		\\
28	&	 2013 SA100	&	6.30524	($\pm	0.07263$)	&	0.36839	($\pm	0.12542$)	&	0.34469	($\pm	0.18486$)	&	0.09865	($\pm	0.12099$)	&	-0.14421	($\pm	0.13887$)	&	0.621	($\pm	0.143$)	&	0.654	($\pm	0.102$)	&	0.796	($\pm	0.113$)	&	   Hot Classical	&	$291_{-34}^{+43}$ &	                 BR		&		&		&		&		\\
29	&	 2013 TZ171*	&	7.71058	($\pm	0.07739$)	&	0.17679	($\pm	0.06203$)	&	0.13537	($\pm	0.05302$)	&	-0.01743	($\pm	0.07681$)	&	0.24026	($\pm	0.10805$)	&	0.510	($\pm	0.101$)	&	0.665	($\pm	0.120$)	&	1.020	($\pm	0.166$)	&	   Hot Classical	&	$148_{-17}^{+21}$ &	                 BR		&		&		&		&		\\
30	&	 2013 RS109	&	7.37979	($\pm	0.05889$)	&	0.14294	($\pm	0.12170$)	&	-0.05004	($\pm	0.09235$)	&	-0.00846	($\pm	0.11723$)	&	-0.14598	($\pm	0.08187$)	&	0.518	($\pm	0.087$)	&	0.708	($\pm	0.101$)	&	1.023	($\pm	0.086$)	&	   Resonant (11:6)	&	$124_{-11}^{+14}$ &	                 BR		&		&		&		&		\\
31	&	 2007 RW10	&	7.12967	($\pm	0.07410$)	&	0.10114	($\pm	0.06339$)	&	0.11567	($\pm	0.08207$)	&	0.10778	($\pm	0.07492$)	&	0.07811	($\pm	0.08708$)	&	0.570	($\pm	0.120$)	&	0.710	($\pm	0.103$)	&	0.789	($\pm	0.126$)	&	   SDO	&	$174_{-50}^{+74}$ &	                 BR		&		&		&		&		\\
32	&	 2013 SP102	&	6.47372	($\pm	0.08709$)	&	0.12799	($\pm	0.13515$)	&	0.17615	($\pm	0.11970$)	&	0.12303	($\pm	0.07671$)	&	0.03422	($\pm	0.09583$)	&	0.626	($\pm	0.123$)	&	0.719	($\pm	0.109$)	&	0.860	($\pm	0.123$)	&	   Resonant (3:2)	&	$262_{-29}^{+35}$ &	                 BR		&		&		&		&		\\
33	&	 2013 RN124	&	7.05427	($\pm	0.07582$)	&	0.31776	($\pm	0.10988$)	&	0.25068	($\pm	0.08854$)	&	0.26507	($\pm	0.09388$)	&	0.05383	($\pm	0.25931$)	&	0.627	($\pm	0.104$)	&	0.719	($\pm	0.091$)	&	0.952	($\pm	0.132$)	&	   Hot Classical	&	$206_{-23}^{+30}$ &	                 BR		&		&		&		&		\\
34	&	 2013 TF172	&	7.43152	($\pm	0.06973$)	&	0.12909	($\pm	0.11614$)	&	0.14169	($\pm	0.08224$)	&	0.02634	($\pm	0.08169$)	&	-0.18530	($\pm	0.22888$)	&	0.496	($\pm	0.088$)	&	0.722	($\pm	0.090$)	&	0.894	($\pm	0.147$)	&	   Resonant (7:4)	&	$121_{-12}^{+14}$ &	                 BR		&		&		&		&		\\
35	&	 2016 QV89*	&	6.39664	($\pm	0.05486$)	&	0.16738	($\pm	0.07945$)	&	0.17847	($\pm	0.06285$)	&	-0.09890	($\pm	0.06446$)	&	0.10003	($\pm	0.05564$)	&	0.582	($\pm	0.072$)	&	0.725	($\pm	0.074$)	&	0.952	($\pm	0.072$)	&	   Detached	&	$208_{-32}^{+52}$ &	                 BR		&		&		&		&		\\
36	&	 2012 TD324	&	7.43607	($\pm	0.06240$)	&	0.23411	($\pm	0.09130$)	&	0.17770	($\pm	0.08499$)	&	0.01180	($\pm	0.11587$)	&	0.11494	($\pm	0.10664$)	&	0.614	($\pm	0.096$)	&	0.739	($\pm	0.104$)	&	1.061	($\pm	0.126$)	&	   Resonant (3:2)	&	$167_{-16}^{+20}$ &	                 BR		&		&		&		&		\\
37	&	 2014 SP373*	&	7.51920	($\pm	0.07602$)	&	0.03741	($\pm	0.07048$)	&	-0.05661	($\pm	0.08064$)	&	0.05655	($\pm	0.15122$)	&	-0.07885	($\pm	0.12508$)	&	0.457	($\pm	0.122$)	&	0.744	($\pm	0.151$)	&	0.806	($\pm	0.160$)	&	   Resonant (3:2)	&	$156_{-17}^{+20}$ &	                 BR		&		&		&		&		\\
38	&	 2013 RJ109*	&	7.34073	($\pm	0.05974$)	&	0.30349	($\pm	0.08806$)	&	0.11528	($\pm	0.07569$)	&	-0.23554	($\pm	0.14205$)	&	-0.18847	($\pm	0.22790$)	&	0.599	($\pm	0.084$)	&	0.744	($\pm	0.091$)	&	0.847	($\pm	0.146$)	&	   Resonant (7:4)	&	$129_{-12}^{+14}$ &	                 BR		&		&		&		&		\\
39	&	 2014 LQ28	&	6.35506	($\pm	0.11186$)	&	0.25684	($\pm	0.11604$)	&	0.03198	($\pm	0.08926$)	&	0.18095	($\pm	0.11476$)	&	0.11819	($\pm	0.20734$)	&	0.517	($\pm	0.134$)	&	0.812	($\pm	0.138$)	&	1.025	($\pm	0.158$)	&	   Resonant (7:4)	&	$199_{-23}^{+28}$ &	                 BR		&		&		&		&		\\
40	&	 2016 SS55	&	7.01610	($\pm	0.08117$)	&	0.25785	($\pm	0.10543$)	&	-0.08950	($\pm	0.08934$)	&	0.13763	($\pm	0.09655$)	&	0.12718	($\pm	0.18871$)	&	0.533	($\pm	0.114$)	&	0.910	($\pm	0.114$)	&	1.084	($\pm	0.188$)	&	   Detached	&	$155_{-26}^{+41}$ &	                 BR		&		&		&		&		\\
41	&	 2016 TT94*	&	7.08401	($\pm	0.25875$)	&	0.06904	($\pm	0.20402$)	&	0.02449	($\pm	0.15199$)	&	0.19905	($\pm	0.17602$)	&	0.01389	($\pm	0.18146$)	&	0.491	($\pm	0.321$)	&	0.915	($\pm	0.344$)	&	0.743	($\pm	0.334$)	&	   Resonant (5:3)	&	$142_{-26}^{+33}$ &	                 BR		&		&		&		&		\\
42	&	 2016 SQ58	&	8.30793	($\pm	0.09946$)	&	0.23484	($\pm	0.09725$)	&	-0.04697	($\pm	0.07313$)	&	0.20972	($\pm	0.07887$)	&	-0.00353	($\pm	0.09801$)	&	0.387	($\pm	0.135$)	&	0.863	($\pm	0.132$)	&	0.974	($\pm	0.152$)	&	   Resonant (4:3)	&	$79_{-9}^{+11}$ &	                 BR		&		&		&		&		\\
43	&	 2013 RW124*	&	7.72249	($\pm	0.08829$)	&	0.42706	($\pm	0.09349$)	&	0.35756	($\pm	0.10068$)	&	0.13714	($\pm	0.12350$)	&	0.29263	($\pm	0.19948$)	&	0.458	($\pm	0.126$)	&	0.606	($\pm	0.143$)	&	0.896	($\pm	0.219$)	&	   Resonant (4:3)	&	$105_{-11}^{+13}$ &	                 BR		&		&		&		&		\\
44	&	 2014 ON6	&	12.56718	($\pm	0.10342$)	&	0.05668	($\pm	0.02323$)	&	0.05312	($\pm	0.02360$)	&	0.05423	($\pm	0.02367$)	&	0.02946	($\pm	0.08508$)	&	0.827	($\pm	0.147$)	&	1.073	($\pm	0.150$)	&	0.986	($\pm	0.483$)	&	   Centaur	&	$21_{-2}^{+2}$ &	                 IR		&		&		&		&		\\
45	&	 2010 SB41	&	8.65504	($\pm	0.07955$)	&	0.03918	($\pm	0.10535$)	&	-0.35459	($\pm	0.13533$)	&	0.09692	($\pm	0.08870$)	&	0.04525	($\pm	0.05385$)	&	0.471	($\pm	0.134$)	&	1.126	($\pm	0.113$)	&	1.193	($\pm	0.102$)	&	   Resonant (3:2)	&	$92_{-10}^{+12}$ &	                 IR		&		&		&		&		\\
46	&	 2013 SF106*	&	5.31641	($\pm	0.24701$)	&	0.37959	($\pm	0.24500$)	&	0.14496	($\pm	0.21724$)	&	0.22248	($\pm	0.20380$)	&	0.42497	($\pm	0.25601$)	&	0.491	($\pm	0.353$)	&	0.877	($\pm	0.336$)	&	1.251	($\pm	0.381$)	&	   Detached	&	$335_{-81}^{+132}$ &	                 IR		&		&		&		&		\\
47	&	 2016 SE56	&	7.50792	($\pm	0.07028$)	&	0.24665	($\pm	0.08252$)	&	0.02750	($\pm	0.11003$)	&	0.40117	($\pm	0.14520$)	&	0.07099	($\pm	0.08371$)	&	0.497	($\pm	0.132$)	&	1.144	($\pm	0.162$)	&	1.058	($\pm	0.116$)	&	   SDO	&	$197_{-24}^{+31}$ &	                 IR		&		&		&		&		\\
48	&	 2010 RF188	&	5.75169	($\pm	0.10288$)	&	0.04667	($\pm	0.32891$)	&	0.29100	($\pm	0.14347$)	&	-0.16908	($\pm	0.16117$)	&	0.20370	($\pm	0.24326$)	&	0.742	($\pm	0.126$)	&	0.880	($\pm	0.125$)	&	1.094	($\pm	0.149$)	&	   SDO	&	$470_{-61}^{+79}$ &	                 IR		&		&		&		&		\\
49	&	 2013 TQ187*	&	7.96942	($\pm	0.11787$)	&	-0.36609	($\pm	0.14149$)	&	-0.00996	($\pm	0.17522$)	&	-0.32048	($\pm	0.14275$)	&	-0.08442	($\pm	0.10704$)	&	0.883	($\pm	0.201$)	&	0.862	($\pm	0.198$)	&	1.296	($\pm	0.162$)	&	   Resonant (7:4)	&	$103_{-13}^{+15}$ &	                 IR		&		&		&		&		\\
50	&	 2013 RM109*	&	7.73851	($\pm	0.17566$)	&	-0.31635	($\pm	0.22828$)	&	-0.05306	($\pm	0.14919$)	&	-0.49404	($\pm	0.14717$)	&	-0.42629	($\pm	0.15969$)	&	0.885	($\pm	0.219$)	&	0.984	($\pm	0.215$)	&	1.353	($\pm	0.215$)	&	   Resonant (9:4)	&	$115_{-16}^{+20}$ &	                 IR		&		&		&		&		\\
51	&	 2006 SG415	&	8.38376	($\pm	0.10006$)	&	-0.13023	($\pm	0.17507$)	&	0.38107	($\pm	0.10280$)	&	-0.05064	($\pm	0.13726$)	&	-0.17049	($\pm	0.11802$)	&	0.943	($\pm	0.119$)	&	1.027	($\pm	0.129$)	&	1.189	($\pm	0.128$)	&	   SDO	&	$146_{-19}^{+24}$ &	                 IR		&		&		&		&		\\
52	&	 2014 SK349	&	8.17786	($\pm	0.12311$)	&	0.16852	($\pm	0.28659$)	&	0.15217	($\pm	0.17162$)	&	0.10560	($\pm	0.09901$)	&	0.14135	($\pm	0.58645$)	&	0.607	($\pm	0.156$)	&	0.880	($\pm	0.143$)	&	1.132	($\pm	0.761$)	&	   Resonant (3:2)	&	$119_{-15}^{+19}$ &	                 IR		&		&		&		&		\\
53	&	 2014 QL441	&	7.08887	($\pm	0.06869$)	&	0.21788	($\pm	0.05537$)	&	0.22335	($\pm	0.05166$)	&	0.02583	($\pm	0.03960$)	&	0.22688	($\pm	0.03351$)	&	0.754	($\pm	0.092$)	&	0.925	($\pm	0.083$)	&	1.165	($\pm	0.082$)	&	   SDO	&	$255_{-29}^{+38}$ &	                 IR		&		&		&		&		\\
54	&	 2014 QS510*	&	8.52427	($\pm	0.08618$)	&	-0.14293	($\pm	0.11708$)	&	0.28089	($\pm	0.16209$)	&	0.03059	($\pm	0.07873$)	&	0.00367	($\pm	0.13703$)	&	0.868	($\pm	0.134$)	&	0.931	($\pm	0.109$)	&	1.380	($\pm	0.139$)	&	   Resonant (7:2)	&	$80_{-8}^{+10}$ &	                 IR		&		&		&		&		\\
55	&	 2014 UD241*	&	7.59668	($\pm	0.09057$)	&	0.29146	($\pm	0.15147$)	&	0.04755	($\pm	0.15054$)	&	0.00239	($\pm	0.16052$)	&	-0.00851	($\pm	0.11953$)	&	0.536	($\pm	0.150$)	&	0.955	($\pm	0.156$)	&	1.201	($\pm	0.127$)	&	   Resonant (5:3)	&	$113_{-12}^{+15}$ &	                 IR		&		&		&		&		\\
56	&	 2014 UP250*	&	6.77534	($\pm	0.20012$)	&	0.04773	($\pm	0.30013$)	&	0.10565	($\pm	0.12656$)	&	0.08576	($\pm	0.15230$)	&	-0.45206	($\pm	0.22593$)	&	0.853	($\pm	0.217$)	&	0.965	($\pm	0.226$)	&	1.117	($\pm	0.220$)	&	   Resonant (7:3)	&	$177_{-27}^{+34}$ &	                 IR		&		&		&		&		\\
57	&	 2014 OD394	&	6.06076	($\pm	0.07262$)	&	0.25957	($\pm	0.12571$)	&	0.07364	($\pm	0.12415$)	&	0.31568	($\pm	0.13230$)	&	0.22842	($\pm	0.16208$)	&	0.539	($\pm	0.113$)	&	0.986	($\pm	0.107$)	&	1.060	($\pm	0.128$)	&	   Resonant (9:5)	&	$229_{-23}^{+27}$ &	                 IR		&		&		&		&		\\
58	&	 2016 SP56	&	6.91744	($\pm	0.07657$)	&	0.04823	($\pm	0.11071$)	&	0.04328	($\pm	0.09797$)	&	0.35095	($\pm	0.13948$)	&	0.20107	($\pm	0.07518$)	&	0.660	($\pm	0.111$)	&	0.986	($\pm	0.127$)	&	1.076	($\pm	0.097$)	&	   Resonant (19:9)	&	$159_{-16}^{+19}$ &	                 IR		&		&		&		&		\\
59	&	 2008 SO266	&	7.03613	($\pm	0.07555$)	&	0.21294	($\pm	0.08021$)	&	0.00246	($\pm	0.07305$)	&	-0.05013	($\pm	0.07844$)	&	0.06763	($\pm	0.09531$)	&	0.722	($\pm	0.109$)	&	1.005	($\pm	0.115$)	&	1.256	($\pm	0.130$)	&	   Resonant (3:2)	&	$207_{-22}^{+27}$ &	                 IR		&		&		&		&		\\
60	&	 2014 OR394	&	7.83466	($\pm	0.06019$)	&	0.10224	($\pm	0.10821$)	&	0.22379	($\pm	0.10151$)	&	0.04876	($\pm	0.15723$)	&	0.09719	($\pm	0.06468$)	&	0.724	($\pm	0.084$)	&	1.009	($\pm	0.114$)	&	1.332	($\pm	0.087$)	&	   SDO	&	$179_{-20}^{+26}$ &	                 IR		&		&		&		&		\\
61	&	 2013 TV158	&	7.01409	($\pm	0.04382$)	&	0.17877	($\pm	0.04564$)	&	0.13430	($\pm	0.03620$)	&	0.12989	($\pm	0.03697$)	&	0.15421	($\pm	0.02743$)	&	0.721	($\pm	0.058$)	&	1.018	($\pm	0.064$)	&	1.196	($\pm	0.055$)	&	   Detached	&	$162_{-25}^{+39}$ &	                 IR		&		&		&		&		\\
62	&	 2013 SJ106*	&	7.25011	($\pm	0.07742$)	&	0.09808	($\pm	0.09645$)	&	0.14624	($\pm	0.07402$)	&	0.06128	($\pm	0.07642$)	&	0.14322	($\pm	0.13298$)	&	0.588	($\pm	0.095$)	&	1.024	($\pm	0.098$)	&	1.348	($\pm	0.122$)	&	   Resonant (9:5)	&	$134_{-13}^{+16}$ &	                 IR		&		&		&		&		\\
63	&	 2014 SP363*	&	6.80930	($\pm	0.14508$)	&	0.39107	($\pm	0.17135$)	&	0.14206	($\pm	0.15268$)	&	0.09505	($\pm	0.12932$)	&	0.01666	($\pm	0.18247$)	&	0.709	($\pm	0.197$)	&	1.046	($\pm	0.189$)	&	1.208	($\pm	0.229$)	&	   Resonant (3:1)	&	$170_{-22}^{+27}$ &	                 IR		&		&		&		&		\\
64	&	 2016 SG58	&	7.99395	($\pm	0.06520$)	&	0.12398	($\pm	0.08480$)	&	0.29567	($\pm	0.08021$)	&	0.16460	($\pm	0.07137$)	&	-0.15654	($\pm	0.08421$)	&	0.918	($\pm	0.092$)	&	1.056	($\pm	0.101$)	&	1.073	($\pm	0.091$)	&	   SDO	&	$174_{-20}^{+25}$ &	                 IR		&		&		&		&		\\
65	&	 2014 QM441*	&	9.19004	($\pm	0.10748$)	&	-0.14194	($\pm	0.12386$)	&	0.06065	($\pm	0.07507$)	&	0.11043	($\pm	0.05196$)	&	-0.03523	($\pm	0.05158$)	&	0.735	($\pm	0.131$)	&	1.068	($\pm	0.121$)	&	1.093	($\pm	0.121$)	&	   Resonant (3:2)	&	$77_{-9}^{+11}$ &	                 IR		&		&		&		&		\\
66	&	 2005 QU182	&	4.23740	($\pm	0.16935$)	&	0.33052	($\pm	0.31043$)	&	0.29559	($\pm	0.31541$)	&	0.38298	($\pm	0.14802$)	&	0.23680	($\pm	0.13276$)	&	0.756	($\pm	0.242$)	&	1.074	($\pm	0.195$)	&	1.117	($\pm	0.192$)	&	   Detached	&	$395_{-100}^{+135}$ &	                 IR		&		&		&		&		\\
67	&	 2014 SK378*	&	8.85179	($\pm	0.09550$)	&	0.17264	($\pm	0.06666$)	&	0.14695	($\pm	0.07125$)	&	0.22220	($\pm	0.07250$)	&	0.13274	($\pm	0.09802$)	&	0.724	($\pm	0.156$)	&	1.104	($\pm	0.154$)	&	1.206	($\pm	0.196$)	&	   SDO	&	$112_{-15}^{+19}$ &	                 IR		&		&		&		&		\\
68	&	 2003 QM91	&	7.14349	($\pm	0.04743$)	&	0.17407	($\pm	0.08075$)	&	0.10981	($\pm	0.08295$)	&	0.06269	($\pm	0.07845$)	&	0.08417	($\pm	0.13187$)	&	0.811	($\pm	0.080$)	&	1.105	($\pm	0.076$)	&	1.274	($\pm	0.125$)	&	   Cold Classical	&	$163_{-11}^{+12}$ &	                 IR		&		&		&		&		\\
69	&	 2014 VV39*	&	7.24348	($\pm	0.05767$)	&	0.23724	($\pm	0.09338$)	&	0.12951	($\pm	0.09042$)	&	0.11731	($\pm	0.08847$)	&	-0.02061	($\pm	0.10987$)	&	0.769	($\pm	0.090$)	&	1.117	($\pm	0.089$)	&	1.377	($\pm	0.091$)	&	   Cold Classical	&	$154_{-11}^{+12}$ &	                 IR		&		&		&		&		\\
70	&	 2013 SN106*	&	7.43946	($\pm	0.05793$)	&	0.02632	($\pm	0.11313$)	&	0.14245	($\pm	0.08162$)	&	0.18850	($\pm	0.10094$)	&	-0.06302	($\pm	0.26905$)	&	0.614	($\pm	0.086$)	&	1.119	($\pm	0.087$)	&	1.296	($\pm	0.198$)	&	   Resonant (7:4)	&	$124_{-11}^{+14}$ &	                 IR		&		&		&		&		\\
71	&	 2013 UT22*	&	7.61362	($\pm	0.08220$)	&	0.44333	($\pm	0.08456$)	&	0.25126	($\pm	0.07383$)	&	0.25882	($\pm	0.10060$)	&	-0.07212	($\pm	0.06793$)	&	0.752	($\pm	0.113$)	&	1.122	($\pm	0.134$)	&	1.085	($\pm	0.103$)	&	   Resonant (5:3)	&	$118_{-12}^{+15}$ &	                 IR		&		&		&		&		\\
72	&	 2015 TW361	&	7.84039	($\pm	0.06708$)	&	0.08134	($\pm	0.09436$)	&	-0.00967	($\pm	0.11180$)	&	0.15226	($\pm	0.11546$)	&	0.00537	($\pm	0.10658$)	&	0.673	($\pm	0.106$)	&	1.128	($\pm	0.130$)	&	1.146	($\pm	0.119$)	&	   Detached	&	$109_{-18}^{+28}$ &	                 IR		&		&		&		&		\\
73	&	 2014 QZ510*	&	8.42014	($\pm	0.10453$)	&	-0.00437	($\pm	0.08327$)	&	0.12132	($\pm	0.08542$)	&	0.22111	($\pm	0.09666$)	&	-0.05746	($\pm	0.07594$)	&	0.851	($\pm	0.150$)	&	1.134	($\pm	0.151$)	&	0.994	($\pm	0.137$)	&	   Resonant (3:2)	&	$113_{-13}^{+16}$ &	                 IR		&		&		&		&		\\
74	&	 2005 TB190	&	5.02302	($\pm	0.04387$)	&	0.28297	($\pm	0.08237$)	&	0.18411	($\pm	0.08585$)	&	0.23048	($\pm	0.09003$)	&	0.10647	($\pm	0.13469$)	&	0.809	($\pm	0.088$)	&	1.148	($\pm	0.073$)	&	1.212	($\pm	0.101$)	&	   Detached	&	$414_{-64}^{+102}$ &	                 IR		&		&		&		&	IR	\\
75	&	 1999 RB216	&	7.84352	($\pm	0.07891$)	&	0.07828	($\pm	0.10738$)	&	0.15992	($\pm	0.09904$)	&	0.14231	($\pm	0.08502$)	&	0.22330	($\pm	0.06836$)	&	0.890	($\pm	0.108$)	&	1.150	($\pm	0.101$)	&	1.301	($\pm	0.100$)	&	   Resonant (2:1)	&	$109_{-11}^{+13}$ &	                 IR		&	IR,BR	&		&	BR	&	IR,BR	\\
76	&	 2014 WD536*	&	8.33849	($\pm	0.14872$)	&	0.08464	($\pm	0.12899$)	&	0.08495	($\pm	0.07618$)	&	0.23418	($\pm	0.07077$)	&	0.04106	($\pm	0.09456$)	&	0.635	($\pm	0.173$)	&	1.159	($\pm	0.174$)	&	1.049	($\pm	0.191$)	&	   Resonant (3:2)	&	$111_{-15}^{+19}$ &	                 IR		&		&		&		&		\\
77	&	 2013 RC109*	&	8.78482	($\pm	0.11312$)	&	-0.10835	($\pm	0.13707$)	&	0.00419	($\pm	0.19297$)	&	0.31933	($\pm	0.17571$)	&	0.09193	($\pm	0.25023$)	&	0.706	($\pm	0.193$)	&	1.183	($\pm	0.214$)	&	1.042	($\pm	0.378$)	&	   Resonant (3:2)	&	$92_{-12}^{+14}$ &	                 IR		&		&		&		&		\\
78	&	 2014 QU441*	&	8.41596	($\pm	0.10709$)	&	-0.00952	($\pm	0.09712$)	&	0.02195	($\pm	0.06251$)	&	0.25896	($\pm	0.06051$)	&	-0.05211	($\pm	0.07384$)	&	0.613	($\pm	0.127$)	&	1.213	($\pm	0.125$)	&	1.123	($\pm	0.132$)	&	   Hot Classical	&	$110_{-14}^{+18}$ &	                 IR		&		&		&		&		\\
79	&	 2013 SK106	&	7.45439	($\pm	0.08728$)	&	0.08580	($\pm	0.11164$)	&	0.00155	($\pm	0.10272$)	&	0.45042	($\pm	0.10697$)	&	0.10998	($\pm	0.15457$)	&	0.757	($\pm	0.121$)	&	1.221	($\pm	0.113$)	&	1.181	($\pm	0.146$)	&	   Resonant (9:5)	&	$127_{-13}^{+16}$ &	                 IR		&		&		&		&		\\
80	&	 2014 SS373*	&	7.02399	($\pm	0.20099$)	&	0.49994	($\pm	0.20732$)	&	0.38971	($\pm	0.20718$)	&	0.10494	($\pm	0.25439$)	&	0.20217	($\pm	0.20397$)	&	0.936	($\pm	0.283$)	&	1.036	($\pm	0.310$)	&	1.503	($\pm	0.266$)	&	   Resonant (8:3)	&	$161_{-25}^{+32}$ &	                 RR		&		&		&		&		\\
81	&	 2014 UC225*	&	8.45318	($\pm	0.14340$)	&	-0.05789	($\pm	0.31379$)	&	0.09379	($\pm	0.16612$)	&	-0.00608	($\pm	0.13236$)	&	0.01357	($\pm	0.17069$)	&	0.961	($\pm	0.172$)	&	1.290	($\pm	0.183$)	&	1.507	($\pm	0.254$)	&	   Resonant (5:4)	&	$84_{-11}^{+13}$ &	                 RR		&		&		&		&		\\
82	&	 2013 TM187*	&	9.01161	($\pm	0.09293$)	&	-0.03112	($\pm	0.09615$)	&	0.12577	($\pm	0.08818$)	&	0.33943	($\pm	0.08677$)	&	0.14368	($\pm	0.12924$)	&	1.019	($\pm	0.120$)	&	1.141	($\pm	0.133$)	&	1.574	($\pm	0.189$)	&	   Resonant (4:3)	&	$66_{-7}^{+8}$ &	                 RR		&		&		&		&		\\
83	&	 1999 OX3	&	7.90379	($\pm	0.09565$)	&	0.11721	($\pm	0.06200$)	&	0.11778	($\pm	0.07767$)	&	0.12010	($\pm	0.05547$)	&	0.07600	($\pm	0.09578$)	&	1.027	($\pm	0.135$)	&	1.443	($\pm	0.164$)	&	1.570	($\pm	0.176$)	&	   SDO	&	$155_{-22}^{+26}$ &	                 RR		&	RR	&	RR	&	RR	&	RR	\\
84	&	 2012 WF37	&	8.97819	($\pm	0.10971$)	&	0.21865	($\pm	0.11352$)	&	0.10666	($\pm	0.11046$)	&	-0.13733	($\pm	0.13153$)	&	0.20115	($\pm	0.06953$)	&	0.952	($\pm	0.141$)	&	1.073	($\pm	0.181$)	&	1.643	($\pm	0.130$)	&	   Resonant (3:2)	&	$89_{-11}^{+13}$ &	                 RR		&		&		&		&		\\
85	&	 2014 QP441	&	9.91427	($\pm	0.07407$)	&	-0.12757	($\pm	0.06009$)	&	0.20392	($\pm	0.04570$)	&	0.18069	($\pm	0.03319$)	&	0.19331	($\pm	0.04637$)	&	0.929	($\pm	0.093$)	&	1.105	($\pm	0.085$)	&	1.520	($\pm	0.094$)	&	   SDO	&	$72_{-8}^{+11}$ &	                 RR		&		&		&		&		\\
86	&	 2014 PM82	&	7.50616	($\pm	0.05174$)	&	0.08940	($\pm	0.07016$)	&	0.27837	($\pm	0.09126$)	&	-0.00244	($\pm	0.09914$)	&	0.09846	($\pm	0.09862$)	&	1.066	($\pm	0.087$)	&	1.188	($\pm	0.098$)	&	1.443	($\pm	0.110$)	&	   Detached	&	$139_{-22}^{+35}$ &	                 RR		&		&		&		&		\\
87	&	 2010 PC88	&	8.37575	($\pm	0.13297$)	&	0.18560	($\pm	0.10356$)	&	0.18578	($\pm	0.09313$)	&	0.10124	($\pm	0.11225$)	&	0.10442	($\pm	0.08760$)	&	0.913	($\pm	0.185$)	&	1.190	($\pm	0.181$)	&	1.467	($\pm	0.171$)	&	   Centaur	&	$147_{-17}^{+20}$ &	                 RR		&		&		&		&		\\
88	&	 2015 TP362	&	7.66003	($\pm	0.08789$)	&	0.39502	($\pm	0.09650$)	&	0.42974	($\pm	0.08860$)	&	0.10009	($\pm	0.12963$)	&	0.20468	($\pm	0.14161$)	&	1.005	($\pm	0.110$)	&	1.198	($\pm	0.143$)	&	1.641	($\pm	0.140$)	&	   Detached	&	$128_{-22}^{+35}$ &	                 RR		&		&		&		&		\\
89	&	 2010 RO64	&	5.80413	($\pm	0.12786$)	&	0.28007	($\pm	0.14396$)	&	0.24748	($\pm	0.11718$)	&	0.19143	($\pm	0.17015$)	&	0.24713	($\pm	0.18936$)	&	0.897	($\pm	0.165$)	&	1.222	($\pm	0.195$)	&	1.467	($\pm	0.197$)	&	   Hot Classical	&	$390_{-53}^{+68}$ &	                 RR		&		&		&		&		\\
90	&	 2011 SO277	&	8.63262	($\pm	0.12013$)	&	-0.10799	($\pm	0.11696$)	&	0.08280	($\pm	0.12396$)	&	0.00316	($\pm	0.09721$)	&	-0.01310	($\pm	0.06807$)	&	1.059	($\pm	0.181$)	&	1.256	($\pm	0.174$)	&	1.450	($\pm	0.148$)	&	   SDO	&	$133_{-19}^{+24}$ &	                 RR		&		&		&		&		\\
91	&	 2014 QN441*	&	7.58242	($\pm	0.09016$)	&	0.05794	($\pm	0.06989$)	&	0.10772	($\pm	0.05485$)	&	0.01858	($\pm	0.05602$)	&	0.02517	($\pm	0.03993$)	&	1.019	($\pm	0.110$)	&	1.282	($\pm	0.110$)	&	1.529	($\pm	0.102$)	&	   SDO	&	$214_{-26}^{+34}$ &	                 RR		&		&		&		&		\\
92	&	 2014 XY40*	&	5.91629	($\pm	0.22480$)	&	-0.18751	($\pm	0.24255$)	&	0.02813	($\pm	0.15334$)	&	0.19122	($\pm	0.15498$)	&	0.36161	($\pm	0.18278$)	&	0.967	($\pm	0.262$)	&	1.287	($\pm	0.259$)	&	1.525	($\pm	0.278$)	&	   Hot Classical	&	$376_{-67}^{+87}$ &	                 RR		&		&		&		&		\\
93	&	 2012 VS113*	&	8.07284	($\pm	0.07740$)	&	-0.21159	($\pm	0.06725$)	&	0.03310	($\pm	0.05917$)	&	0.10683	($\pm	0.05374$)	&	0.25136	($\pm	0.06497$)	&	1.029	($\pm	0.102$)	&	1.319	($\pm	0.103$)	&	1.678	($\pm	0.118$)	&	   Resonant (5:2)	&	$101_{-10}^{+12}$ &	                 RR		&		&		&		&		\\
94	&	 2013 RF109	&	7.56425	($\pm	0.05301$)	&	-0.06597	($\pm	0.10076$)	&	0.02492	($\pm	0.06690$)	&	0.04009	($\pm	0.09677$)	&	0.07059	($\pm	0.12774$)	&	1.022	($\pm	0.072$)	&	1.457	($\pm	0.078$)	&	1.619	($\pm	0.123$)	&	   Hot Classical	&	$178_{-18}^{+23}$ &	                 RR		&		&		&		&		\\\\
\hline\\

95	&	 2009 HH36	&	11.13146	($\pm	0.18318$)	&	0.14264	($\pm	0.07521$)	&	0.02633	($\pm	0.07417$)	&	0.00121	($\pm	0.07150$)	&	0.05624	($\pm	0.04031$)	&	0.202	($\pm	0.259$)	&	0.387	($\pm	0.245$)	&	0.654	($\pm	0.208$)	&	   Centaur	&	$30_{-6}^{+8}$ &	                 BB		&		&		&		&		\\
96	&	 2007 UM126	&	10.10765	($\pm	0.28067$)	&	0.28927	($\pm	0.10507$)	&	0.05857	($\pm	0.13264$)	&	0.03437	($\pm	0.11227$)	&	-0.11873	($\pm	0.09047$)	&	-0.096	($\pm	0.467$)	&	0.041	($\pm	0.409$)	&	-0.418	($\pm	0.380$)	&	   Centaur	&	$53_{-10}^{+13}$ &	                 BB		&		&	BR,BB	&		&	BB,BR	\\
97	&	 2013 TU187*	&	8.28484	($\pm	0.11791$)	&	0.06236	($\pm	0.13017$)	&	0.16660	($\pm	0.10203$)	&	-0.27424	($\pm	0.11147$)	&	-0.19154	($\pm	0.07953$)	&	0.671	($\pm	0.163$)	&	0.441	($\pm	0.181$)	&	0.946	($\pm	0.149$)	&	   Resonant (3:2)	&	$115_{-14}^{+18}$ &	                 BB		&		&		&		&		\\
98	&	 2005 SA278	&	6.50807	($\pm	0.07630$)	&	0.40253	($\pm	0.09222$)	&	0.24743	($\pm	0.12244$)	&	0.32233	($\pm	0.10971$)	&	0.07177	($\pm	0.11854$)	&	0.346	($\pm	0.138$)	&	0.743	($\pm	0.131$)	&	0.680	($\pm	0.112$)	&	   SDO	&	$302_{-37}^{+48}$ &	                 BB		&		&		&		&		\\
99	&	 2014 VW37*	&	7.25938	($\pm	0.21646$)	&	0.25285	($\pm	0.15995$)	&	0.04878	($\pm	0.20274$)	&	0.39309	($\pm	0.19429$)	&	-0.26236	($\pm	0.19671$)	&	0.355	($\pm	0.348$)	&	0.958	($\pm	0.338$)	&	0.202	($\pm	0.342$)	&	   Resonant (5:3)	&	$127_{-22}^{+28}$ &	                 BB		&		&		&		&		\\
100	&	 2010 RF64	&	5.77010	($\pm	0.15879$)	&	0.13619	($\pm	0.13357$)	&	0.18416	($\pm	0.12298$)	&	0.31867	($\pm	0.16879$)	&	0.10410	($\pm	0.10467$)	&	0.546	($\pm	0.220$)	&	0.783	($\pm	0.270$)	&	0.481	($\pm	0.203$)	&	   SDO	&	$444_{-70}^{+92}$ &	                 BB		&		&		&		&		\\
101	&	 2014 SR303	&	11.90960	($\pm	0.14104$)	&	0.07776	($\pm	0.06546$)	&	0.01968	($\pm	0.03938$)	&	0.00346	($\pm	0.04986$)	&	0.19640	($\pm	0.08431$)	&	0.440	($\pm	0.182$)	&	0.675	($\pm	0.230$)	&	1.209	($\pm	0.255$)	&	   Centaur	&	$26_{-3}^{+4}$ &	                 BR		&		&		&		&		\\
102	&	 2010 RD188	&	6.34221	($\pm	0.15417$)	&	-0.03667	($\pm	0.13220$)	&	0.08208	($\pm	0.13344$)	&	-0.04117	($\pm	0.24511$)	&	-0.11941	($\pm	0.18924$)	&	0.682	($\pm	0.213$)	&	0.711	($\pm	0.342$)	&	0.690	($\pm	0.280$)	&	   Hot Classical	&	$290_{-44}^{+56}$ &	                 BR		&		&		&		&		\\
103	&	 2013 TD172*	&	7.44839	($\pm	0.10075$)	&	-0.12368	($\pm	0.20436$)	&	-0.07411	($\pm	0.16785$)	&	0.26828	($\pm	0.13348$)	&	-0.05238	($\pm	0.14042$)	&	0.381	($\pm	0.161$)	&	1.078	($\pm	0.153$)	&	0.942	($\pm	0.161$)	&	   SDO	&	$197_{-26}^{+34}$ &	                 BR		&		&		&		&		\\
104	&	 2016 TS94*	&	8.16462	($\pm	0.15059$)	&	-0.05832	($\pm	0.11160$)	&	-0.10245	($\pm	0.13747$)	&	0.24318	($\pm	0.10329$)	&	-0.06073	($\pm	0.14338$)	&	0.415	($\pm	0.252$)	&	1.043	($\pm	0.202$)	&	0.982	($\pm	0.230$)	&	   Resonant (3:2)	&	$114_{-17}^{+21}$ &	                 BR		&		&		&		&		\\
105	&	 2004 TY364	&	4.87563	($\pm	0.12971$)	&	0.26776	($\pm	0.15074$)	&	0.29450	($\pm	0.10991$)	&	-0.11842	($\pm	0.32482$)	&	-0.16583	($\pm	0.39106$)	&	0.872	($\pm	0.165$)	&	0.862	($\pm	0.261$)	&	0.962	($\pm	0.322$)	&	   Hot Classical	&	$527_{-76}^{+82}$ &	                 BR		&		&	IR,RR,BR	&		&	BR,IR	\\
106	&	 2013 TC188	&	7.64033	($\pm	0.09910$)	&	0.13430	($\pm	0.14497$)	&	0.17810	($\pm	0.08920$)	&	-0.03183	($\pm	0.14201$)	&	0.16121	($\pm	0.10944$)	&	0.542	($\pm	0.121$)	&	0.719	($\pm	0.158$)	&	1.124	($\pm	0.144$)	&	   Resonant (9:5)	&	$111_{-12}^{+15}$ &	                 BR		&		&		&		&		\\
107	&	 2014 QR441*	&	6.99793	($\pm	0.16296$)	&	0.27863	($\pm	0.14752$)	&	0.33627	($\pm	0.14317$)	&	0.36988	($\pm	0.14228$)	&	-0.07925	($\pm	0.11648$)	&	0.610	($\pm	0.236$)	&	0.865	($\pm	0.233$)	&	0.467	($\pm	0.212$)	&	   Detached	&	$159_{-33}^{+52}$ &	                 BR		&		&		&		&		\\
108	&	 2015 BP519*	&	4.77642	($\pm	0.47824$)	&	0.43731	($\pm	0.50491$)	&	0.20154	($\pm	0.34963$)	&	0.04958	($\pm	0.41804$)	&	-0.27483	($\pm	0.69193$)	&	0.653	($\pm	0.588$)	&	0.781	($\pm	0.622$)	&	0.608	($\pm	0.846$)	&	   SDO	&	$719_{-208}^{+309}$ &	                 BR		&		&		&		&		\\
109	&	 2008 UA332	&	6.70766	($\pm	0.14577$)	&	0.28098	($\pm	0.17624$)	&	-0.04997	($\pm	0.21521$)	&	-0.08509	($\pm	0.17405$)	&	0.28059	($\pm	0.28039$)	&	0.349	($\pm	0.245$)	&	0.581	($\pm	0.217$)	&	1.235	($\pm	0.271$)	&	   Detached	&	$171_{-35}^{+55}$ &	                 BR		&		&		&		&		\\
110	&	 2015 VJ181*	&	7.54155	($\pm	0.24512$)	&	0.12861	($\pm	0.18374$)	&	-0.01029	($\pm	0.21047$)	&	0.33610	($\pm	0.15660$)	&	0.04686	($\pm	0.19278$)	&	0.438	($\pm	0.373$)	&	1.100	($\pm	0.312$)	&	0.840	($\pm	0.346$)	&	   SDO	&	$191_{-39}^{+52}$ &	                 BR		&		&		&		&		\\
111	&	 2015 RU245	&	10.23198	($\pm	0.06368$)	&	0.05550	($\pm	0.05182$)	&	0.01728	($\pm	0.06598$)	&	0.19885	($\pm	0.07163$)	&	0.07239	($\pm	0.06434$)	&	0.679	($\pm	0.119$)	&	1.327	($\pm	0.117$)	&	1.321	($\pm	0.111$)	&	   SDO	&	$59_{-7}^{+9}$ &	                 IR		&		&		&		&		\\
112	&	 2012 VR113*	&	6.88142	($\pm	0.08057$)	&	0.33389	($\pm	0.08797$)	&	0.19725	($\pm	0.06057$)	&	0.27807	($\pm	0.06483$)	&	0.39552	($\pm	0.06026$)	&	0.594	($\pm	0.101$)	&	0.773	($\pm	0.104$)	&	1.198	($\pm	0.101$)	&	   Resonant (2:1)	&	$159_{-16}^{+19}$ &	                 IR		&		&		&		&		\\
113	&	 2013 VM46*	&	8.87923	($\pm	0.21163$)	&	0.04615	($\pm	0.19270$)	&	-0.03704	($\pm	0.18402$)	&	0.41799	($\pm	0.17639$)	&	-0.13874	($\pm	0.14727$)	&	0.319	($\pm	0.300$)	&	1.174	($\pm	0.291$)	&	1.007	($\pm	0.230$)	&	   SDO	&	$101_{-18}^{+25}$ &	                 IR		&		&		&		&		\\
114	&	 2005 RN43	&	4.22843	($\pm	0.08305$)	&	0.16590	($\pm	0.15250$)	&	-0.38141	($\pm	0.20488$)	&	0.19205	($\pm	0.19586$)	&	0.29326	($\pm	0.18347$)	&	0.415	($\pm	0.156$)	&	1.109	($\pm	0.138$)	&	1.367	($\pm	0.198$)	&	   Hot Classical	&	$639_{-101}^{+99}$ &	                 IR		&		&		&		 &	IR	\\
115	&	 2011 SW281	&	7.01408	($\pm	0.06653$)	&	0.45723	($\pm	0.11308$)	&	0.09904	($\pm	0.13025$)	&	-0.34001	($\pm	0.14568$)	&	0.20054	($\pm	0.11791$)	&	0.715	($\pm	0.115$)	&	0.747	($\pm	0.123$)	&	1.374	($\pm	0.103$)	&	   Cold Classical	&	$169_{-13}^{+14}$ &	                 IR		&		&		&		&		\\
116	&	 2013 SM102*	&	7.96137	($\pm	0.05917$)	&	-0.20272	($\pm	0.08716$)	&	0.20635	($\pm	0.07829$)	&	-0.26886	($\pm	0.10549$)	&	0.03442	($\pm	0.16347$)	&	0.823	($\pm	0.089$)	&	0.809	($\pm	0.100$)	&	1.242	($\pm	0.126$)	&	   Resonant (7:3)	&	$102_{-9}^{+11}$ &	                 IR		&		&		&		&		\\
117	&	 2014 XC48*	&	8.03891	($\pm	0.12644$)	&	0.10661	($\pm	0.12469$)	&	0.00977	($\pm	0.11548$)	&	0.13238	($\pm	0.16405$)	&	-0.38598	($\pm	0.22224$)	&	0.906	($\pm	0.173$)	&	1.329	($\pm	0.218$)	&	0.972	($\pm	0.273$)	&	   SDO	&	$170_{-24}^{+31}$ &	                 IR		&		&		&		&		\\
118	&	 2014 QX510*	&	7.84977	($\pm	0.05919$)	&	0.24343	($\pm	0.09737$)	&	0.40639	($\pm	0.08268$)	&	0.13567	($\pm	0.08815$)	&	-0.22804	($\pm	0.11309$)	&	0.979	($\pm	0.076$)	&	1.253	($\pm	0.081$)	&	1.110	($\pm	0.108$)	&	   Hot Classical	&	$155_{-16}^{+21}$ &	                 IR		&		&		&		&		\\
119	&	 2013 RK109*	&	7.84696	($\pm	0.09301$)	&	-0.23249	($\pm	0.12784$)	&	0.22869	($\pm	0.09491$)	&	-0.04375	($\pm	0.09845$)	&	-0.28238	($\pm	0.10332$)	&	0.938	($\pm	0.125$)	&	0.995	($\pm	0.130$)	&	0.915	($\pm	0.131$)	&	   SDO	&	$186_{-23}^{+30}$ &	                 IR		&		&		&		&		\\
120	&	 2013 TH172*	&	8.43834	($\pm	0.09793$)	&	0.12432	($\pm	0.10832$)	&	-0.18675	($\pm	0.09085$)	&	0.16780	($\pm	0.09010$)	&	0.32608	($\pm	0.18273$)	&	0.549	($\pm	0.123$)	&	1.269	($\pm	0.134$)	&	1.525	($\pm	0.221$)	&	   Resonant (5:3)	&	$77_{-8}^{+10}$ &	                 RR		&		&		&		&		\\
121	&	 2014 QY510*	&	8.42844	($\pm	0.25049$)	&	0.18405	($\pm	0.20338$)	&	0.07023	($\pm	0.08794$)	&	0.12151	($\pm	0.08220$)	&	0.42011	($\pm	0.16101$)	&	0.629	($\pm	0.269$)	&	1.148	($\pm	0.267$)	&	1.738	($\pm	0.314$)	&	   SDO	&	$133_{-26}^{+34}$ &	                 RR		&		&		&		&		\\
122	&	 2006 QQ180	&	7.18241	($\pm	0.17232$)	&	0.33306	($\pm	0.20829$)	&	0.09405	($\pm	0.11516$)	&	0.01374	($\pm	0.08035$)	&	0.27839	($\pm	0.07250$)	&	0.716	($\pm	0.191$)	&	1.083	($\pm	0.187$)	&	1.708	($\pm	0.188$)	&	   Resonant (5:3)	&	$143_{-20}^{+25}$ &	                 RR		&		&		&		&		\\
123	&	 2013 SJ102*	&	8.15772	($\pm	0.06189$)	&	0.21658	($\pm	0.07520$)	&	0.09055	($\pm	0.06163$)	&	0.19530	($\pm	0.07239$)	&	0.19981	($\pm	0.11375$)	&	0.788	($\pm	0.087$)	&	1.259	($\pm	0.091$)	&	1.457	($\pm	0.139$)	&	   Resonant (7:4)	&	$93_{-9}^{+10}$ &	                 RR		&		&		&		&		\\
124	&	 2014 OQ394	&	7.73271	($\pm	0.12962$)	&	-0.17927	($\pm	0.15120$)	&	0.12493	($\pm	0.07998$)	&	0.08763	($\pm	0.07970$)	&	0.04666	($\pm	0.09162$)	&	0.973	($\pm	0.147$)	&	1.156	($\pm	0.148$)	&	1.347	($\pm	0.150$)	&	   Resonant (3:1)	&	$117_{-14}^{+17}$ &	                 RR		&		&		&		&		\\
125	&	 2012 VU113*	&	8.32495	($\pm	0.08192$)	&	0.13496	($\pm	0.07322$)	&	0.03201	($\pm	0.08600$)	&	0.39899	($\pm	0.06626$)	&	0.29868	($\pm	0.06107$)	&	1.031	($\pm	0.107$)	&	0.992	($\pm	0.117$)	&	1.402	($\pm	0.101$)	&	   Resonant (3:2)	&	$122_{-13}^{+16}$ &	                 RR		&		&		&		&		\\
126	&	 2003 QW90	&	6.03464	($\pm	0.15597$)	&	-0.14490	($\pm	0.21634$)	&	0.16077	($\pm	0.25583$)	&	-0.02442	($\pm	0.22887$)	&	0.44590	($\pm	0.21435$)	&	1.035	($\pm	0.225$)	&	1.584	($\pm	0.248$)	&	1.973	($\pm	0.253$)	&	   Resonant (7:4)	&	$361_{-72}^{+97}$ &	                 RR		&		&		&		&	RR	\\
127	&	 2015 RT245	&	7.77257	($\pm	0.09257$)	&	-0.13295	($\pm	0.13054$)	&	0.37514	($\pm	0.10633$)	&	0.10991	($\pm	0.12076$)	&	0.24832	($\pm	0.10972$)	&	1.076	($\pm	0.142$)	&	1.600	($\pm	0.128$)	&	1.663	($\pm	0.132$)	&	   Resonant (9:5)	&	$117_{-13}^{+15}$ &	                 RR		&		&		&		&		\\
128	&	 2012 TC324	&	7.58670	($\pm	0.07077$)	&	-0.01449	($\pm	0.08219$)	&	0.33912	($\pm	0.12069$)	&	0.17521	($\pm	0.08544$)	&	-0.20961	($\pm	0.08168$)	&	1.187	($\pm	0.114$)	&	1.859	($\pm	0.164$)	&	1.638	($\pm	0.114$)	&	   Resonant (5:3)	&	$131_{-13}^{+16}$ &	                 RR		&		&		&		&		\\
129	&	 2002 PA149	&	7.14549	($\pm	0.07824$)	&	-0.01780	($\pm	0.09478$)	&	0.22672	($\pm	0.12006$)	&	0.43168	($\pm	0.24248$)	&	-0.25049	($\pm	0.12383$)	&	1.230	($\pm	0.248$)	&	1.261	($\pm	0.128$)	&	1.117	($\pm	0.129$)	&	   Resonant (7:4)	&	$162_{-19}^{+23}$ &	                 RR		&		&		&		&	RR	\\
130	&	 2005 PU21	&	7.37677	($\pm	0.05876$)	&	-0.02808	($\pm	0.07140$)	&	0.30808	($\pm	0.09069$)	&	0.25046	($\pm	0.08137$)	&	0.09846	($\pm	0.13428$)	&	1.239	($\pm	0.089$)	&	1.613	($\pm	0.092$)	&	1.703	($\pm	0.107$)	&	   SDO	&	$246_{-27}^{+35}$ &	                 RR		&		&		&		&	RR	\\
131	&	 2013 RR98*	&	8.09384	($\pm	0.27116$)	&	0.00245	($\pm	0.21748$)	&	0.20928	($\pm	0.16832$)	&	0.26246	($\pm	0.25017$)	&	0.15989	($\pm	0.20194$)	&	1.297	($\pm	0.411$)	&	1.481	($\pm	0.344$)	&	1.759	($\pm	0.388$)	&	   SDO	&	$179_{-37}^{+51}$ &	                 RR		&		&		&		&		\\
132	&	 2014 RH70*	&	8.57994	($\pm	0.43915$)	&	-0.34780	($\pm	0.32519$)	&	0.15094	($\pm	0.26758$)	&	0.26370	($\pm	0.24346$)	&	0.32660	($\pm	0.13929$)	&	1.330	($\pm	0.548$)	&	1.468	($\pm	0.564$)	&	1.983	($\pm	0.475$)	&	   SDO	&	$144_{-39}^{+57}$ &	                 RR		&		&		&		&		\\
133	&	 2013 TY171*	&	8.41303	($\pm	0.17785$)	&	-0.08261	($\pm	0.31527$)	&	0.04556	($\pm	0.08066$)	&	0.31421	($\pm	0.11230$)	&	0.06254	($\pm	0.08523$)	&	1.340	($\pm	0.198$)	&	1.509	($\pm	0.192$)	&	1.852	($\pm	0.195$)	&	   Resonant (3:2)	&	$125_{-18}^{+23}$ &	                 RR		&		&		&		&		\\
134	&	 2013 WG114*	&	7.74538	($\pm	0.06968$)	&	-0.08350	($\pm	0.11049$)	&	0.29646	($\pm	0.08426$)	&	0.45620	($\pm	0.14208$)	&	0.10434	($\pm	0.09891$)	&	1.347	($\pm	0.146$)	&	1.591	($\pm	0.099$)	&	1.670	($\pm	0.111$)	&	   Resonant (9:5)	&	$126_{-13}^{+15}$ &	                 RR		&		&		&		&		\\
135	&	 2015 TK363	&	8.21054	($\pm	0.12384$)	&	-0.32897	($\pm	0.19124$)	&	-0.02689	($\pm	0.10371$)	&	0.31576	($\pm	0.11834$)	&	-0.34879	($\pm	0.13633$)	&	1.480	($\pm	0.164$)	&	1.538	($\pm	0.152$)	&	1.604	($\pm	0.182$)	&	   Hot Classical	&	$145_{-20}^{+25}$ &	                 RR		&		&		&		&		\\
136	&	 2015 TJ363	&	7.57186	($\pm	0.09089$)	&	-0.15699	($\pm	0.09980$)	&	0.06295	($\pm	0.09142$)	&	0.10119	($\pm	0.10154$)	&	0.23573	($\pm	0.11220$)	&	0.900	($\pm	0.116$)	&	1.162	($\pm	0.123$)	&	1.800	($\pm	0.134$)	&	   Hot Classical	&	$173_{-21}^{+26}$ &	                 RR		&		&		&		&		\\
137	&	 2013 RO109*	&	7.52846	($\pm	0.07390$)	&	0.13707	($\pm	0.10428$)	&	0.43767	($\pm	0.09636$)	&	0.15893	($\pm	0.07503$)	&	0.39126	($\pm	0.09408$)	&	1.144	($\pm	0.115$)	&	1.202	($\pm	0.093$)	&	1.715	($\pm	0.108$)	&	   Resonant (9:5)	&	$133_{-13}^{+16}$ &	                 RR		&		&		&		&		\\
138	&	 2014 VC41*	&	8.73171	($\pm	0.18004$)	&	0.06211	($\pm	0.10961$)	&	0.12607	($\pm	0.12362$)	&	0.26359	($\pm	0.12805$)	&	0.37510	($\pm	0.11819$)	&	0.779	($\pm	0.279$)	&	1.218	($\pm	0.276$)	&	1.620	($\pm	0.261$)	&	   SDO	&	$120_{-20}^{+27}$ &	                 RR		&		&		&		&		\\
139	&	 2001 QO297	&	7.36105	($\pm	0.05795$)	&	0.04855	($\pm	0.08874$)	&	0.18791	($\pm	0.11222$)	&	0.11284	($\pm	0.08301$)	&	0.46892	($\pm	0.11125$)	&	0.939	($\pm	0.096$)	&	1.313	($\pm	0.083$)	&	1.730	($\pm	0.098$)	&	   Cold Classical	&	$151_{-11}^{+12}$ &	                 RR		&		&		&		&		\\
140	&	 2013 RE109*	&	7.18544	($\pm	0.07823$)	&	0.00182	($\pm 0.12441$)	&	-0.03549	($\pm	0.15556$)	&	0.23395	($\pm	0.17616$)	&	0.26907	($\pm	0.09862$)	&	0.931	($\pm	0.130$)	&	1.410	($\pm	0.143$)	&	1.746	($\pm	0.104$)	&	   Resonant (7:4)	&	$149_{-15}^{+19}$ &	                 RR		&		&		&		&		\\
141	&	 2013 RP109*	&	7.58177	($\pm	0.06829$)	&	0.00285	($\pm	0.09160$)	&	-0.07386	($\pm	0.09479$)	&	0.25922	($\pm	0.08660$)	&	0.13102	($\pm	0.09534$)	&	0.841	($\pm	0.107$)	&	1.473	($\pm	0.097$)	&	1.584	($\pm	0.105$)	&	   Resonant (11:6)	&	$122_{-12}^{+14}$ &	                 RR		&		&		&		&		\\
142	&	 2003 QC112	&	9.99768	($\pm	0.11867$)	&	-0.16660	($\pm	0.10936$)	&	0.15659	($\pm	0.09410$)	&	-0.01817	($\pm	0.10714$)	&	0.07303	($\pm	0.13321$)	&	1.167	($\pm	0.169$)	&	1.478	($\pm	0.177$)	&	1.864	($\pm	0.191$)	&	   Centaur	&	$73_{-8}^{+9}$ &	                 RR		&		&		&		&		\\
143	&	 2013 TJ159*	&	7.96805	($\pm	0.07000$)	&	-0.05751	($\pm	0.08132$)	&	-0.17781	($\pm	0.07195$)	&	0.11357	($\pm	0.05653$)	&	0.12205	($\pm	0.11111$)	&	0.905	($\pm	0.095$)	&	1.499	($\pm	0.087$)	&	1.835	($\pm	0.117$)	&	   SDO	&	$175_{-20}^{+26}$ &	                 RR		&		&		&		&		\\
144	&	 2013 VX30*	&	9.51218	($\pm	0.06394$)	&	-0.06209	($\pm	0.05767$)	&	0.02875	($\pm	0.04816$)	&	-0.06744	($\pm	0.04833$)	&	-0.01428	($\pm	0.03865$)	&	1.112	($\pm	0.088$)	&	1.529	($\pm	0.085$)	&	1.810	($\pm	0.082$)	&	   SDO	&	$90_{-10}^{+13}$ &	                 RR		&		&		&		&		\\

\hline
\caption{TNOs and Centaurs from DES photometry that were classified into four groups, from neutral (BB) to very red (RR) colours, following the taxonomic scheme presented by \citet{2005AJ....130.1291B} and \citet{2008ssbn.book..181F} . Column 1: Sequential numbering. From 1 to 94: defining objects as explained in the text. From 95 to 144: remaining ones; column 2: object ID; column 3: absolute magnitudes in the \textit{g} band and respective uncertainties; columns 4 to 7: Phase slopes and uncertainties in the band \textit{g}, \textit{r}, \textit{i} and \textit{z} bands, respectively; columns 8 to 10: colours from absolute magnitudes, and respective uncertainties, used to classify the objects into taxonomy; column 11: dynamical class as provided by SkyBoT; column 12: estimated diameters and respective uncertainties; column 13: taxonomic classification obtained in this work; columns 14 to 17: taxonomy for common objects found in the literature ([1]: \citet{2008ssbn.book..181F}, [2]: \citet{2010A&A...510A..53P}, [3]:\citet{2005AJ....130.1291B}), [4]: \citet{2015Icar..250..482B}. Asterisks represent those objects with initial reported observation by DECam \citep[see also][]{2022ApJS..258...41B}. The discovery credit is given when the object is numbered, as defined by the Minor Planet Center (MPC, See \url{https://www.minorplanetcenter.net/db_search})}.
\label{tab:taxonclass}
\end{longtable}
\end{center}
\end{landscape}
\twocolumn



\bibliographystyle{mnras}
\bibliography{bibliography} 





\bsp	
\label{lastpage}
\end{document}